\documentclass[aps,prd,reprint,amsmath,amssymb,graphicx,longbibliography,floatfix,nofootinbib]{revtex4-1}

\usepackage{bm,amsmath,amssymb,braket,longtable,docmute,xcolor,graphicx,comment,grffile}
\usepackage[caption=false]{subfig}
\usepackage{bbm}
\usepackage[multiple]{footmisc}
\usepackage[utf8x]{inputenc}

\usepackage{mathtools}
\usepackage{bbold}
\usepackage{ulem}

\usepackage{appendix}
\usepackage{float}
\usepackage{slashed}

\usepackage{tikz} 
\usetikzlibrary{patterns}

\definecolor{darkred}{rgb}{0.4,0.0,0.0}
\definecolor{darkgreen}{rgb}{0.0,0.3,0.0}
\definecolor{darkblue}{rgb}{0.0,0.0,0.7}
\usepackage[bookmarks,linktocpage,colorlinks,
linkcolor = darkred,
urlcolor  = darkgreen,
citecolor = darkblue]{hyperref}
\def\beq{\begin{equation}}
  \def\enq{\end{equation}}
\def\zq{${\mathbb Z}_q\ $}

\usepackage{comment}
\definecolor{winered}{rgb}{0.8,0,0}
\definecolor{darkb}{rgb}{0,0,0.8}

\usepackage[english]{babel}

\def\rar{\rightarrow}

\def\zq{${\mathbb Z}_q\ $}

\begin{document}

\title{Clock model interpolation and symmetry breaking in O(2) models}
\author{Leon Hostetler$^{1,2}$}
\author{Jin Zhang$^{3}$}
\email{jin-zhang@uiowa.edu}
\author{Ryo Sakai$^{3}$}
\author{Judah Unmuth-Yockey$^{4}$}
\author{Alexei Bazavov$^{2,1}$}
\author{Yannick Meurice$^{3}$}

\affiliation{$^1$ Department of Physics and Astronomy, Michigan State University, East Lansing, Michigan 48824, USA}
\affiliation{$^2$ Department of Computational Mathematics, Science and Engineering, Michigan State University, East Lansing, Michigan 48824, USA}
\affiliation{$^3$ Department of Physics and Astronomy, The University of Iowa, Iowa City, Iowa 52242, USA }
\affiliation{$^4$ Fermilab, Batavia, Illinois 60510, USA}
\def\lt{\lambda ^t}
\def\note{note}
\def\beq{\begin{equation}}
  \def\enq{\end{equation}}

\date{\today}

\begin{abstract}

Motivated by recent attempts to quantum simulate lattice models with continuous Abelian symmetries using discrete approximations, we define an extended-O(2) model by adding a $\gamma \cos(q\varphi)$ term to the ordinary O(2) model with angular values restricted to a $2\pi$ interval. In the $\gamma \rar \infty$ limit, the model becomes an extended $q$-state clock model that reduces to the ordinary $q$-state clock model when $q$ is an integer and otherwise is a continuation of the clock model for noninteger $q$. By shifting the $2\pi$ integration interval, the number of angles selected can change discontinuously and two cases need to be considered. What we call case $1$ has one more angle than what we call case $2$. We investigate this class of clock models in two space-time dimensions using Monte Carlo and tensor renormalization group methods. Both the specific heat and the magnetic susceptibility show a double-peak structure for fractional $q$. In case $1$, the small-$\beta$ peak is associated with a crossover, and the large-$\beta$ peak is associated with an Ising critical point, while both peaks are crossovers in case $2$. When $q$ is close to an integer by an amount $\Delta q$ and the system is close to the small-$\beta$ Berezinskii-Kosterlitz-Thouless transition, the system has a magnetic susceptibility that scales as $\sim 1 / (\Delta q)^{1 - 1/\delta'}$ with $\delta'$ estimates consistent with the magnetic critical exponent $\delta = 15$. The crossover peak and the Ising critical point move to Berezinskii-Kosterlitz-Thouless transition points with the same power-law scaling. A phase diagram for this model in the $(\beta, q)$ plane is sketched. These results are possibly relevant for configurable Rydberg-atom arrays where the interpolations among phases with discrete symmetries can be achieved by varying  continuously the distances among atoms and the detuning frequency.

\end{abstract}

\maketitle

\section{Introduction}
\label{sec_intro}

In recent years, the idea of using quantum computers or quantum simulation experiments to approach the real-time evolution or the finite-density behavior of lattice models of interest for high-energy physics has gained considerable interest \cite{preskill2018,banulsreview,treview,klco2021,davoudi2020,davoudi2019,bender2018,zohar2015,wiese2013,PhysRevD.100.054505}. As the current noisy intermediate scale quantum (NISQ) devices that are available to implement this research program have a very limited number of quantum computing units, such as qubits, trapped ions or Rydberg atoms, it is essential to optimize the discretization procedure. Starting from the standard Lagrangian formulation of lattice field theory models with continuous field variables, one can either discretize the field variables \cite{lamm2018,klco18,hackett2018} used in the path integral, expand the Boltzmann weights using character expansions \cite{treview,celi2014,Liu:2013nsa}, or use the quantum link method \cite{wiese97,Brower97}.  

Models with continuous Abelian symmetries are of great physical interest. Besides the electromagnetic interactions of charged particles in 3+1 dimensions,  this also includes models where a mass gap is dynamically generated \cite{schwinger62,polyakov1975} or a Berezinskii-Kosterlitz-Thouless (BKT) transition \cite{Berezinskii:1971, Kosterlitz_1973, Kosterlitz_1974} occurs. For models with a $U(1)$ symmetry, the character expansion mentioned above is simply the Fourier series. It has been shown \cite{meurice2019,meurice2020} that the truncation of these series preserves the original symmetry. On the other hand, the \zq clock approximation of the integration over the circle only preserves the \zq discrete subgroup. A recent proposal applies the $\mathbb{Z}_q$ clock approximation to the simulation of the Abelian gauge theory in $2+1$ dimensions, where transformations between the electric representation and the magnetic representation can significantly reduce the required computational resources \cite{Haase2021resourceefficient}. In order to decide how good the \zq approximation is in a variety of situations, it is useful to build a continuous family of models interpolating among the various possibilities. 

In this article, we focus on the case of the O(2) nonlinear sigma model in 1+1 dimensions. This model was key to understanding the BKT transition \cite{Berezinskii:1971, Kosterlitz_1973, Kosterlitz_1974,kogut79} and the corresponding \zq clock model has been studied extensively \cite{PhysRevD.19.3698, PhysRevB.23.1362, PhysRevB.26.6201, PhysRevB.28.5371, Murty:1984, PhysRevB.33.437, Leroyer:1991, PhysRevB.65.184405, PhysRevLett.96.140603, PhysRevE.74.041106, PhysRevE.80.060101, PhysRevE.80.042103, PhysRevE.82.031102, PhysRevE.83.041120, PhysRevE.85.021114, Ortiz:2012, Chen_2017, Chen_2018, PhysRevE.98.032109, Surungan_2019, PhysRevE.101.060105}. We propose to interpolate among these models by starting with the standard O(2) action and introducing a symmetry-breaking term,
\begin{equation}
\Delta S (\gamma,q)=-\gamma \sum_{x} \cos(q\varphi_x).   
\label{eq:dels}
\end{equation}
When $q$ is an integer, if we take the limit $\gamma \rightarrow \infty$, we recover the \zq clock model. For the rest of the discussion, it is important to realize that the O(2)-symmetric  action is $2\pi$-periodic for all the $\varphi_x$ variables. In contrast, $\Delta S$ has a $2\pi/q$ periodicity. When $q$ is an integer, if we apply the shift $q$ times we obtain the periodicity of the O(2) action. In order to interpolate among the clock models, we will consider noninteger values of $q$ while keeping a fixed $\varphi$ interval of length $2\pi$. The model and the effect of the symmetry breaking are discussed in Sec. \ref{sec_model} both in the standard Lagrangian and tensor formulations.

The idea of having a doubly continuous set of models is interesting from a theoretical point of view but also from a quantum simulation point of view. If we attempt to quantum simulate these models using Rydberg atoms as in Refs. \cite{Zhang:2018ufj,51qubits,keesling2019}, 
it is possible to tune the ratio $R_b/a$ of the radius for the Rydberg blockade and the lattice spacing, as well as local chemical potentials  continuously. This allowed interpolations among \zq phases for different integer values of $q$ \cite{keesling2019}. Sequences of clock models also appear in  models for nuclear matter 
when the number of colors is varied \cite{pisarski2021}. 

It is often a difficult task to detect BKT transitions in the quantum Hamiltonian approach, as it is hard to find a good indicator of BKT transitions that has a clear discontinuity, peak or dip. The equivalence of the path integral formulation and  statistical mechanics can be used to access universal features and detect phase transitions in statistical mechanics based on Monte Carlo (MC) simulations and tensor renormalization group (TRG) calculations. The Markov chain MC (MCMC) simulations efficiently explore the typical set of the physical configurations. MC calculations use the universal jump in the helicity modulus \cite{PhysRevA.8.1111} as an indicator for BKT transitions. But ambiguities in the definition of the helicity modulus in the $\mathbb{Z}_5$ clock model can result in controversial conclusions \cite{PhysRevE.82.031102, PhysRevB.88.104427}. The TRG \cite{Levin:2006jai, 2012PhRvB..86d5139X} calculations provide a coarse-grained theory where the size of the lattice spacing doubles at each step. If the truncations performed are under control, one can go to the thermodynamic limit quickly. Calculation of the magnetic susceptibility in the presence of a weak external field is a universal method to detect critical points that can be easily implemented in the TRG. However, it does not show a peak to indicate the large-$\beta$ BKT transition in the five-state clock model \cite{Chen_2018, crossderiv_2020}. The study of the $\gamma \rightarrow \infty$ limit with fractional $q$ not only provides us a clear picture of what phases the symmetry-breaking term will drive the $XY$ model to, paving the way to discussions for the full phase diagram at finite $\gamma$, but also brings us a new tool to detect BKT transitions in $\mathbb{Z}_n$ models. In contrast, the calculation of the specific heat at increasing volume allows us to discriminate between a second-order phase transition---where it diverges logarithmically with the volume in the Ising case---and a BKT transition or a crossover. 

This paper is organized as follows. Section \ref{sec_model} introduces the definition of the extended-O(2) model, the extended $q$-state clock model and thermodynamic quantities. The MC and TRG methods are introduced in Sec. \ref{sec_methods}. The MC method is used to validate the TRG at small volume. The symmetry breaking is discussed in tensorial language. We discuss the behaviors of thermodynamic quantities and point out the crossover peak and the Ising peak in both the specific heat and the magnetic susceptibility in Sec. \ref{sec_results_thermo}. We analyze the crossover peak and the Ising critical point in Secs. \ref{sec_results_peak1} and \ref{sec_results_peak2} respectively. We change the integration interval and discuss a new case in Sec. \ref{sec_results_integration}. The phase diagram in the $(\beta, q)$ plane is sketched in Sec. \ref{sec_results_phasediag}. We summarize our results and give outlooks in Sec. \ref{sec_summary}.

\section{The model}
\label{sec_model}
To define the models that we consider in the following we start with the two-dimensional classical O(2) nonlinear sigma model, or $XY$ model, where the spin degrees of freedom $\vec{\sigma}$ are unit vectors whose possible directions are confined to a plane. They reside on the sites of a two-dimensional lattice of volume $V=N_x\times N_t$ (we prefer to label the second dimension as $t$ to maintain the connection between two-dimensional classical models and 1+1-dimensional quantum field theories). The action is
\begin{align}
	S_{O(2)} = - \sum_{x=1}^{V} \left( \beta \sum_{\mu=1}^{2}\vec{\sigma}_{x}\cdot\vec{\sigma}_{x+\hat\mu}
	+\vec{h}\cdot\vec{\sigma}_x \right)
\end{align}
where the sum on $x$ is over the sites of the two-dimensional lattice, and on $\mu=1$, $2$ over the directions.  The field, $\vec{h}$, is a uniform constant external magnetic field. It is convenient to parametrize the spins $\vec{\sigma}$ with a single angle $\varphi\in[\varphi_0,\varphi_0+2\pi)$. The action then takes the form
\begin{equation}
\label{eq_act_O2}
S_{O(2)} = -\beta \sum_{x,\mu}\cos(\varphi_{x+\hat\mu}-\varphi_{x})
-h \sum_x\cos(\varphi_x-\varphi_h)
\end{equation}
where $h=|\vec{h}|$ and $\varphi_h$ is the direction of the external field that in the absence of other symmetry-breaking terms can be set to zero for convenience.

\begin{figure}
	\centering
	\includegraphics[scale=1]{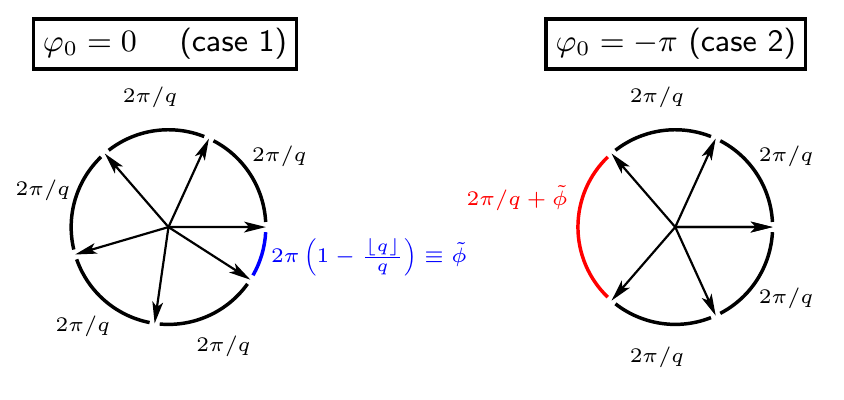}
	\caption{Arrows indicate the allowed spin orientations for the extended $q$-state clock model with the choice $\varphi_0 = 0$ (left) and $\varphi_0 = -\pi$ (right). In this example, $q=5.5$.\label{allowed_angles_example}}
\end{figure}

Next, let us extend the model by introducing a term that can favor certain values of the angle:
\begin{align}
	\label{eq_extO2}
	S_{\mbox{\scriptsize ext-}O(2)} = &-\beta \sum_{x,\mu}\cos(\varphi_{x+\hat\mu}-\varphi_{x})
	\nonumber\\
	&- \gamma\sum_x\cos(q\varphi_x)
	- h \sum_x\cos(\varphi_x-\varphi_h).
\end{align}
We call the model with the action (\ref{eq_extO2}) the ``extended-O(2)'' model.
For integer $q$ the limit $\gamma\to\infty$ forces the spin angles to take the values $\varphi^{(k)}_{x}=2\pi k/q$ with $k\in\mathbb{Z}$. Thus, while $\gamma=0$ corresponds to the O(2) model, $\gamma\to\infty$ corresponds to the $q$-state clock model. The action defined in Eq.~(\ref{eq_extO2}) is also valid for noninteger $q$ and we therefore consider Eq.~(\ref{eq_extO2}) as our definition of the extension of the $q$-state clock model to noninteger $q$ in the $\gamma\to\infty$ limit. In this case the angle $\varphi^{(k)}_{x}$ takes the values 
\begin{equation}
\label{eq_qfrac}
\varphi_0 \leq \varphi^{(k)}_{x}= \frac{2\pi k}{q} < \varphi_0 + 2\pi,
\end{equation}
with $k\in\mathbb{Z}$ and some choice of domain $[\varphi_0,\varphi_0+2\pi)$. By varying $\varphi_0$, we can obtain different sets of angles that are equivalent to either $k=0,1,\ldots,\lfloor q\rfloor$ (case $1$) or $k=0,1,\ldots,\lfloor q\rfloor-1$ (case $2$), since in the $\gamma \rightarrow \infty$ limit---in the absence of an external field---the action only depends on the relative angle between nearest-neighbor sites $\Delta \varphi^{(k)}_{x, \mu} = \varphi^{(k)}_{x+\hat{\mu}} - \varphi^{(k)}_{x}$ (see Appendix \ref{app_angle_cut}). Case $2$ just has one fewer angle than case $1$. As shown in Fig.~\ref{allowed_angles_example}, the angular distance between two adjacent values of $\varphi_{x}$ on a circle takes two values: $2\pi(q - \lfloor q \rfloor) / q < 2\pi/q$ for case $1$, and  $2\pi/q < 2\pi(1 + q - \lfloor q \rfloor) / q$ for case $2$. These values including $0$ have the largest Boltzmann weights in the partition function. The small angular distance is
\begin{equation}
\label{small_angle}
\tilde{\phi} \equiv 2\pi \left(1 - \frac{\lfloor q\rfloor}{q} \right),
\end{equation}
in case $1$ and $2\pi/q$ in case $2$. With the choice $\varphi_0=0$, we have case $1$, while choosing $\varphi_0=-\pi$ is equivalent to case $2$ ($1$) for odd (even) $\lfloor q \rfloor$. At noninteger $q$ the $\mathbb{Z}_q$ symmetry is explicitly broken since the action is not invariant under the operation $k \rightarrow \mod(k+1, \lfloor q \rfloor)$. But there is still a $\mathbb{Z}_2$ symmetry because the action is invariant under the operation $k \rightarrow \lfloor q \rfloor - k$.

We also consider the limit $\gamma\to\infty$ directly by simply restricting the values of the originally continuous angle $\varphi$ to the values given in Eq.~(\ref{eq_qfrac}):
\begin{equation}
\label{eq_extq}
S_{\mbox{\scriptsize ext-}q} = -\beta \sum_{x,\mu}\cos(\varphi^{(k)}_{x+\hat\mu}-\varphi^{(k)}_{x})
-h\sum_x\cos(\varphi^{(k)}_x-\varphi_h).
\end{equation}
We call the model (\ref{eq_extq}) the ``extended $q$-state'' clock model for all values of $q$ and the
``fractional-$q$-state'' clock  model for fractional values of $q$. For integer $q$ the extended $q$-state clock model reduces to the ordinary $q$-state clock model. Numerical results presented in later sections are from the extended $q$-state clock model.

The partition function is
\begin{equation}
\label{eq_Z}
Z=\int\prod_x\frac{d\varphi _x}{2\pi}e^{-S},
\end{equation}
where the $\int$ corresponds to $\int_{\varphi_0}^{\varphi_0+2\pi}$ for the continuous angles in the O(2) and extended-O(2) models and to $\sum_{\varphi^{(k)}}$ for the discrete angles in the extended $q$-state clock model.

With the models defined we turn to observables.
The main observables that we compute to study the critical behavior are the internal energy, magnetization, and their corresponding susceptibilities.  These quantities are defined in the same way for both the continuous and discrete angle cases. The internal energy is defined as
\begin{equation}
\label{eq_internalenergy}
\langle E\rangle= \langle -\sum_{x,\mu} \cos(\varphi_{x+\hat{\mu}} - \varphi_{x}) \rangle = -\frac{\partial}{\partial\beta}\ln Z,
\end{equation}
where $\langle\cdots\rangle$ denotes the ensemble average. The specific heat is
\begin{equation}
\label{eq_specificheat}
C= \frac{-\beta^{2}}{V} 
\frac{\partial \langle E \rangle}{\partial \beta} = \frac{\beta^{2}}{V} (\langle E^{2} \rangle - \langle E \rangle ^{2}). 
\end{equation}
In addition we consider the magnetization.  The magnetization of a given spin configuration is
\begin{equation}
\vec{M}=\sum_x\vec{\sigma}_x
\end{equation}
and the ensemble average is then
\begin{equation}
\label{eq_mag}
\langle\vec{M}\rangle=
\frac{\partial}{\partial\vec{h}}\ln Z=
\left\langle
\sum_x\vec{\sigma}_x
\right\rangle.
\end{equation}
The magnetic susceptibility defined in a manifestly O(2)-invariant way is
\begin{equation}
\label{eq_mag_sus}
\chi_{\vec{M}}=\frac{1}{V}\,
\frac{\partial\langle\vec{M}\rangle}{\partial\vec{h}}=
\frac{1}{V}\left(\langle \vec{M} \cdot \vec{M} \rangle-
\langle\vec{M}\rangle \cdot \langle\vec{M}\rangle \right).
\end{equation}
We note that in Monte Carlo simulations at zero external field in a finite system in the absence
of explicit symmetry breaking terms the definition of the spontaneous magnetization (\ref{eq_mag})
gives $\langle\vec{M}\rangle=0$. In such situations one often resorts to using a proxy observable \cite{Murty:1984, PhysRevB.26.6201, PhysRevB.33.437, PhysRevE.98.032109, PhysRevLett.96.140603, Chen_2018}
\begin{equation}
\label{eq_proxymag}
\langle|\vec{M}|\rangle=
\left\langle\left|
\sum_x\vec{\sigma}_x
\right|\right\rangle
\end{equation}
in place of $\langle\vec{M}\rangle$. The corresponding susceptibility is
\begin{equation}
\label{eq_proxyc}
\chi_{|\vec{M}|}=
\frac{1}{V}\left(\langle |\vec{M}|^2\rangle-
\langle|\vec{M}|\rangle^2\right).
\end{equation}
While one expects that $\langle|\vec{M}|\rangle$ indicates the same critical behavior, in general, $\langle|\vec{M}|\rangle$ is numerically different from $\langle\vec{M}\rangle$ except deep in the ordered phase.  Nevertheless, we expect both definitions of the magnetic susceptibility---Eqs.~\eqref{eq_mag_sus} and~\eqref{eq_proxyc}---possess the same critical behavior, and can be relied upon to extract universal features.  In the next section we detail the methods used to study the observables defined above.

\section{Methods}
\label{sec_methods}

The allowed spin orientations in the extended $q$-state clock model, given by Eq.~(\ref{eq_qfrac}), are discrete, and the model can be studied using a heatbath algorithm. The heatbath algorithm is a MCMC algorithm that drives the lattice toward equilibrium configurations by choosing the new spin at each update according to the probability distribution defined by its neighboring spins. We adapted \textsc{Fortran} code developed by Bernd Berg for the standard Potts model \cite{bergcode}.

Initial exploration of the extended $q$-state clock model was performed via MC on a $4\times 4$ lattice with zero external magnetic field. For this model, the heatbath approach suffers from a slowdown that makes it difficult to study the large-$\beta$ regime already on very small lattices. An alternative approach, the TRG, which does not suffer from this slowdown, was used to study the model on much larger lattices and in the thermodynamic limit. This allows us to perform finite-size scaling and characterize the phase transitions in the extended $q$-state clock models. The TRG results are validated by comparison with exact and Monte Carlo results on small lattices (see Appendix~\ref{sec_appvalidation}). The TRG methods are not exact. Because the bond dimension, $D_{\rm{bond}}$, of the coarse-grained tensor increases exponentially with the renormalization-group (RG) steps, a truncation for $D_{\rm{bond}}$ must be applied to avoid uncontrolled growth of memory needs on classical computers. For noncritical phases, the fixed-point tensor of the RG flow has small $D_{\rm{bond}}$, but a larger bond dimension is needed to have the correct RG flow near the critical point. It has been reported that TRG methods using $D_{\rm{bond}} = 40$ can locate the phase transition point with an error of order $10^{-4}$ for the Ising model \cite{crossderiv_2020}, and of order $10^{-3}$ for the O(2) model \cite{Yu:2013sbi, crossderiv_2020} and the clock models \cite{Chen_2018, crossderiv_2020}.

To perform TRG calculations, we need to express the partition function as a contraction of a tensor network. We rewrite the weight of each link by a singular value decomposition (SVD):
\begin{eqnarray}
\label{eq:linksvd}
e^{\beta \cos \left[\frac{2 \pi}{q}\left(k_{x+\hat{\mu}}-k_{x}\right) \right]} &=& \sum_{n_{x,\mu}=0}^{\lfloor q \rfloor} U_{k_{x+\hat{\mu}} n_{x,\mu}} G_{n_{x,\mu}} V_{k_x n_{x,\mu}}. \nonumber \\
\end{eqnarray}
Then we sum over the original $k$ indices and the partition function can be expressed in the dual space from the expansion in terms of $n$ indices
\begin{eqnarray}
\label{eq:tensorz}
Z = \sum_{\{n\}} \prod_x T_{lrdu},
\end{eqnarray}
where $l = n_{x-\hat{s},s}, r = n_{x,s}, d = n_{x-\hat{\tau},\tau}, u = n_{x,\tau}$ for each site $x$, and the local rank-four tensor is defined as
\begin{eqnarray}
\label{eq:rank4tensor}
\nonumber \hspace*{-0.3cm} T_{lrdu} &=& \sqrt{G_l G_r G_d G_u} C_{lrdu}, \\
\hspace*{-0.3cm} C_{lrdu} &=& \sum_{k_x = 0}^{\lfloor q \rfloor} e^{h \cos\left(\frac{2\pi}{q}k_x - \psi_h \right)} U_{k_x l} V_{k_x r} U_{k_x d} V_{k_x u}.
\end{eqnarray}
Notice that for integral $q$, the matrix $U = V^{-1}$ can be chosen as $U_{k n} = \exp(i 2\pi k n / q)$, then if $h = 0$, the tensor $C_{lrdu}$ becomes a $\delta$-function that gives a $\mathbb{Z}_q$ selection rule for values of $n$:
\begin{eqnarray}
\label{eq:zqselectionrule}
\mod\left(n_{x-\hat{s},s} + n_{x-\hat{\tau},\tau} - n_{x,s} - n_{x,\tau}, q \right) = 0.
\end{eqnarray}

The tensor reformulation of the expectation value of a local observable can be obtained in the same way. For example, the first component of the magnetization is equal to the expectation value of $\cos\left(\varphi \right)$ at an arbitrary site $x_0$, $m_{1} = \langle \cos(\varphi^{(k)}_{x_0}) \rangle$, which can be expressed as
\begin{eqnarray}
\label{eq:expcosphix0}
m_{1} = \frac{\sum_{\{n\}} T^{\mathrm{i}}_{x_0,lrdu} \prod_{x \neq x_0} T_{lrdu}}{\sum_{\{n\}} \prod_{x} T_{lrdu}},
\end{eqnarray}
where $T^{\mathrm{i}}_{x_0,lrdu} = \sqrt{G_l G_r G_d G_u} C^{\mathrm{i}}_{x_0,lrdu}$ is an impure tensor residing at site $x_0$, and
\begin{eqnarray}
\label{eq:Cilrdu}
\nonumber C^{\mathrm{i}}_{x_0,lrdu} = \sum_{k_{x_0} = 0}^{\lfloor q \rfloor} &&  \cos\left(\frac{2\pi}{q} k_{x_0}\right) e^{h \cos\left(\frac{2\pi}{q}k_{x_0} - \psi_h \right)} \\ && \times U_{k_{x_0} l} V_{k_{x_0} r} U_{k_{x_0} d} V_{k_{x_0} u}.
\end{eqnarray}
To compute the internal energy, we need to calculate the expectation values of link interactions $\epsilon_{\mu} = \langle \cos(\varphi^{(k)}_{x_0+\hat{\mu}} - \varphi^{(k)}_{x_0} ) \rangle$. Taking $\mu = s$ as an example, we perform another SVD for the target link
\begin{eqnarray}
\label{eq:x0linksvd}
\nonumber && \cos \left[\frac{2 \pi}{q}\left(k_{x_0+\hat{s}}-k_{x_0}\right)\right] e^{\beta \cos \left[\frac{2 \pi}{q}\left(k_{x_0+\hat{s}}-k_{x_0}\right) \right]} \\ = && \sum_{n_{x_0,s}=0}^{\lfloor q \rfloor} U^{\mathrm{i}}_{k_{x_0+\hat{s}} n_{x_0,s}} G^{\mathrm{i}}_{n_{x_0,s}} V^{\mathrm{i}}_{k_{x_0} n_{x_0,s}},
\end{eqnarray}
and introduce two impure tensors $\tilde{T}^{\mathrm{i}}_{x_0+\hat{s},lrdu}, \tilde{T}^{\mathrm{i}}_{x_0,lrdu}$ residing at nearest neighbor sites $x_0+\hat{s}, x_0$, by replacing $U_{k_{x_0+\hat{s}},l}, G_l$ with $U^{\mathrm{i}}_{k_{x_0+\hat{s}},l}, G^{\mathrm{i}}_l$ and replacing $V_{k_{x_0},r}, G_r$ with $V^{\mathrm{i}}_{k_{x_0},r}, G^{\mathrm{i}}_r$ in Eq.~(\ref{eq:rank4tensor}), respectively. Thus the tensor reformulation of the expectation value of the link interaction is written as
\begin{eqnarray}
\label{eq:expcosdeltaphix0}
\epsilon_{\mu} = \frac{\sum_{\{n\}} \tilde{T}^{\mathrm{i}}_{x_0+\hat{\mu},lrdu} \tilde{T}^{\mathrm{i}}_{x_0,lrdu} \prod_{x \neq x_0+\hat{\mu},x_0} T_{lrdu}}{\sum_{\{n\}} \prod_{x} T_{lrdu}}.
\end{eqnarray}
In the following, we use TRG and higher-order TRG (HOTRG) to contract tenser networks with impure tensors \cite{MORITA201965,PhysRevD.100.054510} up to a volume $V = L^2 = 2^{24} \times 2^{24}$, calculate the first component of the magnetization $\vec{m} = (m_1, m_2)$ and internal energy $E = \epsilon_s + \epsilon_\tau$, and take derivatives of $\vec{m}$ and $E$ with respect to $\vec{h}$ and $\beta$, respectively, to find the magnetic susceptibility and specific heat.\footnote{
	One can assume that the TRG and the HOTRG return the same results if the bond dimension is sufficiently large.
}\textsuperscript{,}\footnote{
	If it is not declared in the main text or in the captions, the bond dimension is set to be sufficiently large so that the outputs converges.
} The locations and heights of the peaks of $\chi_M$ and $C_v$ are obtained via a spline interpolation on datasets with $\Delta \beta = 10^{-3}$. The tensor contraction in HOTRG is performed with \textsc{ITensors Julia Library} \cite{itensor}.

\section{Results}
\label{sec_results}

\subsection{Thermodynamics}
\label{sec_results_thermo}

\begin{figure*}
	\includegraphics[scale=1]{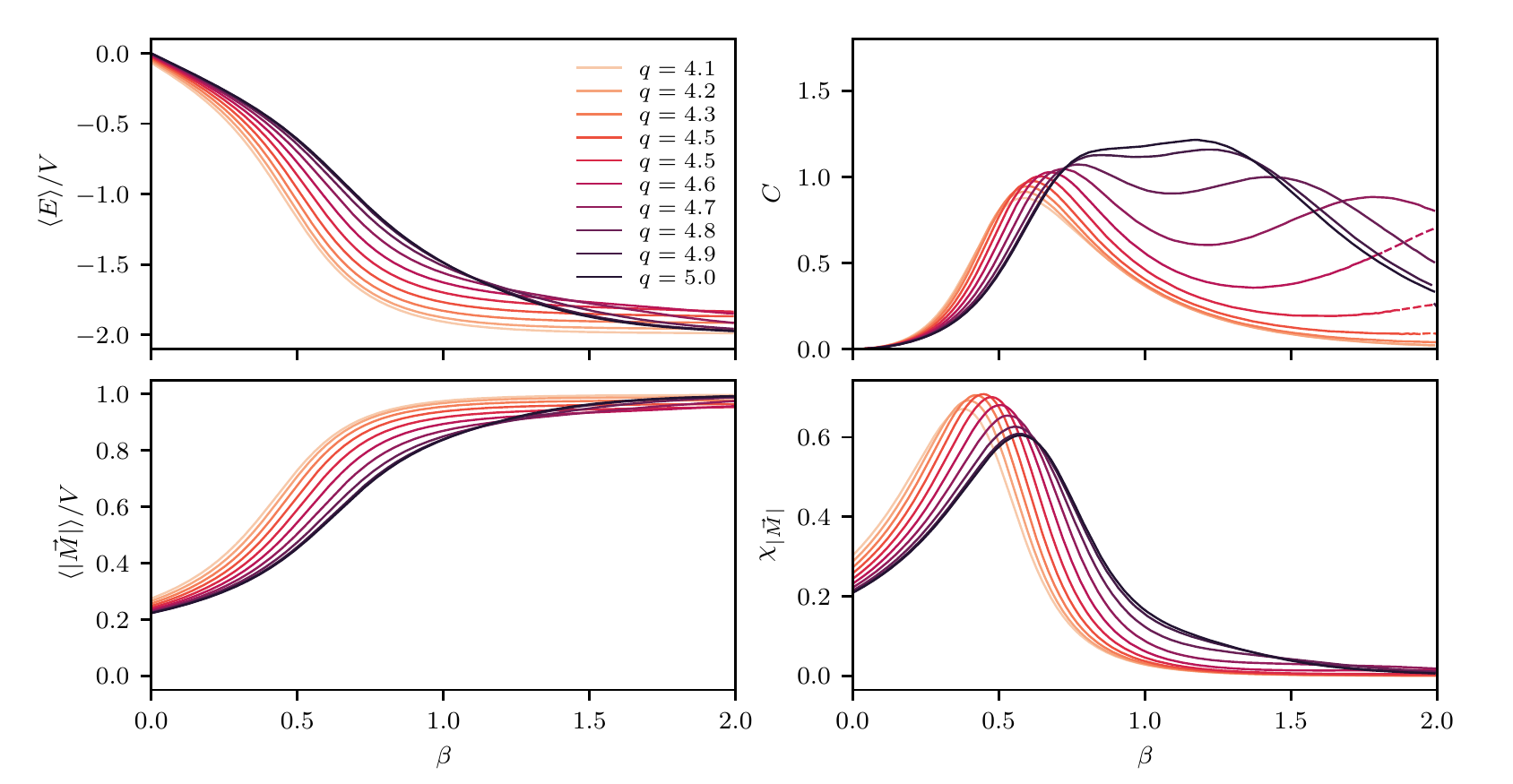}
	\caption{Monte Carlo results for the extended $q$-state clock model on a $4\times 4$ lattice for $4.1 \leq q \leq 5.0$. The top panel shows energy density and specific heat, and the bottom panel shows proxy magnetization and magnetic susceptibility. Statistical error bars are omitted since they are smaller than the line thickness. Dashed lines indicate regions where we have data but we do not have the uncertainty fully under control. \label{zz_fours_all_0004_0p00_23_1_encv}}
\end{figure*}

In the extended $q$-state clock model, there is a $\mathbb{Z}_q$ symmetry when $q \in \mathbb{Z}$. When $q \notin \mathbb{Z}$, this symmetry is explicitly broken. We choose $\varphi_0=0$, so the allowed spin orientations divide the unit circle into $\lceil q\rceil$ arcs of which $\lceil q\rceil-1$ have measure $2\pi/q$. The remainder has measure $\tilde{\phi}$ given in Eq.~(\ref{small_angle}) and illustrated in Fig.~\ref{allowed_angles_example}. There remains a $\mathbb{Z}_2$ symmetry and an approximate $\mathbb{Z}_{\lceil q\rceil}$ symmetry.

Monte Carlo results obtained with a heatbath algorithm on a $4\times 4$ lattice with zero external field are shown for $4.1 \leq q \leq 5.0$ in Fig.~\ref{zz_fours_all_0004_0p00_23_1_encv}. The four panels show the energy density and the specific heat defined in Eqs.~(\ref{eq_internalenergy}) and~(\ref{eq_specificheat}) as well as the proxy magnetization and susceptibility defined in Eqs.~(\ref{eq_proxymag}) and (\ref{eq_proxyc}). For $q = 5$, the energy density is zero at $\beta = 0$ because there is no linear term in the series expansion of the partition function due to the $\mathbb{Z}_{5}$ symmetry. The nonzero energy density at $\beta = 0$ for $q < 5$ is consistent with the explicit $\mathbb{Z}_{5}$ symmetry breaking. As $q \rar 4^{+}$, the stronger symmetry breaking results in a more negative energy density. There is a double-peak structure in the specific heat, where the large-$\beta$ peak moves toward $\beta = \infty$ as $q$ is decreased. The proxy magnetization also increases for smaller values of $q$ with stronger symmetry breaking, and the peak of the magnetic susceptibility moves toward smaller $\beta$ values. Note that the double-peak structure of this proxy magnetic susceptibility will appear at larger system sizes $L \ge 12$ \cite{Murty:1984}. The true magnetic susceptibility with an external field at large volumes will show the double-peak structure (see below). More MC results and additional details are given in Appendix~\ref{sec_appmc}. 

\begin{figure}
	\centering
	\includegraphics[scale=1]{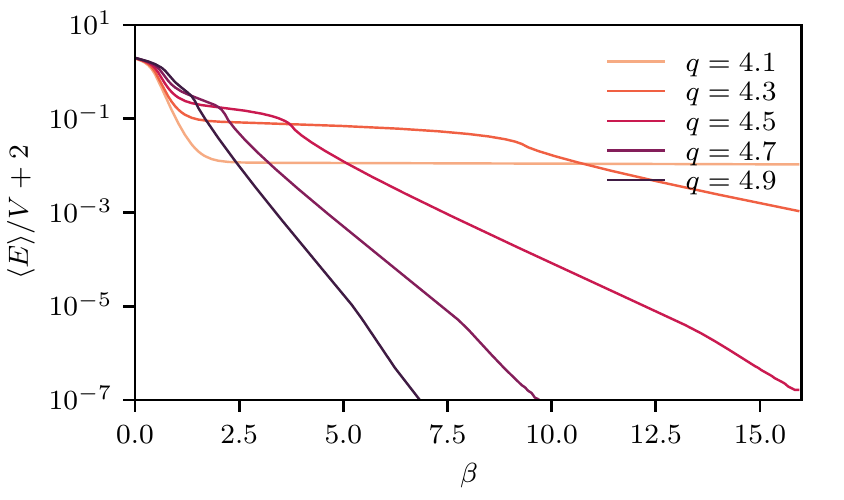}
	\caption{The energy density for the fractional-$q$-state clock model from TRG for $q=4.1$, 4.3, 4.5, 4.7, and 4.9. These results were obtained with a volume $1024 \times 1024$. The energy is shifted vertically by 2 (to make it positive) and plotted on a log scale to better illustrate the difference between the curves. \label{energy_2dclock_q4.1-4.9_Dcut32_iter9}}
\end{figure}

\begin{figure*}
	\centering
	\includegraphics[scale=1]{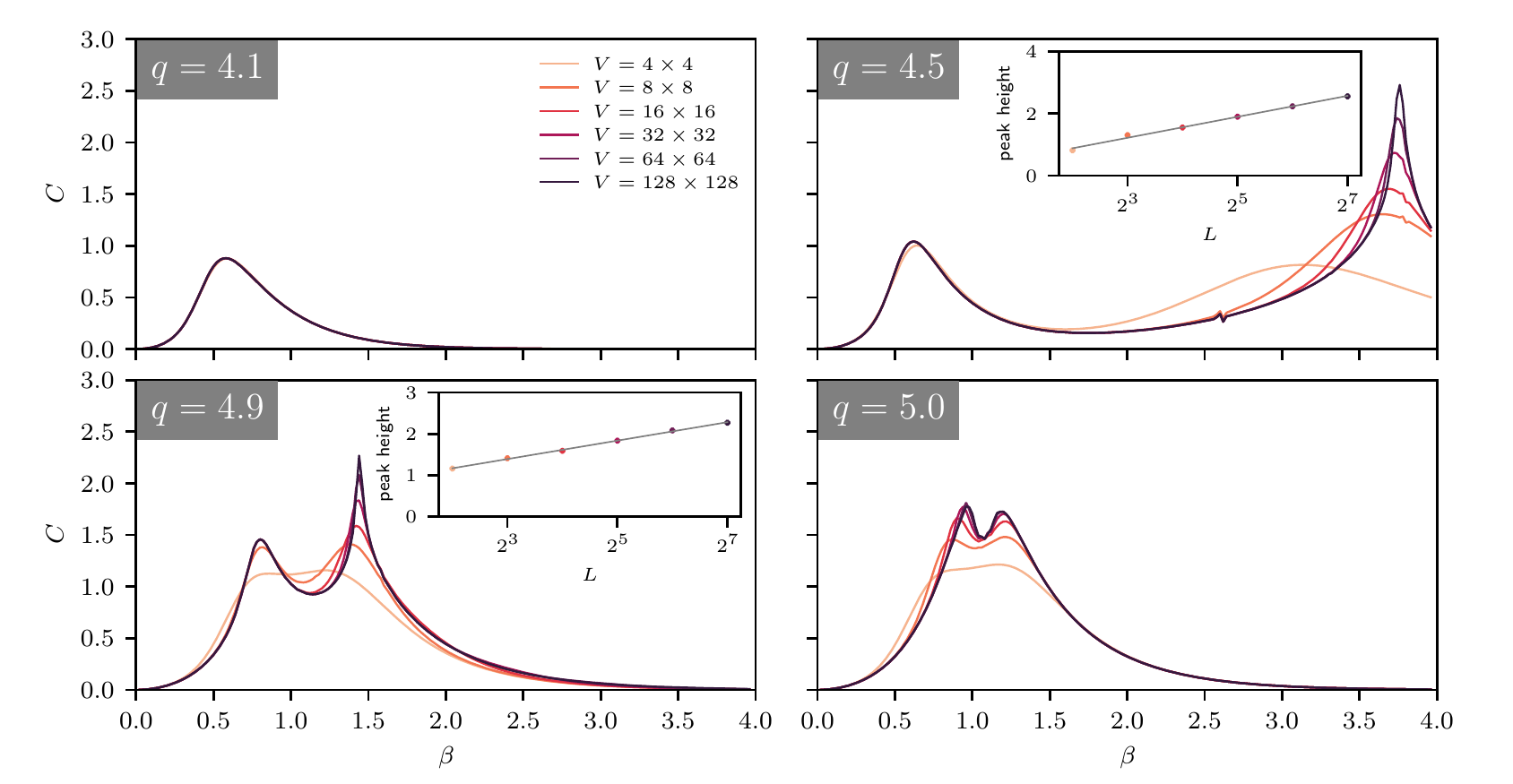}
	\caption{Specific heat of the extended $q$-state clock model from TRG for $q = 4.1$, 4.5, 4.9, and 5.0 at volumes from $4 \times 4$ up to 128 $\times$ 128. All vertical axes use a shared scale, and all horizontal axes use a shared scale. In general, there is a double-peak structure in the specific heat (for $q=4.1$, the second peak is at $\beta \sim 75$). Insets show the height of the second peak versus the linear system size $L = \sqrt{V}$ plotted on a log scale, where the solid line is a linear fit.
	}
	\label{3figs}
\end{figure*}

In Fig.~\ref{energy_2dclock_q4.1-4.9_Dcut32_iter9}, we show the logarithm of the energy density of this model from TRG for $q=4.1$, 4.3, 4.5, 4.7, and 4.9 with $0\leq \beta\leq 16$ and $V = 1024\times 1024$. For large enough $\beta$, we have
\begin{eqnarray}
\label{eq:largebetaZ}
Z &&= e^{2V\beta} \Big[ {\lceil q \rceil} + 2V e^{-4\beta (1-\cos{\tilde{\phi}})} + \cdots \Big],
\end{eqnarray}
so that the energy density converges exponentially with $\beta$. The results in Fig.~\ref{energy_2dclock_q4.1-4.9_Dcut32_iter9} confirm this behavior. We also notice that for smaller $q$, there is a larger range of $\beta$ where the energy density does not change much. In this range, $1-\cos{\tilde{\phi}}$ is close to zero for $q$ close to $4$ from above, and the inverse temperature $\beta$ is large enough that terms containing larger angular distances are negligible, but $\beta$ is still not large enough to change the values of $\exp[-\beta (1-\cos{\tilde{\phi}})]$ significantly from $1$ and ignore higher orders. The result for $q = 4.1$ shown in Fig.~\ref{energy_2dclock_q4.1-4.9_Dcut32_iter9} indicates that the specific heat is almost zero for $\beta > 2.5$, which is confirmed in Fig.~\ref{3figs}.

In Fig.~\ref{3figs}, we show the specific heat for $q = 4.1$, 4.5, 4.9, and 5.0 at volumes ranging from $4\times 4$ to $128\times 128$. For generic $q$, there are two peaks in the specific heat.\footnote{There is only a single peak for $q=2,3,4$ and for fractional $q$ just below these integers.} In Fig.~\ref{3figs} we see only a single peak for $q = 4.1$ since the second peak is at much larger $\beta$. For $q\notin\mathbb{Z}$ and not too close to $5$, the first peak shows little or no dependence on volume. The second peak grows logarithmically with volume, as shown in the insets for $q = 4.5, 4.9$. This is in contrast to the integer case $q=5$ where there are two BKT transitions and both peaks show little dependence on volume for lattice sizes larger than $32 \times 32$. Because the specific heat is the second-order derivative of free energy, the results in Fig.~\ref{3figs} indicate that the first peak is associated with either a crossover or a phase transition with an order larger than $2$, and the second peak is associated with a second-order phase transition. To conclusively characterize the phase transitions, if any, associated with these two peaks in the fractional-$q$-state clock model, we study the magnetic susceptibility in the next two subsections.

We find that the thermodynamic curves vary smoothly for $n<q\leq n+1$ where $n$ is an integer. When $q$ is taken slightly larger than $n$ from below, these curves change abruptly since an additional degree of freedom is introduced. The specific heat exhibits a double-peak structure with the second peak at very large $\beta$. As $q$ is increased further, this second peak moves toward small $\beta$, until at $q=n+1$, the thermodynamic curves of the integer-$(n+1)$-state clock model are recovered.

\begin{figure}
	\centering
	\centering
	\includegraphics[scale=1]{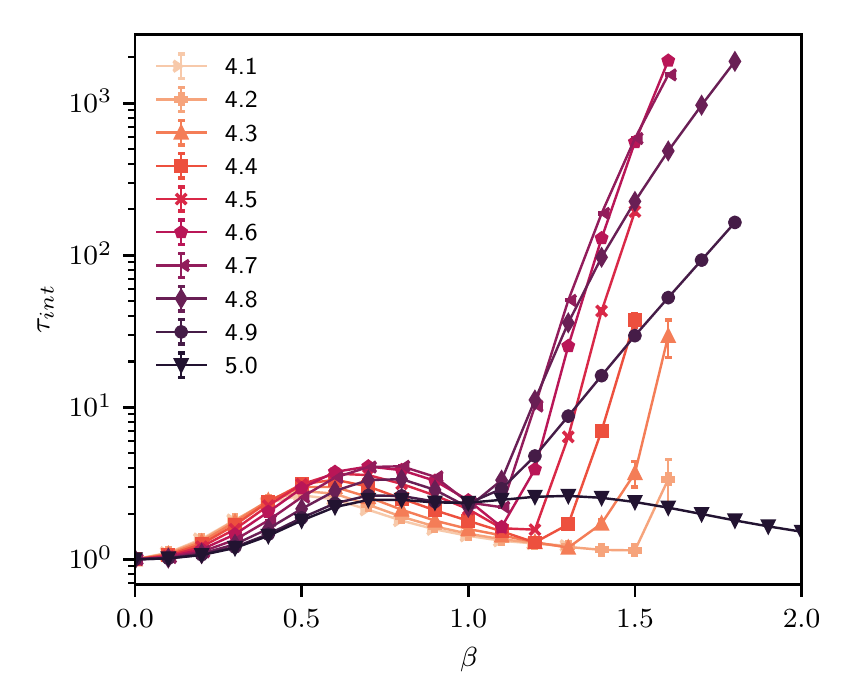} 
	\caption{The integrated autocorrelation time of the energy density for several $q$ on a $4 \times 4$ lattice using a heatbath algorithm. At large $\beta$, the integrated autocorrelation time $\tau_{int}$ grows abruptly when $q \notin\mathbb{Z}$. Note the log scale on the vertical axis. Connecting lines are included to guide the eyes. \label{zz_all_all_0004_0p00_var_1_ac2}}
\end{figure}

In the small-$\beta$ (high temperature) regime, all allowed angles are nearly equally accessible, and the model behaves approximately like a ${\lceil q\rceil}$-state clock model. The model is dominated by the approximate $\mathbb{Z}_{\lceil q\rceil}$ symmetry, and there is a peak in the specific heat. In Fig.~\ref{zz_all_all_0004_0p00_var_1_ac2}, we show that at intermediate $\beta$, an explosion of the integrated autocorrelation time of the energy is observed in the MC simulation as the model quickly reduces to a rescaled Ising model. At large beta, the configuration space separates into thermodynamically distinct sectors, and the Markov chain has trouble adequately sampling both sectors. This is discussed further in Appendix~\ref{sec_appmc}. At large-$\beta$, spin flips across the small angular distance $\tilde{\phi}$ are strongly favored relative to spin flips across the other distances. Thus, in the large-$\beta$ regime, the model behaves as a rescaled Ising model. The existence of an Ising critical point is conclusively established via TRG in Sec.~\ref{sec_results_peak2}.

\begin{figure*}
	\centering
	\includegraphics[scale=1]{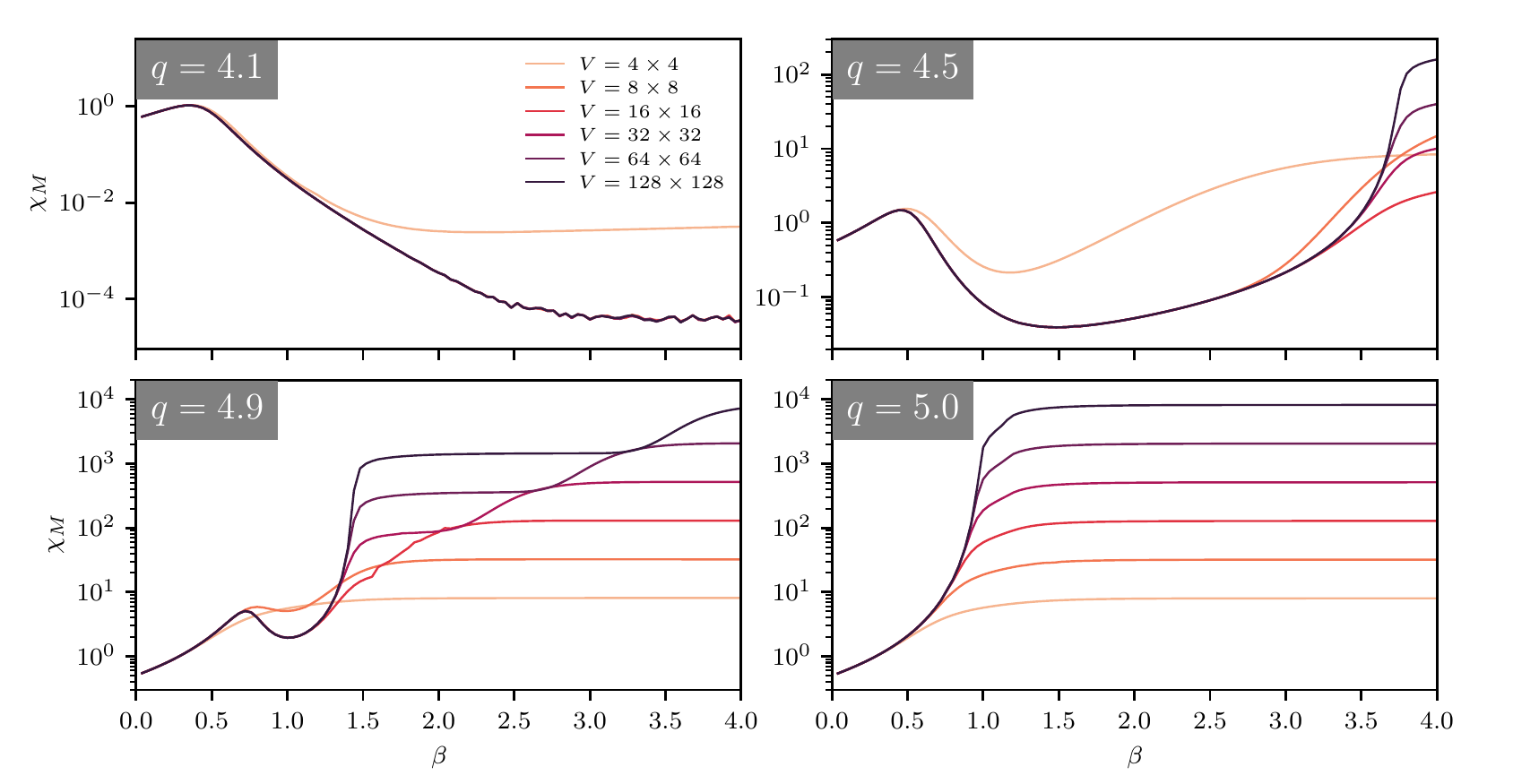}
	\caption{\label{fig:finitesizemsus4qs} The magnetic susceptibility as a function of $\beta$ for finite volumes. $D_{\rm{bond}}=40$. }
\end{figure*}

\begin{figure}
	\includegraphics[width=0.48\textwidth]{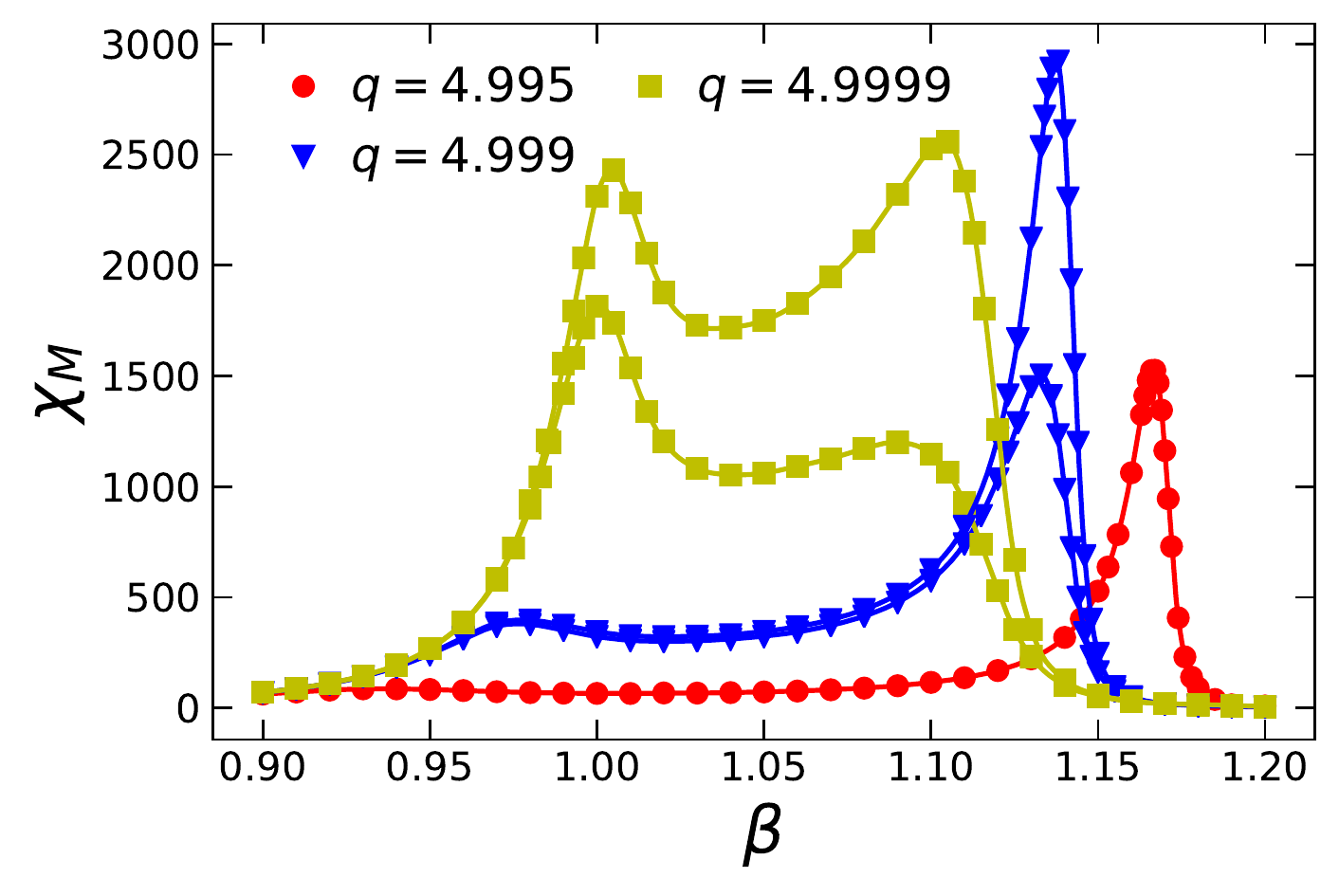}
	\caption{\label{fig:3q5mmsus2peaks} The magnetic susceptibility as a function of $\beta$ at $V = 2^{24}\times 2^{24}$. For $q = 4.995$, the external field is $h = 4 \times 10^{-5}$. For $q = 4.999, 4.9999$, the external field is $h = 4 \times 10^{-5}$ for the lower curve and $h = 2 \times 10^{-5}$ for the upper curve. }
\end{figure}

We next present our TRG results and discuss the phase transitions in the fractional-$q$-state clock model in the rest of this section. We first present the results for the magnetic susceptibility without an external field at small volumes in Fig.~\ref{fig:finitesizemsus4qs}. For $q < 5$, there is a small-$\beta$ peak converging quickly with volume, which means the peak is associated with a crossover. As $q$ is increased, there is a high plateau moving toward small $\beta$. The height of the plateau increases with volume as a power law (notice the logarithmic scale in the $y$-axis). The divergent plateau signals a phase transition. As there is no spontaneous symmetry breaking in small volumes, the response of the system to an external field remains high at low temperatures, so we cannot see the transition from a critical phase to the symmetry-breaking phase where there is a small magnetic susceptibility based on these results for $h = 0$ and small volumes. But notice that for fractional $q$ like $q = 4.9$, there is a higher plateau after the first one for each volume. This is due to an ``approximate $\mathbb{Z}_{\lceil q \rceil}$ symmetry breaking" after a $\mathbb{Z}_2$ symmetry breaking. This ``approximate $\mathbb{Z}_{\lceil q \rceil}$ symmetry breaking" is not a true phase transition so it moves to $\beta = \infty$ in the thermodynamic limit as shown in the results for $q = 4.9$. There is only one $\mathbb{Z}_5$ symmetry breaking for $q = 5$ so there is a single plateau for each volume. We will use the magnetic susceptibility with a weak external field in the thermodynamic limit to detect the phase transitions.

In Fig.~\ref{fig:3q5mmsus2peaks} we present the magnetic susceptibility in the thermodynamic limit as a function of $\beta$ for $q = 4.995, 4.999, 4.9999$ with a small external field $h = 4\times 10^{-5}, 2\times 10^{-5}$. The height of the large-$\beta$ peak is large for all three values of $q$, indicating a phase transition near this peak. The small-$\beta$ peak for $q = 4.995, h = 4\times 10^{-5}$ is invisible because it is very small. For $q = 4.999, h = 4\times 10^{-5}$, closer to $5$, the small-$\beta$ peak is higher, and the large-$\beta$ peak moves toward the large-$\beta$ BKT transition point $\beta^{\mathrm{BKT}}_{q=5,c2}$ for $q = 5$. When the external field is decreased to $h = 2\times 10^{-5}$, the large-$\beta$ peak height is almost doubled, while the small-$\beta$ peak height does not change, which means there is no phase transition near the small-$\beta$ peak. For $q = 4.9999$ (closer to 5) with $h = 4\times 10^{-5}$, one can see that the small-$\beta$ peak becomes higher than the large-$\beta$ one, and the large-$\beta$ peak is fading away, which is consistent with the results in Refs.~\cite{Chen_2018, crossderiv_2020} for the five-state clock model. When the external field is decreased to $h = 2\times 10^{-5}$, the large-$\beta$ peak grows a much larger amount of height than the small-$\beta$ one does and becomes higher than the small-$\beta$ one. We have confirmed that the small-$\beta$ peak will eventually converge for small enough $h$. All these behaviors are evidence that for all fractional $q > 4$, the small-$\beta$ peak in $\chi_M$ does not diverge, and the large-$\beta$ peak diverges at $h = 0$ indicating a phase transition.

In the following, we discuss the behavior of the two peaks in the thermodynamic limit in detail. The main observations are that for $\varphi_0 = 0$ and $q > 4$, there are two peaks in the specific heat and the magnetic susceptibility, the small-$\beta$ one is finite and is associated with a crossover, and the large-$\beta$ one diverges which is characteristic of an Ising critical point. When $q$ is approaching an integer from below, the height of the crossover peak of the magnetic susceptibility diverges as $\lceil q \rceil - q$ goes to zero with a power law:
\begin{eqnarray}
\label{eq:chimpowerlaw}
\chi_M^* \sim \left(\lceil q \rceil - q \right)^{-y}.
\end{eqnarray}
We can formulate the scaling hypothesis with $\Delta q = \lceil q \rceil -q$ and $h$,
\begin{eqnarray}
\label{eq:scalinghypothesis}
f\left(\lambda^{p}\Delta q, \lambda^{r} h \right) = \lambda^d f\left(\Delta q, h \right),
\end{eqnarray}
where $\lambda$ parametrizes a scale transformation, $p$ and $r$ are the scaling dimensions, and $d = 2$ in two-dimensional space. Notice that the reduced temperature should not enter the homogeneous function independently because there is an essential singularity in the correlation length as a function of temperature for BKT transitions. We assume Eq. \eqref{eq:scalinghypothesis} holds for any critical temperature, in particular, the BKT crossover peak position $\beta^{\rm{BKT}}(\Delta q, h)$, which is a power-law function of $\Delta q$ and $h$, considered in the following calculations. Then the magnetization and the magnetic susceptibility satisfy the following relations:
\begin{eqnarray}
\label{eq:magsusscaling}
\lambda^{r} M\left(\lambda^p \Delta q, \lambda^r h \right) &=& \lambda^d M\left(\Delta q, h \right), \\
\lambda^{2r} \chi_M\left(\lambda^p \Delta q, \lambda^r h \right) &=& \lambda^d \chi_M\left(\Delta q, h \right),
\end{eqnarray}
from which one can obtain
\begin{eqnarray}
\label{eq:magsuspowerlaw}
M &\sim& h^{\frac{1}{\delta}} \sim \left(\Delta q \right)^{\frac{1}{\delta'}}, \\
\chi_M &\sim& h^{-1 + \frac{1}{\delta}} \sim \left(\Delta q \right)^{ - \frac{\delta-1}{\delta'}},
\end{eqnarray}
where $1/\delta = (d-r)/r$, $1/\delta' = (d-r)/p$. We then have $\delta' = (\delta-1)/y = 14/y$, where we have used the fact that $\delta = 15$ for BKT transitions. Note that the expansion of the action to the first order in $\Delta q$ is \begin{eqnarray}
\label{eq:expandH}
S_{\mbox{\scriptsize ext-}q} = S_{\mathbb{Z}_{\lceil q \rceil}} + \beta \frac{\Delta q}{q} \Delta \varphi_{x,\mu} \sin(\Delta \varphi_{x,\mu}),
\end{eqnarray}
where $S_{\mathbb{Z}_{\lceil q \rceil}}$ is the action for the integer-$q$ clock model. The perturbation term that breaks the $\mathbb{Z}_{\lceil q \rceil}$ symmetry has a very different form from the one for the external field $h \cos(\varphi_x)$. We numerically show in the following that $\delta'$ is equal to the magnetic critical exponent $\delta = 15$ for the BKT transitions and Ising critical points in two dimensions. Thus if we define a new susceptibility as $\partial M / \partial \Delta q$, which should scale as $(\Delta q)^{-1 + 1/\delta'}$, the exponent $y' = 1 - 1/\delta' = 14/15$, still the same as $y$. However, calculating $\delta'$ from $y'$ requires higher accuracy since $\delta' = 1 / (1-y')$ where one significant digit is subtracted in the denominator. Both the peak position of the crossover and the Ising critical point go to the two BKT transition points at integer $q$ with the same power law, which provides us a new way to locate the phase transitions in models with $\mathbb{Z}_n$ symmetry. However, for $\varphi_0 = -\pi$ and odd $\lfloor q \rfloor$, both peaks of $\chi_M$ are finite for fractional $q$ so there are no critical points. For $q \rightarrow 5^{+}$, only the small-$\beta$ peak can be used to extract the BKT transition of $q = 5$ since the large-$\beta$ peak fades away.

\subsection{Small-\texorpdfstring{$\beta$}{beta} peak: Crossover}
\label{sec_results_peak1}

\begin{figure}
	\includegraphics[width=0.48\textwidth]{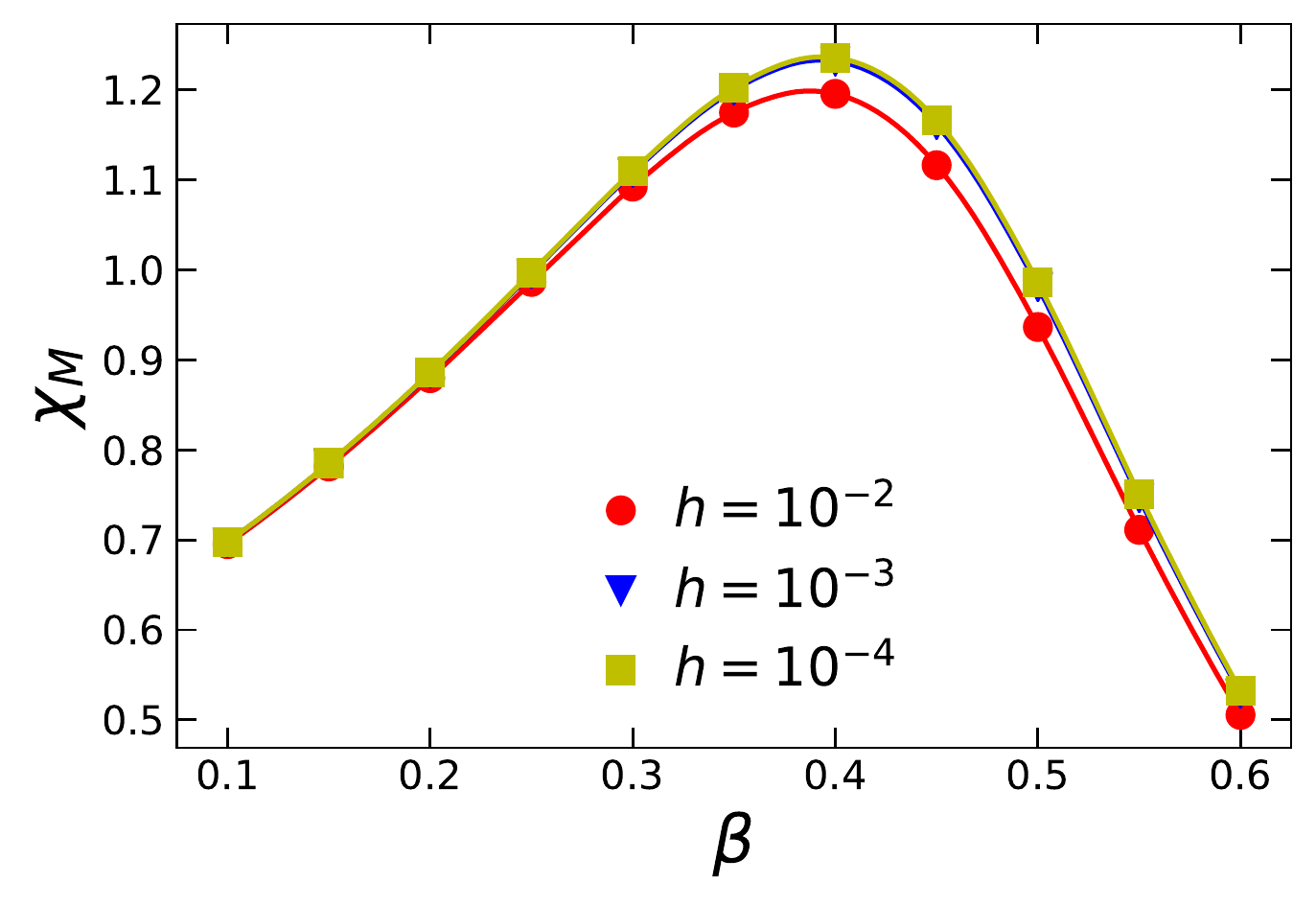}
	\caption{\label{fig:q4p3msus1stpeak} The small-$\beta$ peak of magnetic susceptibility as a function of $\beta$ for $q=4.3$. The peak height converges when the external field is taken to zero.}
\end{figure}

We have shown in Fig.~\ref{fig:3q5mmsus2peaks} that for $q = 4.999$, the height of the small-$\beta$ peak converges to about $400$ for small enough external field. The dependence of the peak height on the external field is larger for values of $q$ closer to an integer from below. In Fig.~\ref{fig:q4p3msus1stpeak}, we show another example for $q = 4.3$. One can see that the $h$ dependence of the peak height is much smaller. The peak height at $h = 10^{-2}$ differs from the value in the $h \rightarrow 0$ limit by only about $0.04$. As the external field is decreased, the peak height converges to a constant $\chi_M \approx 1.2$, implying that there is no phase transition around this peak.  This is true for all fractional $q$. Thus, for fractional $q$, the first peak in the specific heat is associated with a crossover rather than a true phase transition. As $q$ approaches $5$ from below, we expect the small-$\beta$ peak height to diverge because there is a BKT transition for $q = 5$, and we expect the location of the peak to go to the small-$\beta$ BKT transition point. 

\begin{figure}
	\includegraphics[width=0.48\textwidth]{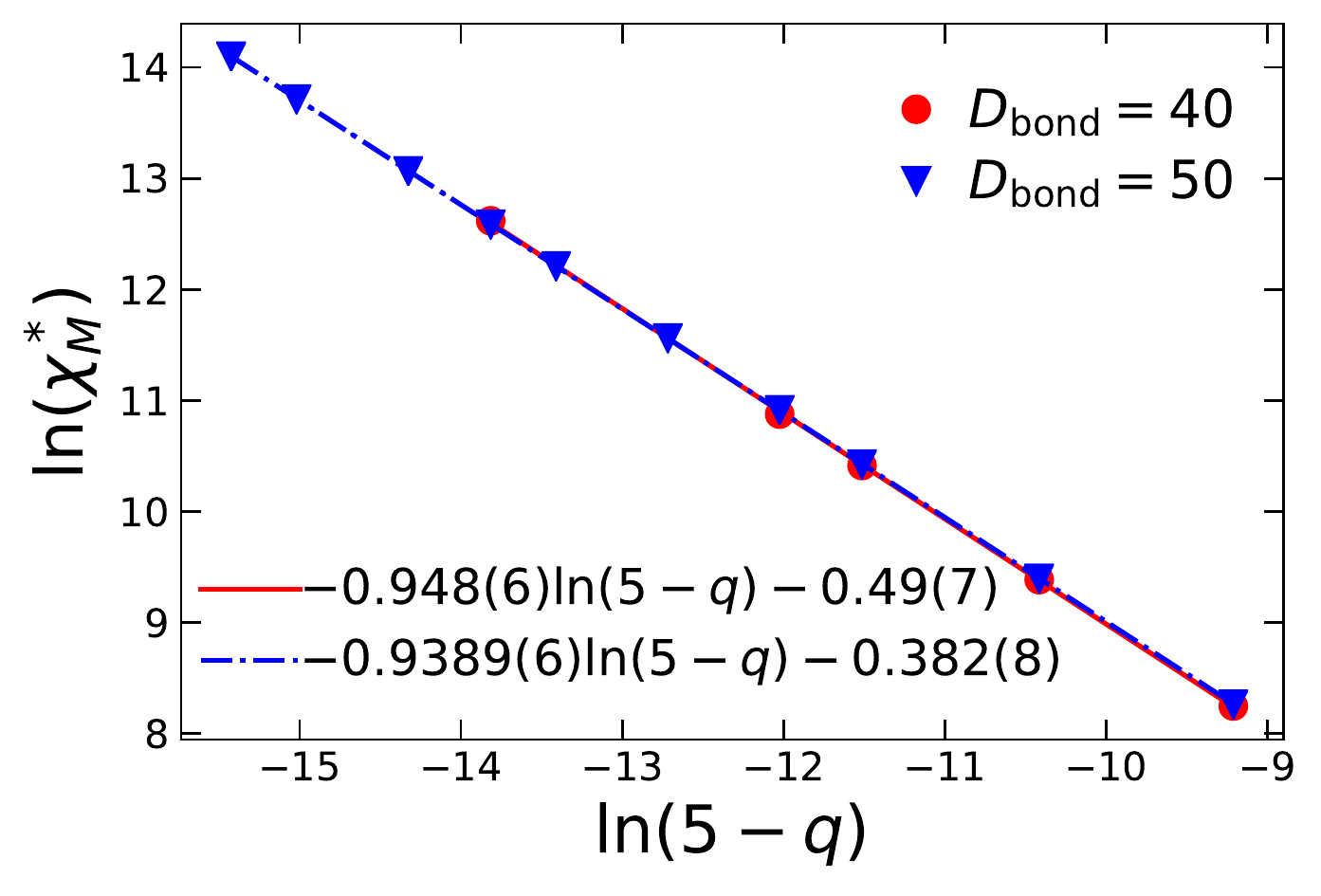}
	\caption{\label{fig:5mqchimheight} The log-log plot of the maximal value of the magnetic susceptibility $\chi_M$ for the small-$\beta$ peak as a function of $5-q$. The peak height diverges with a power law when $q\rar 5^{-}$.}
\end{figure}

\begin{figure}
	\includegraphics[width=0.48\textwidth]{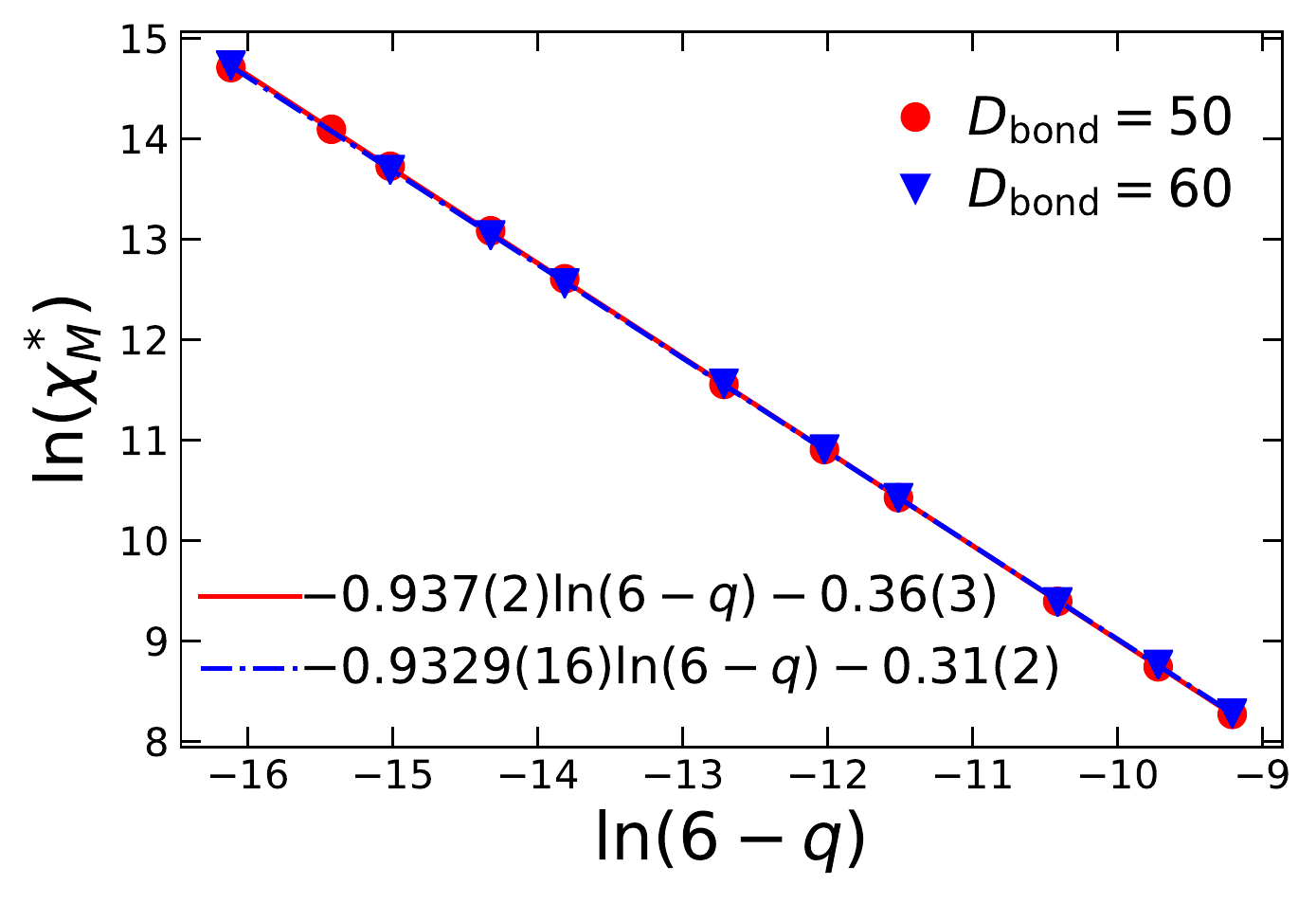}
	\caption{\label{fig:6mqchimheight} Same as Fig.~\ref{fig:5mqchimheight}, but for $q \rar 6^{-}$.}
\end{figure}

To check this, we calculate the converged peak height $\chi^*_M$ of the magnetic susceptibility for values of $5-q \le 10^{-4}$ with $D_{\rm{bond}} = 40$ and $D_{\rm{bond}} = 50$, where we use $d h = 10^{-10}$ in the numerical differentiation. We plot $\chi^*_M$ versus $5-q$ in Fig.~\ref{fig:5mqchimheight}, where we see that the dependence of the peak height on $D_{\rm{bond}}$ is very small for $D_{\rm{bond}} \ge 40$. Applying a linear fit to $\ln(\chi_M^*)$ versus $\ln(5-q)$ shows that the peak height diverges as $q \rightarrow 5^{-}$ with a power law 
\begin{eqnarray}
\label{eq:chimto5powerlawdbond4050}
\chi_M^* \sim 
\begin{cases}
(5-q)^{-0.948(6)} & \text{if $D_{\rm{bond}} = 40$} \\
(5-q)^{-0.9389(6)} & \text{if $D_{\rm{bond}} = 50$},
\end{cases}
\end{eqnarray}
from which we obtain $\delta' = 14.77(9)$ for $D_{\rm{bond}} = 40$ and $\delta' = 14.911(10)$ for $D_{\rm{bond}} = 50$. The value of $\delta'$ is close to the magnetic critical exponent $\delta=15$ for BKT transitions and Ising critical points in two dimensions, but now $5-q$ (rather than the external field $h$) is playing the role of the symmetry-breaking parameter. Since both $h$ and $5-q$ break the $\mathbb{Z}_5$ symmetry to a $\mathbb{Z}_2$ symmetry, the agreement on the critical exponents is reasonable. The value of $\delta'$ is also checked for $q \rightarrow 6^{-}$ in Fig.~\ref{fig:6mqchimheight}, where a larger $D_{\rm{bond}}$ is applied in HOTRG. In this case, the linear fit gives 
\begin{eqnarray}
\label{eq:chimto6powerlawdbond5060}
\chi_M^* \sim 
\begin{cases}
(6-q)^{-0.937(2)} & \text{if $D_{\rm{bond}} = 50$} \\
(6-q)^{-0.9329(16)} & \text{if $D_{\rm{bond}} = 60$},
\end{cases}
\end{eqnarray}
from which we obtain $\delta' = 14.94(3)$ with $D_{\rm{bond}} = 50$ and $\delta' = 15.007(26)$ with $D_{\rm{bond}} = 60$. Within uncertainty, the results of $q \rightarrow 6^{-}, D_{\rm{bond}} = 60$ agree perfectly with the value of the magnetic critical exponent $\delta = 15$.

\begin{figure}
	\includegraphics[width=0.48\textwidth]{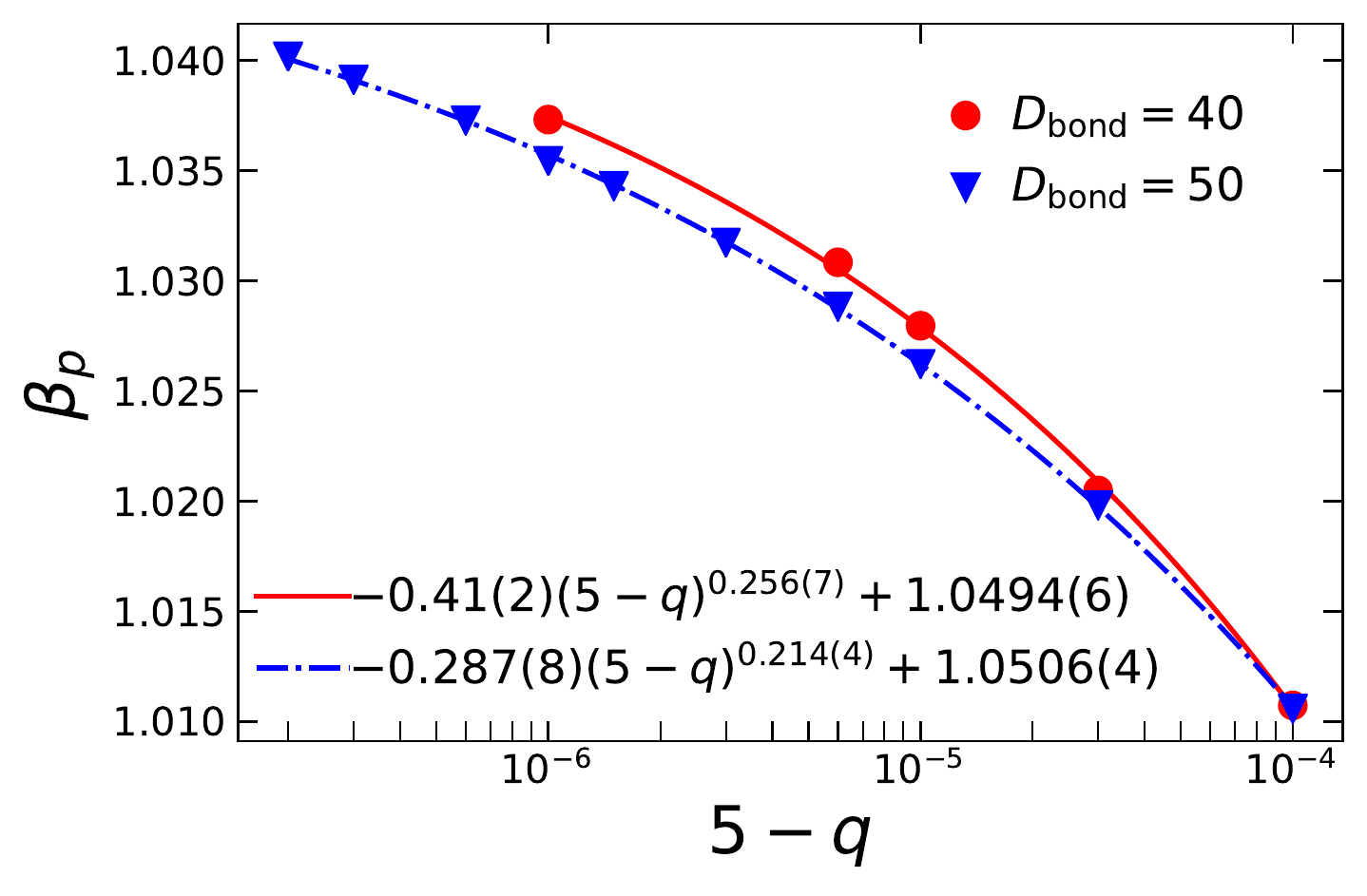}
	\caption{\label{fig:5mqbetapextrapd50} The power-law extrapolation of the small-$\beta$ peak position to $q = 5$ from below. Here the extrapolation gives $\beta_c = 1.0506(4)$.}
\end{figure}

\begin{figure}
	\includegraphics[width=0.48\textwidth]{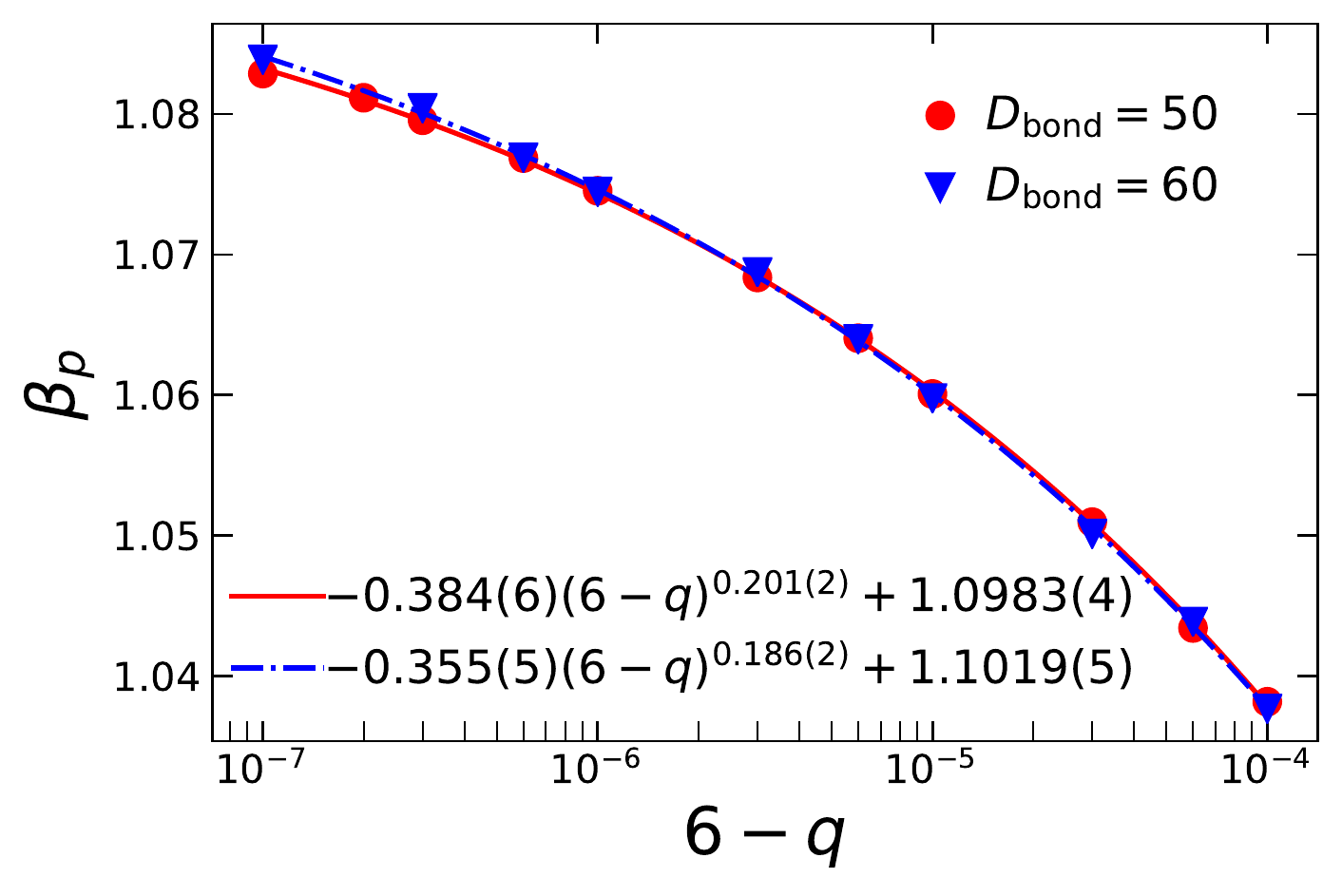}
	\caption{\label{fig:6mqbetapextrapd5060} Same as Fig.~\ref{fig:5mqbetapextrapd50}, but for $q \rar 6^{-}$.}
\end{figure}

The location of the converged peak $\beta_p$ as a function of $5-q$ is depicted in Fig.~\ref{fig:5mqbetapextrapd50}. The discrepancy of the peak positions between $D_{\rm{bond}} = 40$ and $D_{\rm{bond}} = 50$ is invisible at $5-q = 10^{-4}$ and increases as $5-q$ is decreased, which is reasonable because a larger bond dimension is needed for systems closer to a critical point. The overall discrepancy is small. We extrapolate the peak position to $q = 5$ with a power law and obtain the small-$\beta$ BKT transition point $\beta^{\mathrm{BKT}}_{q=5,c1} = 1.0494(6)$ with $D_{\rm{bond}}=40$ and $\beta^{\mathrm{BKT}}_{q=5,c1} = 1.0506(4)$ with $D_{\rm{bond}}=50$ for the five-state clock model. The results are consistent with $1.0503(2)$ in Ref.~\cite{PhysRevE.101.060105}. The same procedure is performed for $q \rightarrow 6^{-}$ in Fig.~\ref{fig:6mqbetapextrapd5060}. The peak positions $\beta_p$ for $D_{\rm{bond}}=50$ and those for $D_{\rm{bond}}=60$ have little discrepancy for $6-q \ge 7\times 10^{-7}$. For $6-q < 7\times 10^{-7}$, $\beta_p$ for $D_{\rm{bond}}=60$ becomes larger than that for $D_{\rm{bond}}=50$. The power-law fit gives us $\beta^{\mathrm{BKT}}_{q=6,c1} = 1.0983(4)$ with $D_{\rm{bond}}=50$ and $\beta^{\mathrm{BKT}}_{q=6,c1} = 1.1019(5)$ for $D_{\rm{bond}}=60$. The results are consistent with $1.101(4)$ in Ref.~\cite{PhysRevE.101.062111}.

We have shown that $\lceil q \rceil - q$ plays the same role as an external magnetic field for the small-$\beta$ BKT transition in integer-$q$-state clock models. The magnetic susceptibility is always finite for fractional $q < \lceil q \rceil$, and diverges as $q \rightarrow \lceil q \rceil^{-}$ with a critical exponent $y = 14/15$. This provides us an alternative way to extract the locations of BKT transitions in clock models. However, the situation is very different for the large-$\beta$ peak of the magnetic susceptibility.

\subsection{Large-\texorpdfstring{$\beta$}{beta} peak: Ising criticality}
\label{sec_results_peak2}
\begin{figure}
	\includegraphics[width=0.48\textwidth]{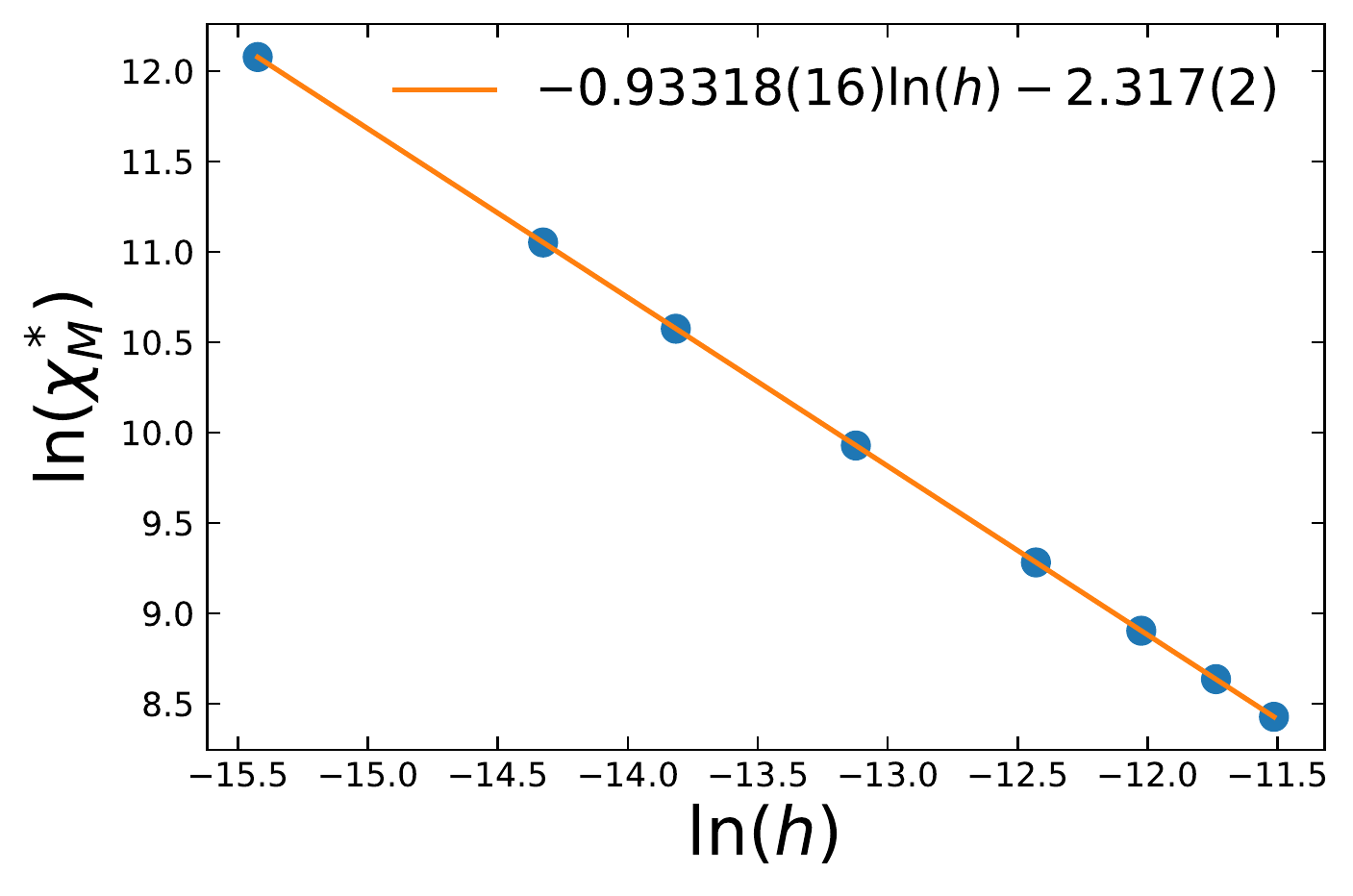}
	\caption{\label{fig:q4p9msusfit} The maximal magnetic susceptibility in Fig.~\ref{fig:q4p9msusextrap} as a function of the external field for $q = 4.9$. A linear fit of the log-log plot gives the exponent $\delta = 14.97(4)$. This is consistent with the value $\delta = 15$ of the Ising universality class. $D_{\rm{bond}} = 40$.}
\end{figure}

\begin{figure}
	\includegraphics[width=0.48\textwidth]{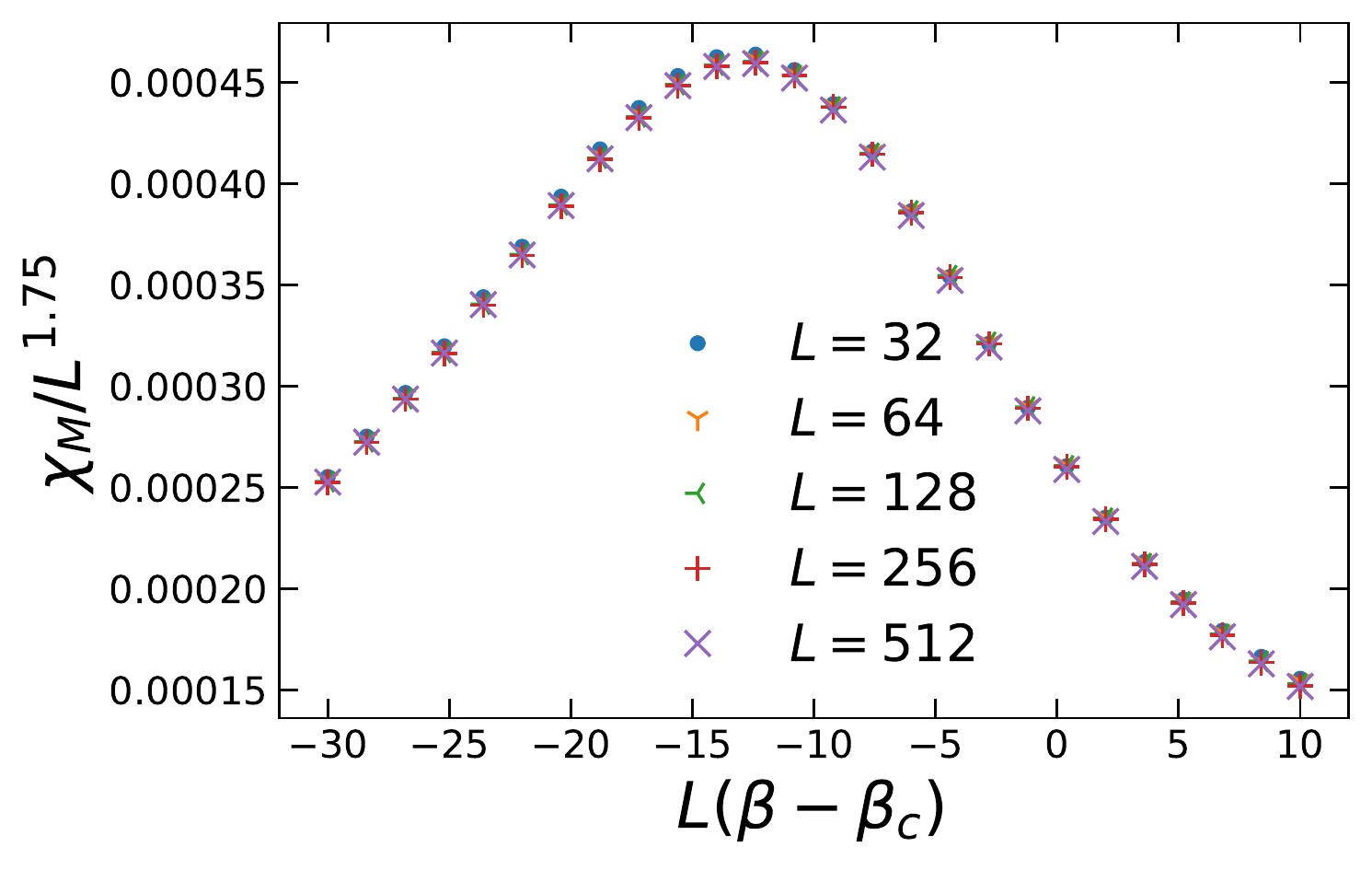}
	\caption{\label{fig:q4p3msuscollapse} Data collapse of the rescaled magnetic susceptibility versus the rescaled inverse temperature for $q = 4.3$. The reduced external field is $hL^{15/8} = 40$, and $\beta_{c} \approx 9.3216$, which is the approximate critical point of the model with $q=4.3$, is obtained from Eq.~(\ref{eq:isingcritical}).}
\end{figure}

To understand the large-$\beta$ peak in the specific heat, we again study the magnetic susceptibility $\chi_M$. For a fixed $q$, the critical point, if any, is given by the location of the peak of $\chi_M$ in the limit $h\rightarrow 0$ where the height $\chi_M^*$ of the peak is infinite. A power-law extrapolation to $h = 0$ is performed on peak positions of $\chi_M$ for small values of $h$. In Fig.~\ref{fig:q4p9msusfit}, we present the peak height $\chi_M^*$ of the susceptibility as a function of the external field $h$. The linear fit of $\ln(\chi^*_M)$ versus $\ln(h)$ gives $\chi_M^* \sim h^{-0.93318(16)}$, from which we obtain the magnetic critical exponent $\delta = 14.97(4)$. This value is consistent with the value $\delta=15$ of the BKT transitions and Ising critical points. We have shown before that there is no phase transition around the small-$\beta$ peak of $\chi_M$. A BKT transition should be accompanied with a continuous critical region, so the divergent large-$\beta$ peak of $\chi_M$ must be an Ising critical point. In the Ising universality class, 
\begin{eqnarray}
\label{eq:isingscaling}
\nonumber \xi &\sim& \left|\beta - \beta_c \right|^{-\nu_e} \sim L, \\
\nonumber M &\sim& \left( \beta - \beta_c \right)^{\beta_e} \sim L^{-1/8} \sim h^{1/15}, \\
\chi_M &\sim& \left| \beta - \beta_c \right|^{-\gamma_e} \sim L^{7/4},
\end{eqnarray}
where $\nu_e = 1, \beta_e = 1/8, \gamma_e = 7/4$ are the universal critical exponents. There is a universal function relating $\chi_M / L^{1.75}$ and $L(\beta-\beta_c)$ with fixed $hL^{15/8}$. In Fig.~\ref{fig:q4p3msuscollapse}, we plot $\chi_M/L^{1.75}$ versus $L(\beta-\beta_{c})$ for various lattice sizes around the large-$\beta$ peak of $\chi_M$, where the value of $\beta_{c}$ is obtained from the Ising approximation in Eq.~(\ref{eq:isingcritical}) described below. One can see that all the data collapse onto a single curve, which gives strong evidence that this is a critical point of the Ising universality class.

\begin{figure}
	\includegraphics[width=0.48\textwidth]{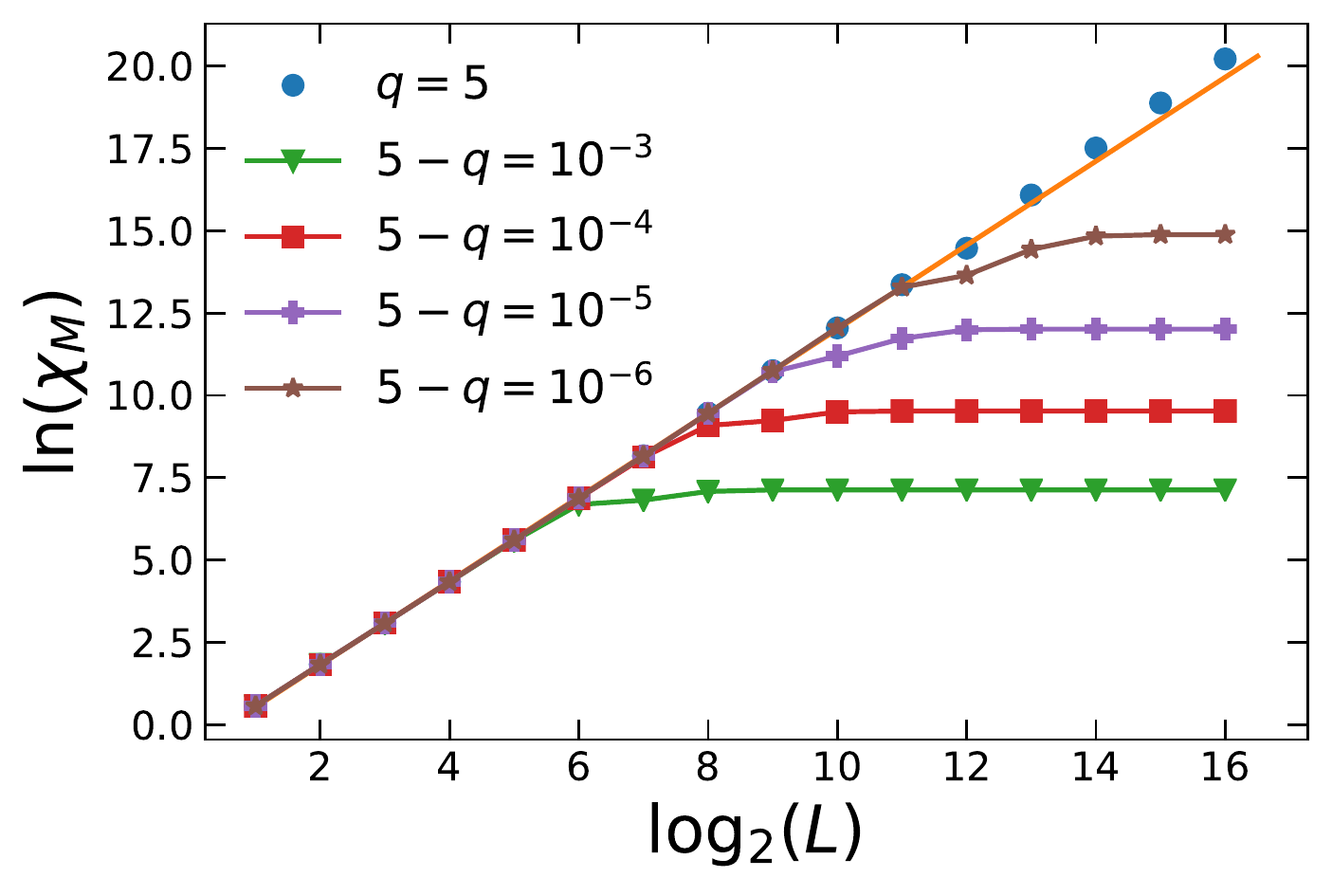}
	\caption{\label{fig:msusvsL} The dependence of magnetic susceptibility on the size of the system $L$ for $\beta = 1.1$ with $D_{\rm{bond}}=40$. The line on the circles is a linear fit for the first ten points: $1.275(3)\log_2(L)-0.74(2)$.}
\end{figure}

In Fig.~\ref{fig:msusvsL}, we show the logarithm of $\chi_M$ as a function of $\log_2(L)$ for $\beta = 1.1$. For $q = 5$, $\beta = 1.1$ is between two BKT points so is a critical point, and $\chi_M$ keeps increasing with a positive power of $L$ as expected. When $q < 5$, even for a very small $5-q = 10^{-6}$, $\chi_M$ always saturates at a large enough volume. These results again prove that there are no BKT transitions in fractional $q$, no matter how close to $\lceil q \rceil$ the $q$ is. Because the maximal value of $\chi_M$ should diverge as a negative power of $5-q$ when $q \rightarrow 5^{-}$, the increment of the height of the plateaus between $5-q = 10^{-n}$ and $5-q = 10^{-n-1}$ should be a constant for any integer $n$, which is also confirmed in Fig.~\ref{fig:msusvsL}.

\begin{figure}
	\includegraphics[width=0.48\textwidth]{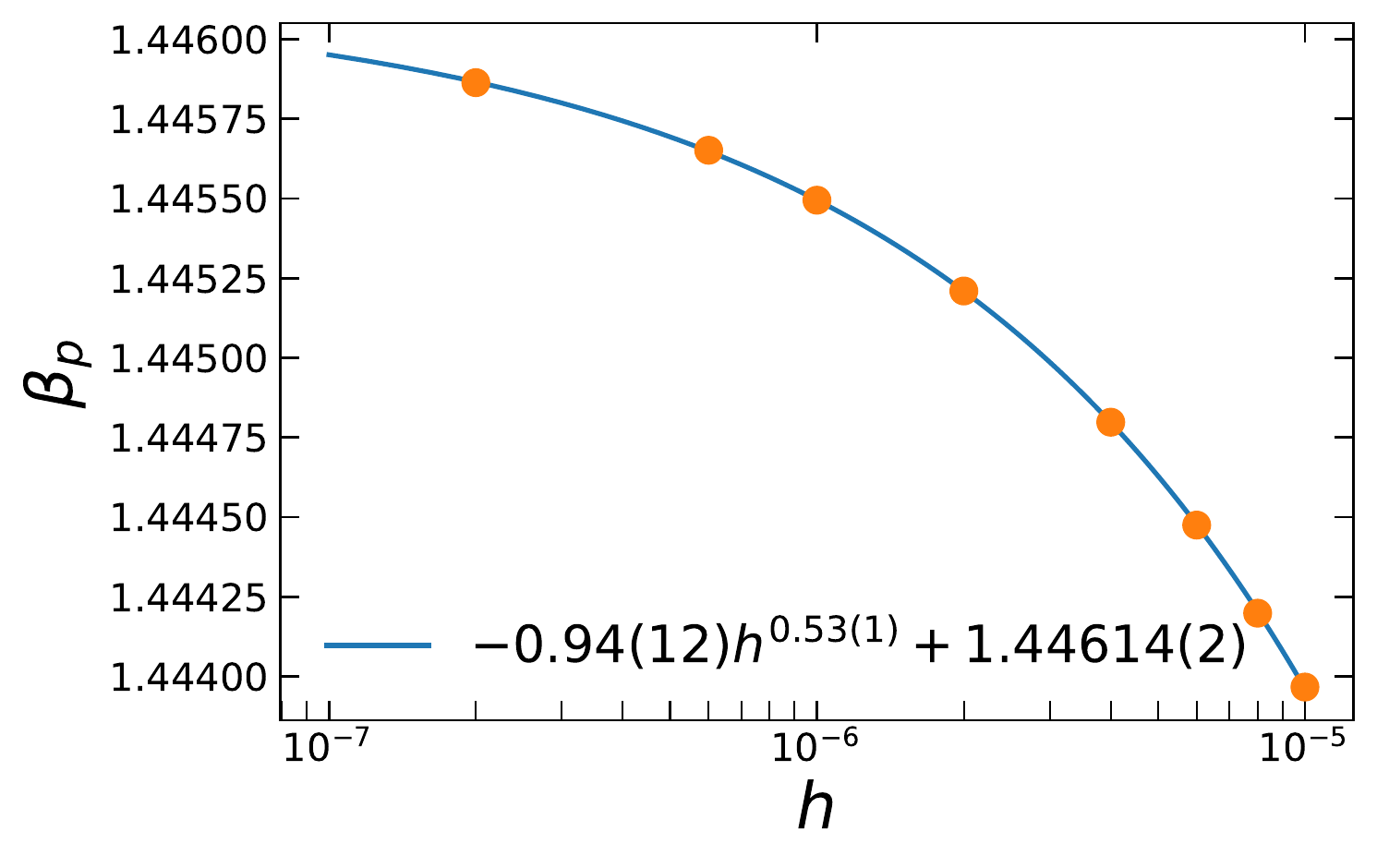}
	\caption{\label{fig:q4p9msusextrap} The power-law extrapolation of the peak position of $\chi_M$ to zero external field for $q = 4.9$. Here the extrapolation gives $\beta_c = 1.44614(2)$. $D_{\rm{bond}} = 40$.} 
\end{figure}

\begin{figure}
	\includegraphics[width=0.48\textwidth]{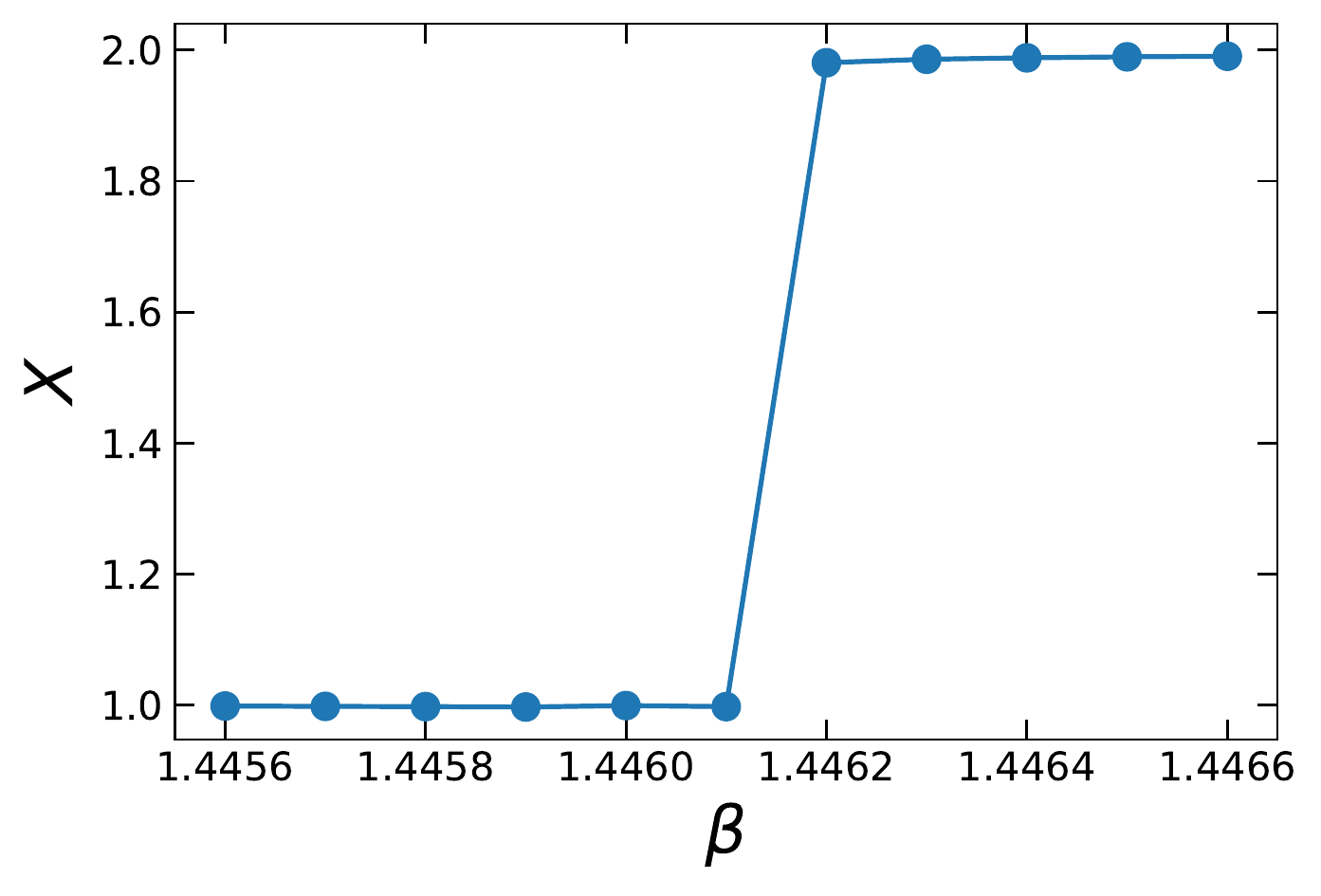}
	\caption{\label{fig:q4p9Xbeta} The $\beta$ dependence of $X$ from the fixed point tensor for $q=4.9$. The discontinuity is located between $1.4461$ and $1.4462$, consistent with the result from the extrapolation of the peak position of $\chi_M$ to zero external field in Fig.~\ref{fig:q4p9msusextrap}. The tensorial bond dimension is $40$.}
\end{figure}

We next obtain the Ising critical point by extrapolating the peak position of $\chi_M$ to $h = 0$. An example for $q=4.9$ is shown in Fig.~\ref{fig:q4p9msusextrap}. The power-law extrapolation gives $\beta_c = 1.44614(2)$. We can repeat the same procedure for other values of $q$. But notice that we need a larger bond dimension in TRG when $q$ is very close to an integer from below, because more degrees of freedom become important, and the critical point is close to a BKT transition point. The phase transition in the Ising universality class is a transition from a disordered phase to a symmetry-breaking phase. The structure of the fixed-point tensor in TRG can easily characterize this phase transition. As proposed in Ref.~\cite{PhysRevB.80.155131}, the symmetry-breaking indicator
\begin{eqnarray}
\label{eq:xfixedpoint}
X = \frac{\left(\sum_{r u} T_{r r u u}\right)^{2}}{\sum_{lrdu } T_{l r u u} T_{r l d d}}
\end{eqnarray}
should be $1$ in the disordered phase and $2$ in the $\mathbb{Z}_2$ symmetry breaking phase. Thus the discontinuity in $X$ for the fixed point tensor as a function of $\beta$ can be used to locate the phase transition. An example for $q = 4.9$ is shown in Fig.~\ref{fig:q4p9Xbeta}, where the value of $X$ changes from $1$ to $2$ at $1.4461 < \beta < 1.4462$, consistent with the result from the extrapolation of the peak position of $\chi_M$ in Fig.~\ref{fig:q4p9msusextrap}. The advantage of this method is that we only need to contract a single tensor network for each value of $\beta$ and scan a $\beta$ range once to locate the phase transition point. This saves us a lot of computational effort when we extrapolate the Ising critical point to $q = \lceil q \rceil$ from below.

\begin{figure}
	\includegraphics[width=0.48\textwidth]{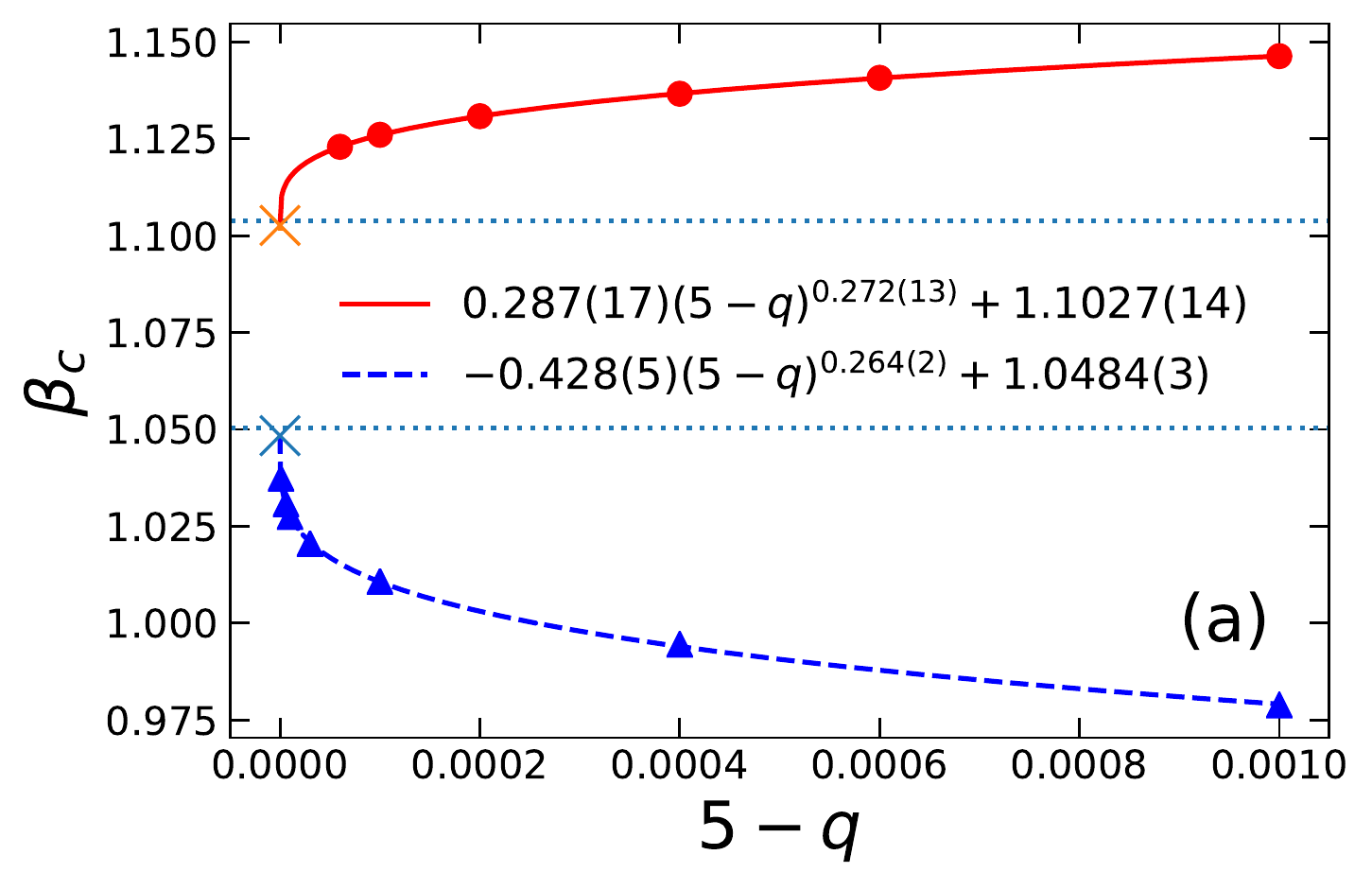}
	\includegraphics[width=0.48\textwidth]{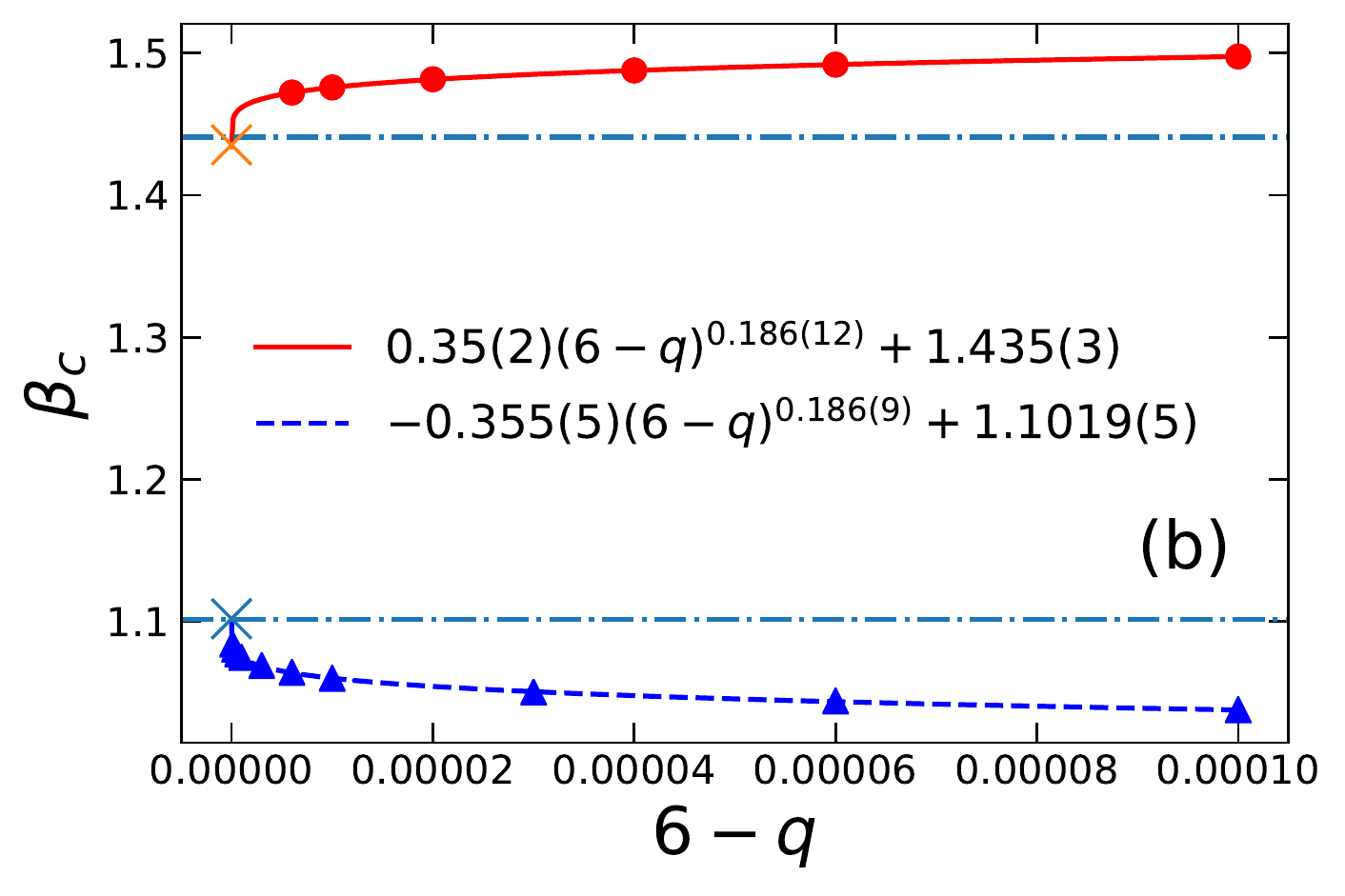}
	\caption{\label{fig:extraptoq5and6} Power law extrapolation of the small-$\beta$ peak position (triangles) of $\chi_M$ with zero external field to $q = 5$ [lower curve in (a)] and to $q = 6$ [lower curve in (b)] . The same procedure is performed for the Ising critical points (circles) at higher $\beta$ [upper curves in (a) and (b)]. The horizontal dashed lines in (a) are the locations of two BKT transitions for $q = 5$ obtained from Ref.~\cite{PhysRevE.101.060105}. The horizontal dash-dotted lines in (b) are the locations of two BKT transitions for $q = 6$ obtained from Ref.~\cite{PhysRevE.101.062111}. $D_{\rm{bond}} = 40$ in (a) and $D_{\rm{bond}} = 60$ in (b). }
\end{figure}

Now we use the value of $X$ to locate the Ising critical point and find the large-$\beta$ BKT transitions for $q = 5, 6$. Notice that for $q = 5$, although both a small external field and a small deviation of $q$ from an integer break the $\mathbb{Z}_{\lceil q \rceil}$ symmetry to a $\mathbb{Z}_{2}$ symmetry, the magnetic susceptibility with a weak external field does not have a peak around the large-$\beta$ BKT transition, so it fails to predict the location of the phase transition \cite{Chen_2018, crossderiv_2020}, but here we always have an Ising critical point for fractional $q$. In Fig.~\ref{fig:extraptoq5and6}(a), we calculate the Ising critical point for $q \rightarrow 5^{-}$ with $D_{\rm{bond}} = 40$ and extrapolate the result to $q = 5$ with a power law. The value of $5-q$ is between $6\times 10^{-4}$ and $10^{-3}$, where the dependence of $\beta_c$ on $D_{\rm{bond}}$ is small. The extrapolated BKT transition point for $q = 5$ is $\beta^{\mathrm{BKT}}_{q=5,c2} = 1.1027(14)$, consistent with the result $1.1039(2)$ obtained in Ref.~\cite{PhysRevE.101.060105}. As a comparison, we also present the extrapolation of the crossover peak with the same $D_{\rm{bond}}$ in the same figure. The exponents of the two power-law scalings are the same within uncertainties, and the values of the exponents are consistent with $0.2677(84)$ obtained in Ref.~\cite{Chen_2018} for magnetic susceptibility with an external field $h \le 10^{-3}$. 

The results for $q \rar 6^{-}$ are shown in Fig.~\ref{fig:extraptoq5and6}(b), where we use a larger bond dimension $D_{\rm{bond}} = 60$ and data with $6-q \le 10^{-4}$. The extrapolated BKT transition point is $\beta^{\mathrm{BKT}}_{q=6,c2} = 1.435(3)$, consistent with $1.441(6)$ in Ref.~\cite{PhysRevE.101.062111}. Comparing the Ising critical points and the crossover peak positions, we find that the power-law scaling parts are exactly the same except for a minus sign within uncertainties, which means they approach the two BKT transition points in the same manner. We believe this behavior can be seen for all $q \rar n^{-}$ for integer $n \ge 5$. To determine the exponent accurately, we need to use a larger $D_{\rm{bond}}$ and data closer to an integer, which is beyond the scope of this work. However, this exponent should be the same as the power-law scaling of $\beta_p$ of $\chi_M$ in a weak external field for all clock models with integer $q \ge 5$, where there is always an emergent O(2) symmetry \cite{Ortiz:2012, PhysRevB.100.094428}. In the limit of O(2) model, this exponent is found to be around $0.162$ \cite{Yu:2013sbi, Jha_2020, zhang2021truncation}.

At large $\beta$, the fractional-$q$-state clock model is a rescaled Ising ($q=2$) model because the link interactions for the two angles 0 and $\tilde{\phi}$ dominate the weights in the partition function. There are two peaks in the specific heat, and if these peaks are sufficiently separated, then the second peak is that of an Ising model where $\beta$ is rescaled as $\beta\rightarrow \alpha \beta$, with a rescaling factor
\begin{equation}
\label{isingscalefactor}
\alpha = \frac{1-\cos \tilde{\phi}}{2},
\end{equation}
where the small angular distance $\tilde{\phi}$ in the model depends on $q$ and is defined in Eq.~(\ref{small_angle}). Thus the critical point $\beta_c$ of the fractional-$q$-state clock model can be approximated by the critical point $\beta_{\mathrm{rIsing}}$ of the two-dimensional rescaled Ising model,
\begin{eqnarray}
\label{eq:isingcritical}
\beta_c \simeq \beta_{\mathrm{rIsing}} \equiv \frac{\ln\left(1+\sqrt{2}\right)}{1-\cos{\tilde{\phi}}}.
\end{eqnarray}
In Table~\ref{tab:isingcriticalpoints} we list some of these critical points for different values of $q$. These points give approximately the location of the large-$\beta$ peak in the specific heat for the extended $q$ clock model.

\begin{table}
	\centering
	\setlength{\tabcolsep}{10pt}
	\begin{tabular}{cccccc}
		$q$ & $\beta_{\mathrm{rIsing}}$ & & & $q$ & $\beta_{\mathrm{rIsing}}$ \\ \hline
		1.1 & 5.5521 & & & 3.6 & 1.7627 \\ \hline
		1.2 & 1.7627 & & & 3.7 & 1.4054 \\ \hline
		1.3 & 1.0022 & & & 3.8 & 1.1681 \\ \hline
		1.4 & 0.7209 & & & 3.9 & 1.0022 \\ \hline
		1.5 & 0.5876 & & & 4.0 & $\infty$ \\ \hline
		1.6 & 0.5163 & & & 4.1 & 75.2052 \\ \hline
		1.7 & 0.4764 & & & 4.2 & 19.8386 \\ \hline
		1.8 & 0.4544 & & & 4.3 & 9.3216 \\ \hline
		1.9 & 0.4437 & & & 4.4 & 5.5521 \\ \hline
		2.0 & $\infty$ & & & 4.5 & 3.7673 \\ \hline
		2.1 & 19.8386 & & & 4.6 & 2.7764 \\ \hline
		2.2 & 5.5521 & & & 4.7 & 2.1665 \\ \hline
		2.3 & 2.7764 & & & 4.8 & 1.7627 \\ \hline
		2.4 & 1.7627 & & & 4.9 & 1.4808 \\ \hline
		2.5 & 1.2755 & & & 5.0 & $\infty$ \\ \hline
		2.6 & 1.0022 & & & 5.1 & 116.284 \\ \hline
		2.7 & 0.8329 & & & 5.2 & 30.3313 \\ \hline
		2.8 & 0.7209 & & & 5.3 & 14.0839 \\ \hline
		2.9 & 0.6433 & & & 5.4 & 8.2861 \\ \hline
		3.0 & $\infty$ & & & 5.5 & 5.5521 \\ \hline
		3.1 & 43.0567 & & & 5.6 & 4.0399 \\ \hline
		3.2 & 11.5787 & & & 5.7 & 3.1120 \\ \hline
		3.3 & 5.5521 & & & 5.8 & 2.4995 \\ \hline
		3.4 & 3.3770 & & & 5.9 & 2.0728 \\ \hline
		3.5 & 2.3409 & & & 6.0 & $\infty$ \\ \hline
	\end{tabular}
	\caption{The critical points of the rescaled Ising model for different values of $q$. They are calculated by Eq.~(\ref{eq:isingcritical}) and give the approximate location of the large-$\beta$ peak in the specific heat of the fractional $q$-state clock model in the infinite-volume limit.}
	\label{tab:isingcriticalpoints}
\end{table}

\begin{figure}
	\includegraphics[width=0.48\textwidth]{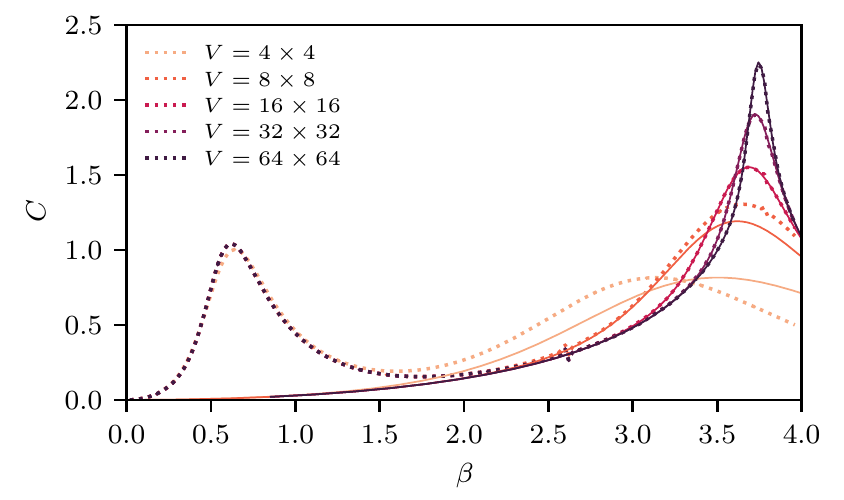}
	\caption{\label{fig:trg_cv_ising} The specific heat for $q=4.5$ for several different lattice sizes. The dotted curves are from TRG and the solid curves are from the rescaled Ising model. For $16\times 16$ and larger lattices, the rescaled Ising model accurately captures the second peak in the specific heat.}
\end{figure}

\begin{figure}
	\includegraphics[width=0.48\textwidth]{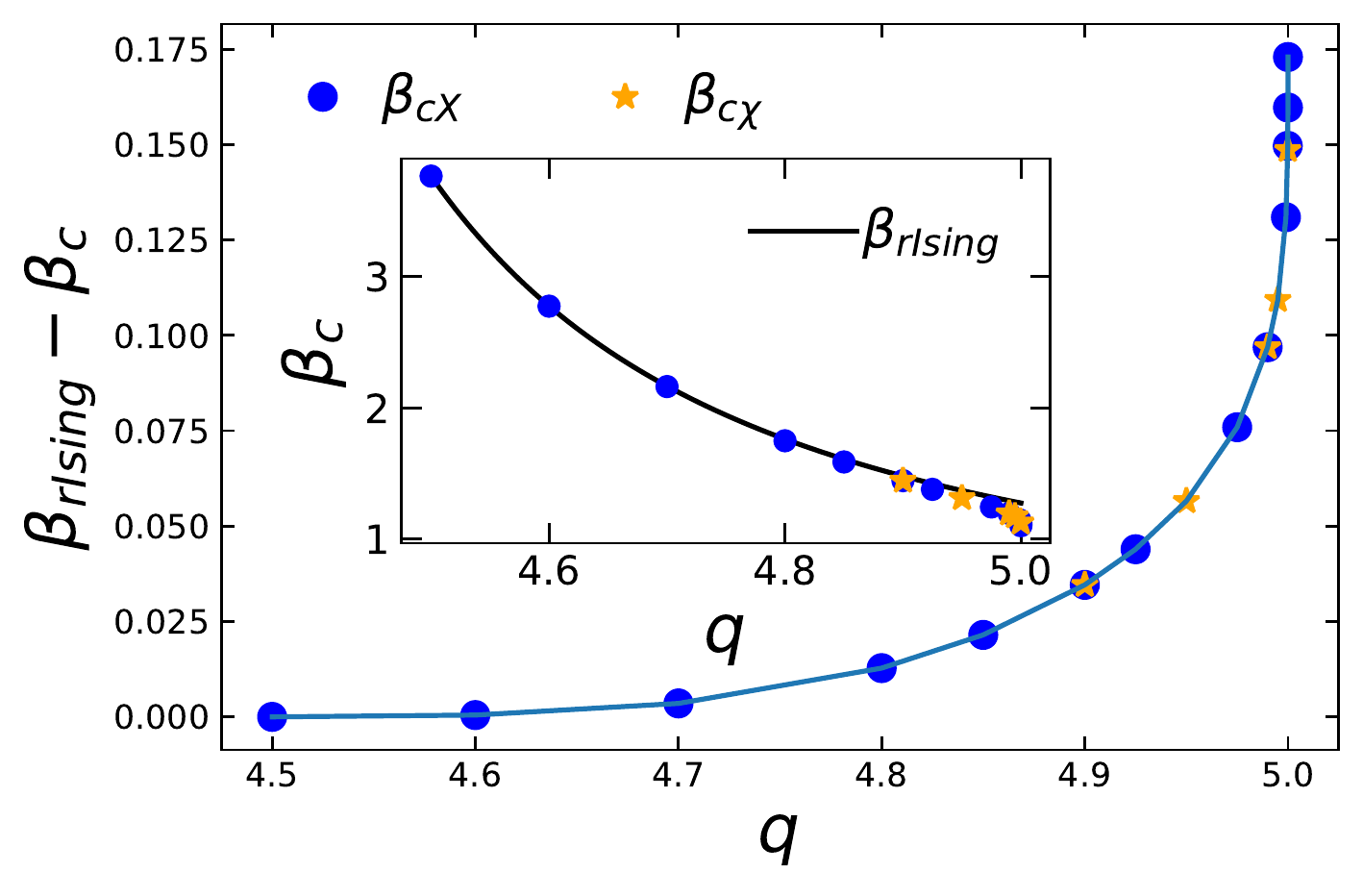}
	\caption{\label{fig:compareisingapprox} The difference between the critical point of the rescaled Ising model $\beta_{\mathrm{rIsing}}$ and the true critical point $\beta_c$ for fractional-$q$-state clock models as a function of $q$. The circles are from the discontinuity of values of $X$, $\beta_{cX}$, the stars are from the extrapolations of the peak position of $\chi_M$ to zero external field, $\beta_{c\chi}$. The inset shows the values of the true critical point as a function of $q$ and the solid line is Eq.~(\ref{eq:isingcritical}).}
\end{figure}

The critical point $\beta_{\mathrm{rIsing}}$ of the rescaled Ising model (see Table~\ref{tab:isingcriticalpoints}) is a good approximation for the true critical point $\beta_c$ for values of $q$ that are not too close to an integer from below. For example, in Fig.~\ref{fig:trg_cv_ising} we compare the specific heat from TRG with the specific heat of the rescaled Ising model for $q=4.5$ and several different lattice sizes. For lattices $16\times 16$ and larger, the specific heat for the rescaled Ising model accurately captures the large-$\beta$ peak of the fractional-$q$-state clock model. As $q$ approaches an integer from below, the approximation begins to fail. In Fig.~\ref{fig:compareisingapprox}, we compare the rescaled Ising critical points with the true critical points as $q \rightarrow 5$ from below. Most of the true critical points are obtained from $X$ defined in Eq.~(\ref{eq:xfixedpoint}), and five points are from $\chi_M$ and agree perfectly with those from $X$. For $4.0 < q \lesssim 4.7$, the difference between the true critical point and the Ising approximation is less than $0.01$. As $q\rightarrow 5^{-}$, the difference becomes larger and is around $0.175$ for $q = 5$.

\subsection{Integration interval}
\label{sec_results_integration}
\begin{figure}
	\includegraphics[width=0.48\textwidth]{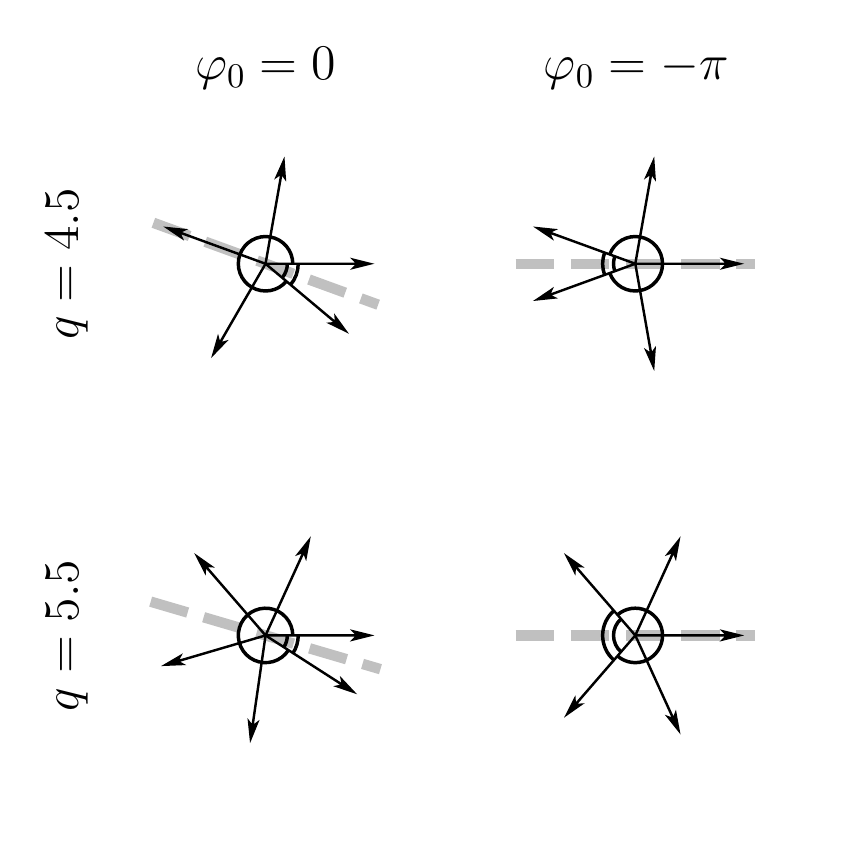}
	\caption{\label{fig:allowed_angles} The allowed spin orientations for the extended $q$-state clock model when $\varphi_0 = 0$ (left column) and $\varphi_0 = -\pi$ (right column) for two different values of $q$. When $q=4.5$ (top row), the number of allowed spin orientations and their relative sizes are the same, and the models are equivalent. When $q=5.5$ (bottom row), there are six allowed spin orientations when $\varphi_0=0$, but only five allowed orientations when $\varphi_0 = -\pi$, so the two models are different. The dashed gray lines indicate the $\mathbb{Z}_2$ symmetry axes.}
\end{figure}

For the extended $q$-state clock model, the allowed angles are restricted to the integration interval $\varphi\in[\varphi_0, \varphi_0+2\pi)$. All results presented so far used $\varphi_0=0$, which is in the so-called case $1$. We also considered the possibility $\varphi_0 = -\pi$. As shown in Fig.~\ref{fig:allowed_angles}, for $q = 4.5$, the number of allowed spin orientations and their relative sizes are the same. In fact, the models with $\varphi_0=0$ and $\varphi_0=-\pi$ are equivalent and in case $1$ for all even $\lfloor q \rfloor$. For odd $\lfloor q \rfloor$, the choice of $\varphi_0=0$ or $\varphi_0=-\pi$ results in a different number of allowed spin orientations and different thermodynamic behaviors. For example, when $5<q<6$, there are six allowed orientations when $\varphi_0=0$, but only five allowed orientations when $\varphi_0 = -\pi$ (see Fig.~\ref{fig:allowed_angles} for $q = 5.5$). One can show that the model with $\varphi_0=-\pi$ is in case $2$ for all odd $\lfloor q \rfloor$. For integer values of $q$, the extended $q$-state clock model reduces to the ordinary $q$-state clock model for both $\varphi_0=0$ and $\varphi_0=-\pi$.

\begin{figure}
	\includegraphics[width=0.48\textwidth]{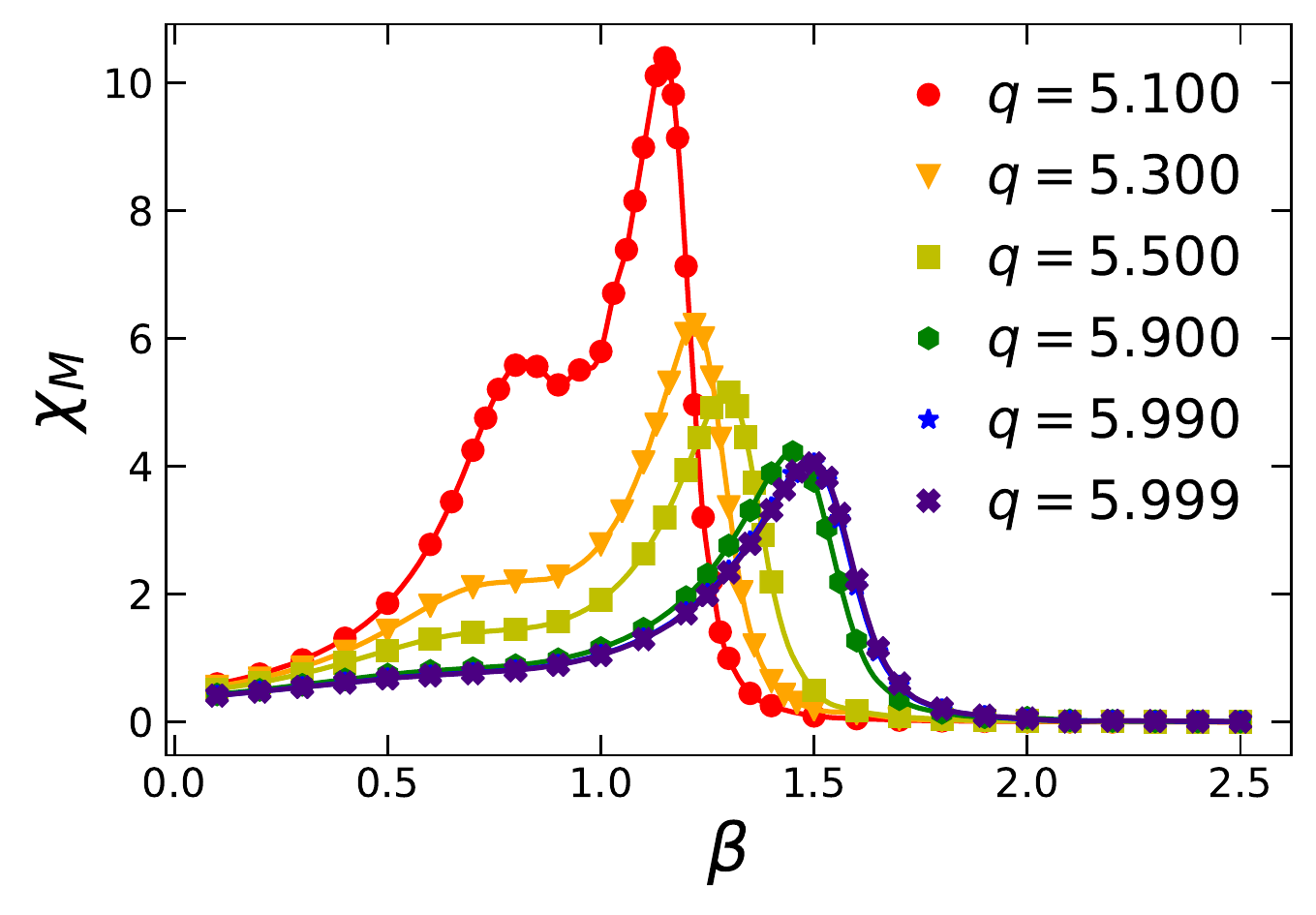}
	\caption{\label{fig:phimpiq5pmsusvsbeta} The magnetic susceptibility at $V = 2^{24} \times 2^{24}$, $h = 0$, and $D_{\rm{bond}} = 40$, as a function of $\beta$ for $\varphi_0 = -\pi$. The susceptibility does not diverge with the volume, which implies there is no phase transition here. This is different from the model with $\varphi_0=0$, which has a divergent peak in the susceptibility corresponding to a second-order phase transition.}
\end{figure}

\begin{figure}
	\includegraphics[width=0.48\textwidth]{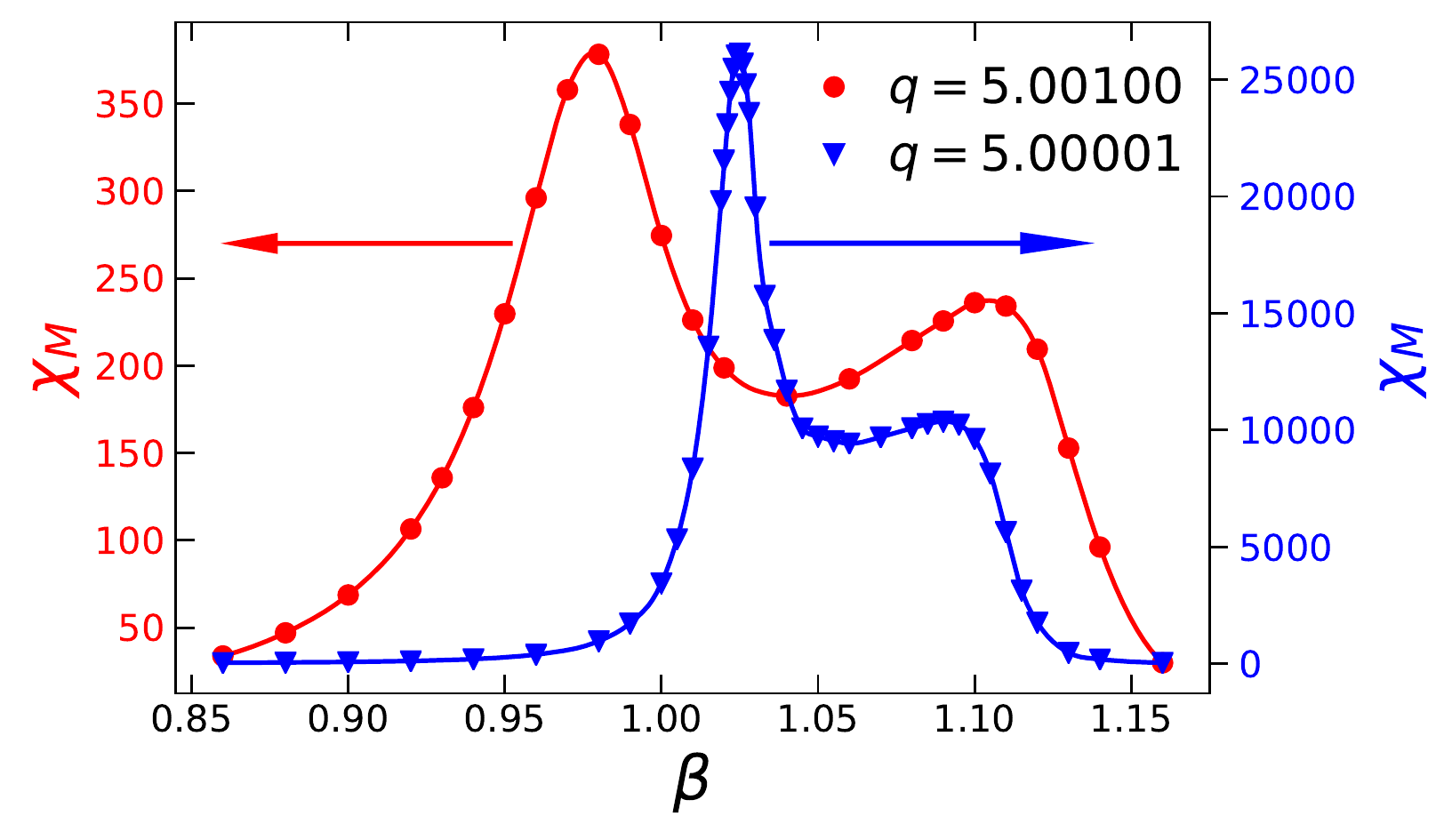}
	\caption{\label{fig:phimpiqcloseto5pmsusvsbeta} Same as Fig.~\ref{fig:phimpiq5pmsusvsbeta}, but for $q = 5.001, 5.00001$. The large-$\beta$ peak fades away and moves across the BKT phase transition point at $q = 5$.}
\end{figure}

We consider $5 < q < 6$ and $\varphi_0 = -\pi$. In Fig.~\ref{fig:phimpiq5pmsusvsbeta}, we show the magnetic susceptibility at $h = 0$ as a function of $\beta$. First of all, one can see that the magnetic susceptibility is finite for all values of $q$ presented here, which means there are no phase transitions for any $5 < q < 6$ and $\varphi_0 = -\pi$. The maximal values of $\chi_M$ are larger for $q$ closer to $5$. For $q=5.1$, we see a double-peak structure in $\chi_M$, with the large-$\beta$ peak higher than the small-$\beta$ peak. As $q$ increases toward $6$, the first crossover peak fades away and disappears around $q = 5.3$. The magnetic susceptibility eventually converges to a single-peak structure with the peak position around $\beta = 1.5$ and the peak height around $4$. In order to see the behavior of $\chi_M$ for $q \rar 5^{+}$, we plot $\chi_M$ versus $\beta$ for $q = 5.001$ and $q = 5.00001$ in Fig.~\ref{fig:phimpiqcloseto5pmsusvsbeta}. We see that the maximal value of $\chi_M$ is much larger than those in Fig.~\ref{fig:phimpiq5pmsusvsbeta}, because we are approaching a $\mathbb{Z}_5$ clock model from above. Unlike $q = 5.1$, the small-$\beta$ peak becomes higher than the large-$\beta$ peak for $q = 5.001$, and the large-$\beta$ peak fades away as we move closer to $q = 5$. The small-$\beta$ peak moves towards the small-$\beta$ BKT transition for $q = 5$ so it can be used to extrapolate the value of $\beta^{\mathrm{BKT}}_{q=5,c1}$, while the large-$\beta$ peak moves across the large-$\beta$ BKT transition for $q = 5$ from right to left so it fails to predict the value of  $\beta^{\mathrm{BKT}}_{q=5,c2}$. This means that the magnetic susceptibility cannot capture the crossover going to the large-$\beta$ BKT transition point for $q \rar 5^{+}$. But the crossover should go to the BKT transition point as the $\mathbb{Z}_{\lfloor q \rfloor}$ symmetry is recovered. The cross derivative of the free energy should be able to capture this \cite{crossderiv_2020}, but we will just focus on the magnetic susceptibility.

\begin{figure}
	\includegraphics[width=0.48\textwidth]{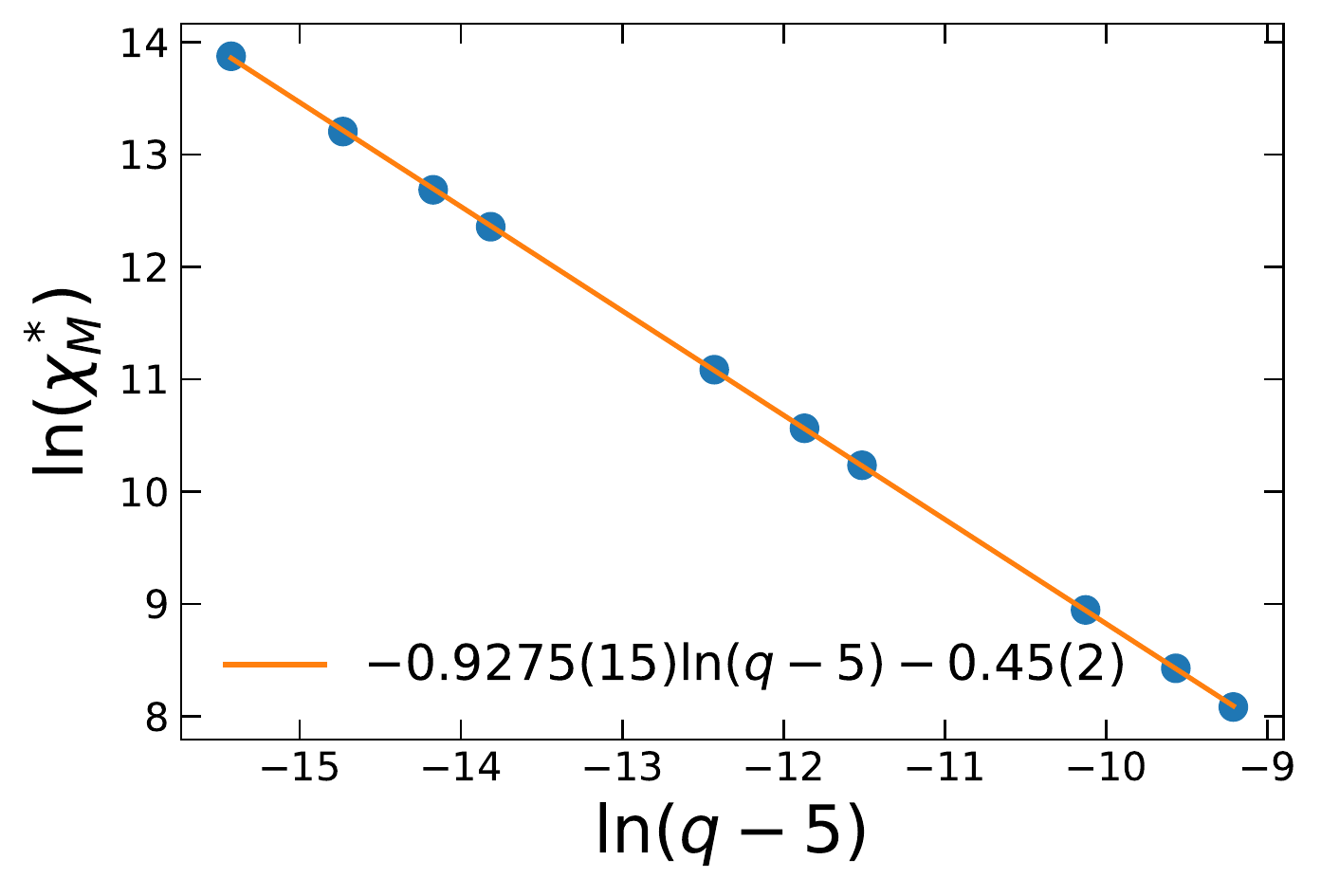}
	\caption{\label{fig:phimpi1stpeakfit} Same as Fig.~\ref{fig:5mqchimheight}, but for $\varphi_0 = -\pi, q \rightarrow 5^{+}$. $D_{\rm{bond}} = 50$. }
\end{figure}

\begin{figure}
	\includegraphics[width=0.48\textwidth]{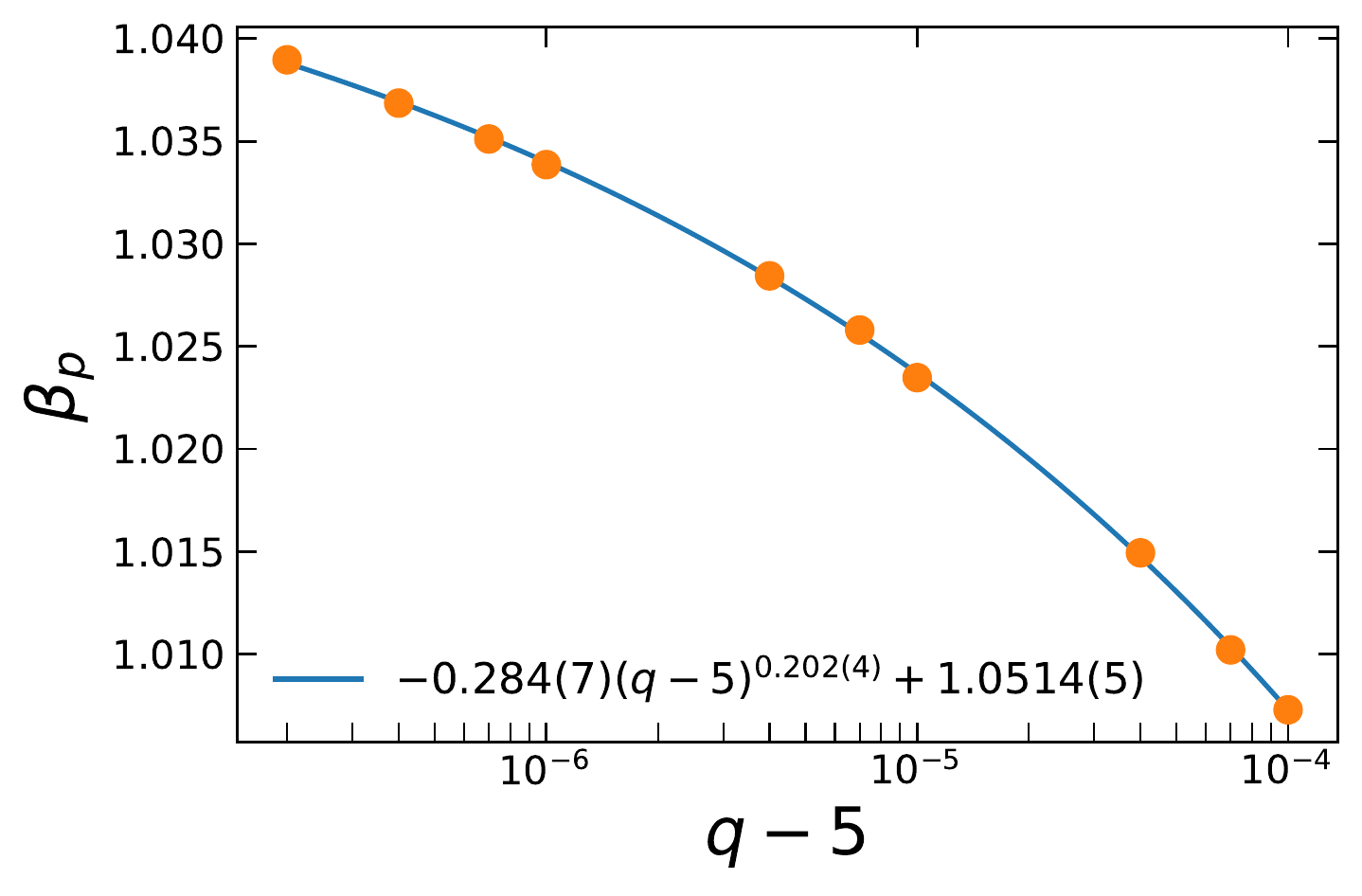}
	\caption{\label{fig:phimpi1stpeakextrap} Same as Fig.~\ref{fig:5mqbetapextrapd50}, but for $\varphi_0 = -\pi, q \rightarrow 5^{+}$. $D_{\rm{bond}} = 50$. }
\end{figure}

We then check the divergent behavior of the small-$\beta$ peak of $\chi_M$ as $q \rar 5^+$ using $D_{\rm{bond}} = 50$. Figure~\ref{fig:phimpi1stpeakfit} shows the linear fit for $\ln(\chi_M^*)$ versus $\ln(q-5)$. We see that the peak height of $\chi_M$ diverges as a power law $\chi_M^* \sim 1/(q-5)^{0.9275(15)}$, which again gives a value of the critical exponent $\delta' = 15.09(2)$, close to the expected value $15$. We extrapolate the peak position to $q = 5$ from above in Fig.~\ref{fig:phimpi1stpeakextrap}. One can see that the power-law scaling is the same as case $1$ where $q \rar 5^{-}, \varphi_0 = 0$, and gives $\beta^{\mathrm{BKT}}_{q=5,c1} = 1.0514(5)$, consistent with the result in Fig.~\ref{fig:5mqbetapextrapd50}.

\subsection{Phase diagram}
\label{sec_results_phasediag}

The clock model with integer $q$ has been studied extensively \cite{PhysRevD.19.3698, PhysRevB.23.1362, PhysRevB.26.6201, PhysRevB.28.5371, Murty:1984, PhysRevB.33.437, Leroyer:1991, PhysRevB.65.184405, PhysRevLett.96.140603, PhysRevE.74.041106, PhysRevE.80.060101, PhysRevE.80.042103, PhysRevE.82.031102, PhysRevE.83.041120, PhysRevE.85.021114, Ortiz:2012, Chen_2017, Chen_2018, PhysRevE.98.032109, Surungan_2019, PhysRevE.101.060105}. For $q=2,3,4$, there is a disordered phase and a $\mathbb{Z}_q$ symmetry-breaking phase separated by a second-order phase transition. For $q \geq 5$, there is a disordered phase at small-$\beta$ and a $\mathbb{Z}_q$ symmetry-breaking phase at large-$\beta$ with a critical phase for intermediate $\beta$ between them. The boundaries of the critical phase are two BKT transition points of infinite order \cite{Ortiz:2012}. In our extended $q$-state clock model, one must make a choice of the integration interval $\varphi\in[\varphi_0,\varphi_0+2\pi)$. For the choice $\varphi_0 = 0$ and fractional $q$, both the specific heat and the magnetic susceptibility have a double-peak structure. We have shown that the small-$\beta$ peak is associated with a crossover, and the large-$\beta$ peak is associated with a phase transition of the Ising universality class. For the choice $\varphi_0 = -\pi$ and fractional $q$, the phase structure is a little more complicated. For even $\lfloor q \rfloor$, we get the same behavior as with $\varphi_0 = 0$, but for odd $\lfloor q \rfloor$, we get a trivial case with no critical point. 

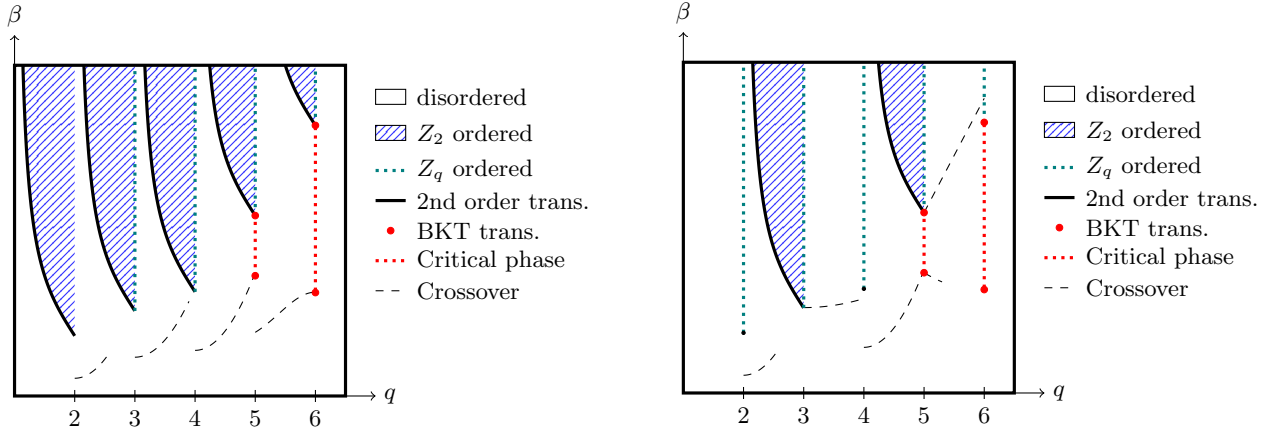
\begin{figure*}%
	\centering
	
	\begin{minipage}{.49\textwidth}
		\begin{tikzpicture}[scale=0.8]
		
		\draw[->] (1,0) -- (7,0) node[anchor=west] {$q$};
		\draw (2,0.1) -- (2,-0.1) node[anchor=north] {2};
		\draw (3,0.1) -- (3,-0.1) node[anchor=north] {3};
		\draw (4,0.1) -- (4,-0.1) node[anchor=north] {4};
		\draw (5,0.1) -- (5,-0.1) node[anchor=north] {5};
		\draw (6,0.1) -- (6,-0.1) node[anchor=north] {6};
		
		
		\draw[dashed, samples=100,domain=2:2.6] plot (\x,{1.27*(\x-2)^2+0.295});
		\draw[dashed, samples=100,domain=3:3.9] plot (\x,{1.15*(\x-3)^2+0.645});
		\draw[dashed, samples=100,domain=4:5.0] plot (\x,{1.29*(\x-4)^2+0.759});
		\draw[dashed, samples=100,domain=5:6.0] plot (\x,{0.89*pow(4,-(\x-6)^2)+0.836});
		
		\fill[pattern color=blue!80!white, pattern=north east lines, samples=100,domain=1.125:2] plot (\x,{-tan(1.571*(\x-0) r)+1.000}) -- (2,6) --cycle;
		\fill[pattern color=blue!80!white, pattern=north east lines, samples=100,domain=2.136:3] plot (\x,{-tan(1.571*(\x-1) r)+1.414}) -- (3,6) --cycle;
		\fill[pattern color=blue!80!white, pattern=north east lines, samples=100,domain=3.146:4] plot (\x,{-tan(1.571*(\x-0) r)+1.732}) -- (4,6) --cycle;
		\fill[pattern color=blue!80!white, pattern=north east lines, samples=100,domain=4.205:5] plot (\x,{-tan(1.571*(\x-1) r)+3.000}) -- (5,6) --cycle;
		\fill[pattern color=blue!80!white, pattern=north east lines, samples=100,domain=5.240:6] plot (\x,{-tan(1.571*(\x-0) r)+4.500}) -- (6,6) --cycle;
		\draw[very thick, samples=100,domain=1.125:2] plot (\x,{-tan(1.571*(\x-0) r)+1.000});
		\draw[very thick, samples=100,domain=2.136:3] plot (\x,{-tan(1.571*(\x-1) r)+1.414});
		\draw[very thick, samples=100,domain=3.146:4] plot (\x,{-tan(1.571*(\x-0) r)+1.732});
		\draw[very thick, samples=100,domain=4.205:5] plot (\x,{-tan(1.571*(\x-1) r)+3.000});
		\draw[very thick, samples=100,domain=5.240:6] plot (\x,{-tan(1.571*(\x-0) r)+4.500});
		
		\draw[teal, dotted, very thick] (3,1.414) -- (3,5.5);
		\draw[teal, dotted, very thick] (4,1.732) -- (4,5.5);
		\draw[teal, dotted, very thick] (5,3.000) -- (5,5.5);
		\draw[teal, dotted, very thick] (6,4.500) -- (6,5.5);
		
		\draw[draw=red, fill=red] (5,2.000) circle (.4ex);
		\draw[draw=red,fill=red] (5,3.000) circle (.4ex);
		\draw[draw=red,fill=red] (6,4.500) circle (.4ex);
		\draw[draw=red,fill=red] (6,1.720) circle (.4ex);
		\draw[red, dotted, very thick] (5,2) -- (5,3);
		\draw[red, dotted, very thick] (6,1.750) -- (6,4.5);
		
		\fill[white] (1,5.5) -- (1,7.1) -- (6.5,7.1) -- (6.5,5.5) --cycle;
		
		\draw[very thick] (1,0) -- (1,5.5) -- (6.5,5.5) -- (6.5,0) --cycle;
		
		\draw[->] (1,0) -- (1,6) node[anchor=south] {$\beta$};
		
		\draw (7,5.1) -- (7.5,5.1) -- (7.5, 4.85) -- (7,4.85) --cycle;
		\filldraw[pattern color=blue!80!white, pattern=north east lines] (7,4.5) -- (7.5,4.5) -- (7.5, 4.25) -- (7,4.25) --cycle;
		\draw[teal, dotted, very thick] (7, 3.75) -- (7.5,3.75);
		\draw[very thick] (7, 3.25) -- (7.5,3.25); 
		\draw[draw=red, fill=red] (7.25, 2.75) circle (.4ex);
		\draw[red, dotted, very thick] (7,2.25) -- (7.5,2.25);
		\draw[dashed] (7,1.75) -- (7.5,1.75);
		
		\node[text width=2.5cm] at (9.25,4.97) {\small disordered};
		\node[text width=2.5cm] at (9.25,4.35) {\small $Z_2$ ordered};
		\node[text width=2.5cm] at (9.25,3.75) {\small $Z_q$ ordered};
		\node[text width=2.5cm] at (9.25,3.25) {\small 2nd order trans.};
		\node[text width=2.5cm] at (9.25,2.75) {\small BKT trans.};
		\node[text width=2.5cm] at (9.25,2.25) {\small Critical phase};
		\node[text width=2.5cm] at (9.25,1.75) {\small Crossover};
		
		\end{tikzpicture}
	\end{minipage}
	\begin{minipage}{.49\textwidth}
		\begin{tikzpicture}[scale=0.8]
		
		\draw[->] (1,0) -- (7,0) node[anchor=west] {$q$};
		\draw (2,0.1) -- (2,-0.1) node[anchor=north] {2};
		\draw (3,0.1) -- (3,-0.1) node[anchor=north] {3};
		\draw (4,0.1) -- (4,-0.1) node[anchor=north] {4};
		\draw (5,0.1) -- (5,-0.1) node[anchor=north] {5};
		\draw (6,0.1) -- (6,-0.1) node[anchor=north] {6};
		
		
		\draw[dashed, samples=100,domain=2:2.6] plot (\x,{1.27*(\x-2)^2+0.295});
		\draw[dashed, samples=100,domain=3:4.0] plot (\x,{0.15*(\x-3)^2+1.42});
		\draw[dashed, samples=100,domain=4:5.0] plot (\x,{1.29*(\x-4)^2+0.759});
		\draw[dashed, samples=100,domain=5:5.3] plot (\x,{0.3*(\x-6)^2+1.7});
		\draw[dashed] (5,3.000) -- (6,4.900);
		
		\fill[pattern color=blue!80!white, pattern=north east lines, samples=100,domain=2.136:3] plot (\x,{-tan(1.571*(\x-1) r)+1.414}) -- (3,6) --cycle;
		\fill[pattern color=blue!80!white, pattern=north east lines, samples=100,domain=4.205:5] plot (\x,{-tan(1.571*(\x-1) r)+3.000}) -- (5,6) --cycle;
		\draw[very thick, samples=100,domain=2.136:3] plot (\x,{-tan(1.571*(\x-1) r)+1.414});
		\draw[very thick, samples=100,domain=4.205:5] plot (\x,{-tan(1.571*(\x-1) r)+3.000});
		
		\draw[teal, dotted, very thick] (2,1) -- (2,5.5);
		\draw[teal, dotted, very thick] (3,1.414) -- (3,5.5);
		\draw[teal, dotted, very thick] (4,1.732) -- (4,5.5);
		\draw[teal, dotted, very thick] (5,3.000) -- (5,5.5);
		\draw[teal, dotted, very thick] (6,4.500) -- (6,5.5);
		
		\draw[fill=black] (4,1.732) circle (.2ex);
		\draw[fill=black] (2,1.000) circle (.2ex);
		
		\draw[draw=red, fill=red] (5,2.000) circle (.4ex);
		\draw[draw=red,fill=red] (5,3.000) circle (.4ex);
		\draw[draw=red,fill=red] (6,4.500) circle (.4ex);
		\draw[draw=red,fill=red] (6,1.720) circle (.4ex);
		\draw[red, dotted, very thick] (5,2) -- (5,3);
		\draw[red, dotted, very thick] (6,1.750) -- (6,4.5);
		
		\fill[white] (1,5.5) -- (1,7) -- (6.5,7) -- (6.5,5.5) --cycle;
		
		\draw[very thick] (1,0) -- (1,5.5) -- (6.5,5.5) -- (6.5,0) --cycle;
		
		\draw[->] (1,0) -- (1,6) node[anchor=south] {$\beta$};
		
		\draw (7,5.1) -- (7.5,5.1) -- (7.5, 4.85) -- (7,4.85) --cycle;
		\filldraw[pattern color=blue!80!white, pattern=north east lines] (7,4.5) -- (7.5,4.5) -- (7.5, 4.25) -- (7,4.25) --cycle;
		\draw[teal, dotted, very thick] (7, 3.75) -- (7.5,3.75);
		\draw[very thick] (7, 3.25) -- (7.5,3.25); 
		\draw[draw=red, fill=red] (7.25, 2.75) circle (.4ex);
		\draw[red, dotted, very thick] (7,2.25) -- (7.5,2.25);
		\draw[dashed] (7,1.75) -- (7.5,1.75);
		
		\node[text width=2.5cm] at (9.25,4.97) {\small disordered};
		\node[text width=2.5cm] at (9.25,4.35) {\small $Z_2$ ordered};
		\node[text width=2.5cm] at (9.25,3.75) {\small $Z_q$ ordered};
		\node[text width=2.5cm] at (9.25,3.25) {\small 2nd order trans.};
		\node[text width=2.5cm] at (9.25,2.75) {\small BKT trans.};
		\node[text width=2.5cm] at (9.25,2.25) {\small Critical phase};
		\node[text width=2.5cm] at (9.25,1.75) {\small Crossover};
		
		\end{tikzpicture}
	\end{minipage}
	\caption{The phase diagram of the extended $q$-state clock model [i.e. the $\gamma=\infty$ plane of the extended-O(2) model] for $\varphi_0=0$ (left) and $\varphi_0=-\pi$ (right). For $q=2,3,4$, there is a second-order phase transition with a $\mathbb{Z}_q$ ordered phase at large $\beta$. For finite integer $q\geq 5$, there is a critical phase between a pair of BKT transitions and a $\mathbb{Z}_q$ ordered phase at large $\beta$. For fractional $q>2$ with $\varphi_0=0$, there is a crossover line, a second-order transition line, and a $\mathbb{Z}_2$ ordered region between every consecutive pair of integers. For $\varphi_0=-\pi$, the same is true for every other consecutive pair of integers.\label{phasediagram}}
\end{figure*}

The phase diagrams for both $\varphi_0 = 0$ and $\varphi_0 = -\pi$ are shown in Fig.~\ref{phasediagram}. For $\varphi_0 = 0$, as $q \rar \lceil q \rceil^-$, the $\mathbb{Z}_{\lceil q \rceil}$ symmetry is recovered. For $q > 4$ and $\varphi_0 = 0$, both the small-$\beta$ crossover line and the large-$\beta$ Ising critical point are smoothly connected to the small-$\beta$ BKT transition point and the large-$\beta$ BKT transition point for integer $q$ from below, respectively. Notice that for $q < 4$ and $\varphi_0 = 0$, only the large-$\beta$ Ising critical point is smoothly connected to the second-order phase transition for integer $q$ from below, while the crossover peak fades away for $q$ close enough (around $3.9$ for $3 < q < 4$) to an integer from below. When $q \rar \lfloor q \rfloor ^{+}$ and $\varphi_0 = 0$, the Ising critical point goes to infinity, while the crossover line goes to a smaller value than the phase transition point for the $\lfloor q \rfloor$-state clock model because there is $1$ more degree of freedom than the $\mathbb{Z}_{\lfloor q \rfloor}$ clock model. For even $\lfloor q \rfloor$ and $\varphi_0 = -\pi$, the phase diagram is the same as that for even $\lfloor q \rfloor$ and $\varphi_0 = 0$. For odd $\lfloor q \rfloor$ and $\varphi_0 = -\pi$, the $\mathbb{Z}_{\lfloor q \rfloor}$ symmetry is recovered when $q \rar \lfloor q \rfloor ^{+}$. For $5 < q < 6$, both crossover lines are smoothly connected to the two BKT transition points for integer $q$ from above. When $q$ is increased toward an even $\lceil q \rceil$ and $\varphi_0 = -\pi$, the small-$\beta$ crossover line fades away and the large-$\beta$ crossover line goes to a larger value than the large-$\beta$ BKT transition for integer $q$ because there is $1$ fewer degree of freedom than the $\lceil q \rceil$-state clock model. For $3 < q < 4$, there is only one crossover line that is smoothly connected to the second-order phase transition point for $q = 3$ and goes to around $0.77$ when $q \rar 4^{-}$.

\begin{figure}
	\centering
	\includegraphics[scale=0.8]{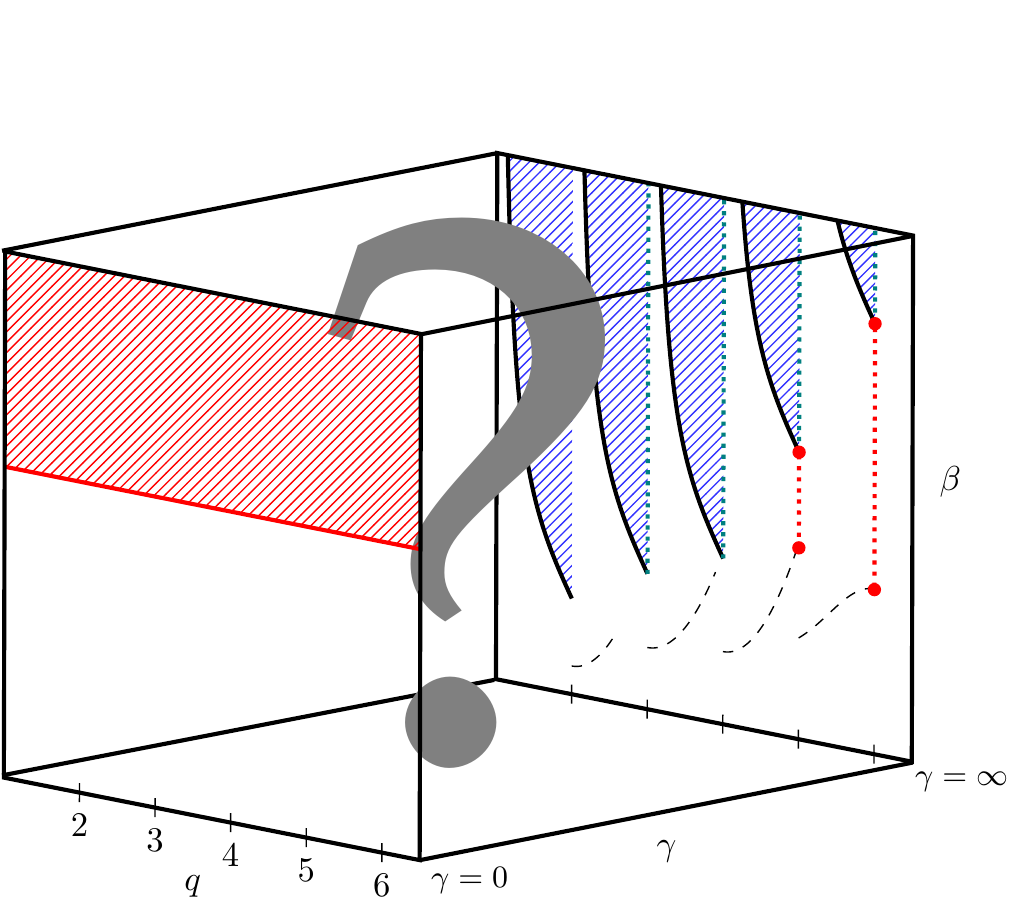}
	\caption{For the extended-O(2) model, the phase diagram is three-dimensional. In the $\gamma=0$ plane, it is the $XY$ model for all values of $q$. The $XY$ model has a disordered phase at small $\beta$, a single BKT transition at $\beta_c = 1.11995(6)$ \cite{Jha_2020}, and a critical phase at large $\beta$. In the $\gamma=\infty$ plane, it is the extended $q$-state clock model, which has the phase diagram shown in Fig.~\ref{phasediagram}. In this example we have $\varphi_0 = 0$. Establishing the phase diagram at finite-$\gamma$ will be addressed in future work.\label{phasediagram_3d}}
\end{figure}

\section{Summary and Outlook}
\label{sec_summary}

Interpolations among $\mathbb{Z}_n$ clock models have been realized experimentally using a simple Rydberg simulator, where $\mathbb{Z}_n$ ($n \ge 2$) symmetries emerge by tuning continuous parameters, the detuning and Rabi frequency of the laser coupling, and the interaction strength between Rydberg atoms \cite{keesling2019}. This paves the way to quantum simulation of lattice field theory with discretized field variables. We are interested in a theory that can interpolate among the O(2) model and $\mathbb{Z}_n$ clock models. We define an extended-O(2) model by adding a symmetry breaking term $\gamma \cos(q\varphi_x)$ to the action of the two-dimensional O(2) model. For integer $q$, $\mathbb{Z}_q$ clock models emerge for large enough $\gamma$. For fractional $q$, we believe there exists a much more interesting phase structure. The first step to graph the full phase diagram in the $(\gamma, q, \beta)$ cube is to consider the limit $\gamma \rar \infty$. In this work, we studied the fractional-$q$-state clock model as the $\gamma \rar \infty$ limit of the extended-O(2) model with angular variables in the domain $[\varphi_0, \varphi_0+2\pi)$. In this limit, the angular variable takes discrete values $\varphi_{x,k} = 2\pi k/q$ with $k$ integral. By varying $\varphi_0$, the set of values of integer $k$ take either case $1$ $(0, 1, \ldots, \lfloor q \rfloor)$ or case $2$ $(0, 1, \ldots, \lfloor q \rfloor - 1)$. In case $1$, $\mathbb{Z}_{\lceil q \rceil}$ symmetry is recovered as $q \rightarrow \lceil q \rceil^{-}$, while $\mathbb{Z}_{\lfloor q \rfloor}$ symmetry is recovered as $q \rightarrow \lfloor q \rfloor^{+}$ in case $2$.

For the integer-$q$-state clock model, there is a single second-order phase transition when $q=2,3,4$. When $q\geq 5$, there are two BKT transitions with a critical phase between them. We studied the fractional-$q$-state clock model using Monte Carlo (MC) and tensor renormalization group (TRG) methods. We establish the phase diagram of the model for both $\varphi_0 = 0$ and $\varphi_0 = -\pi$. When $\varphi_0 = 0$, we are in case $1$, and analysis of the finite-size scaling shows a crossover and a phase transition of the Ising universality class. When $\varphi_0 = -\pi$, we are in case $1$ for even $\lfloor q \rfloor$ and in case $2$ for odd $\lfloor q \rfloor$. There are no critical points for case $2$.

In case $1$, we found that there are two peaks in both the specific heat and the magnetic susceptibility. The height of the small-$\beta$ peak is always finite for fractional $q$. The large-$\beta$ peak diverges and characterizes an Ising critical point. When $q \rar \lceil q \rceil^{-}$ and $q < 4$, the large-$\beta$ Ising critical point is smoothly connected to the second-order phase transition point for $\mathbb{Z}_{\lceil q \rceil}$ clock models, while the small-$\beta$ peak fades away. When $q \rar \lceil q \rceil^{-}$ and $q > 4$, the large-$\beta$ Ising critical point and the position of the small-$\beta$ peak are smoothly connected, with the same power-law scaling $\sim (\lceil q \rceil - q)^b$, to the large and small BKT points respectively for $\mathbb{Z}_{\lceil q \rceil}$ clock models. We also found that the height of the small-$\beta$ peak of the magnetic susceptibility diverges as a power law $1/(\lceil q \rceil - q)^{14/15}$, from which we obtain a critical exponent $\delta' = 15$ in the ansatz of the scaling of the magnetization $M \sim \left( \lceil q \rceil - q \right)^{1/\delta'}$. This critical exponent is equal to $\delta$ associated with the magnetization with an external field $M \sim h^{1/\delta}$. In case $2$, there are no critical points. When $q \rar \lfloor q \rfloor ^{+}$ and $q > 5$, the small-$\beta$ peak also goes to the small BKT point with the same power-law scaling and the same $\delta'$ exponent as case $1$, while the large-$\beta$ peak fades away and cannot be used to extrapolate the large BKT point of $\mathbb{Z}_{\lfloor q \rfloor}$ clock models.

To use the magnetic susceptibility to locate a critical point, a weak external field must be applied for the magnetic susceptibility to be finite, and extrapolate the peak position to $h = 0$. This method works in most cases, but the peak fades away near the large-$\beta$ BKT point of integer-$q$-state clock models. Our procedure provides an alternative approach to locate the BKT transitions of clock models, by breaking the $\mathbb{Z}_{\lceil q \rceil}$ symmetry to a $\mathbb{Z}_2$ symmetry in the $q$ direction instead of $h$ direction. This procedure creates an Ising critical point that can be used to extrapolate the large-$\beta$ BKT point for clock models.

Our results clarify what phases the symmetry-breaking term $\gamma \cos(q\varphi_x)$ will drive the system to. These phases should have boundaries in the finite-$\gamma$ direction. For small enough $\gamma$, the extended-O(2) model should go back to the same universality class as the ordinary $XY$ model, which has been studied extensively \cite{Yu:2013sbi, PhysRevE.100.062136, PhysRevB.45.2883, PhysRevB.47.11969, PhysRevB.52.4526, PhysRevE.79.011107, komura:2012, Jha_2020}. For the ordinary $XY$ model, there is a single BKT transition from a disordered phase to a quasi-long-range-ordered critical phase at $\beta_c = 1.11995(6)$ \cite{Jha_2020}. Figure~\ref{phasediagram_3d} shows the work that remains to be done to figure out the phase diagram in the $(\beta, \gamma, q)$ space interpolating between the known phase diagram at $\gamma = 0$ and the phase diagram at $\gamma = \infty$ discussed here. There should be a rich phase diagram in the finite-$\gamma$ region, which is beyond the scope of this work and will be discussed in future work.

It is interesting to note that the BKT critical point found in the O(2) model---and here at the limit of the extended phase diagram---can also be reached through a completely different interpolation.  By considering the O(3) nonlinear sigma model with an additional symmetry breaking term which breaks the O(3) symmetry down to an O(2) symmetry, one can interpolate between  $\mathbb{Z}_{2}$ to O(3), and from O(3) to O(2) by tuning the sign, and magnitude, of the additional symmetry-breaking term~\cite{KLOMFASS1991264}.  Further additional symmetry breaking terms could be interesting.  Positive-definite worm algorithms have been constructed for the O(3) nonlinear sigma model, and could be used to simulate the model efficiently~\cite{WOLFF2010254,PhysRevD.94.114503}.

Topics currently under study include the autocorrelations at different volumes, dynamical critical exponents, spatial correlations, vortices, density of states, and  zeros of the partition function.

\section*{ACKNOWLEDGMENTS}

We thank Gerardo Ortiz, James Osborne, Nouman Butt, Richard Brower, and  members of the QuLAT collaboration for useful discussions and comments. This work was supported in part by the U.S. Department of Energy (DOE) under Awards No. DE-SC0010113 and No. DE-SC0019139.

%

\appendix

\section{Effect of the Angle Cutoff}
\label{app_angle_cut}

In the extended $q$-state clock model, the spins are allowed to take the angles $\varphi^{(k)} = 2\pi k / q$, where $k$ is an integer. By restricting the values of the angle to the domain $[\varphi_0,\varphi_0+2\pi)$, the values of $k$ must satisfy $\varphi_0 q / 2\pi \le k < \varphi_0 q / 2\pi + q$. Let $q = \lfloor q \rfloor + \delta q, \varphi_0 q / 2\pi = p + \epsilon$, $p$ is an integer and $0 \le \epsilon, \delta q < 1$, then
\begin{eqnarray}
\label{eq:valuesofk}
\nonumber k \in
\begin{cases}
(p, p+1, \ldots, p+\lfloor q \rfloor)  & \text{if $\epsilon = 0$}, \\
(p+1, p+2, \ldots, p+\lfloor q \rfloor) & \text{if $\epsilon > 0, \epsilon+\delta q < 1$ }, \\
(p+1, p+2, \ldots, p+\lfloor q \rfloor+1) & \text{if $\epsilon > 0, \epsilon+\delta q \ge 1$ }.
\end{cases} \\
\end{eqnarray}
Because the action only depends on the angular distance, we actually have two cases: $k = 0, 1, 2, \ldots, \lfloor q \rfloor$ (case 1), $k = 0, 1, 2, \ldots, \lfloor q \rfloor-1$ (case 2). This defines a \textit{particular} model that interpolates between the integer $q$'s of the ordinary $q$-state clock model. If $\varphi_0 = 0$, we are in case 1. For this model, the interpolation is smooth (in the sense that the thermodynamic curves change smoothly) as an integer $q$ is approached from below. In this case, the allowed angles feature a cutoff at $\lfloor q \rfloor 2\pi/q$ that breaks the periodicity. One could choose the allowed angles differently and go to case $2$, where the allowed angles feature a cutoff at $2\pi (\lfloor q \rfloor - 1)/q$ that breaks the periodicity, and the interpolation is smooth as an integer $q$ is approached from above.

One could remove the cutoff and restore a form of periodicity by allowing angles
\[ \varphi^{(k)} = \frac{2\pi k}{q}, \qquad k = 0,1,2,\ldots, \infty.\]
However, in this case, for a rational $q=r/s$, where $r/s$ is a reduced fraction, the model is the ordinary $r$-state clock model since $2\pi k/(r/s) = 2\pi ks/r$ and $ks$ is an integer. Thus, removing the cutoff would destroy the smooth interpolation in $n < q \leq n+1$ for integer $n$. For any irrational $q$, this model without cutoff becomes equivalent to the $\infty$-state clock model, that is, the $XY$ model.

\begin{figure}
	\includegraphics[scale=1]{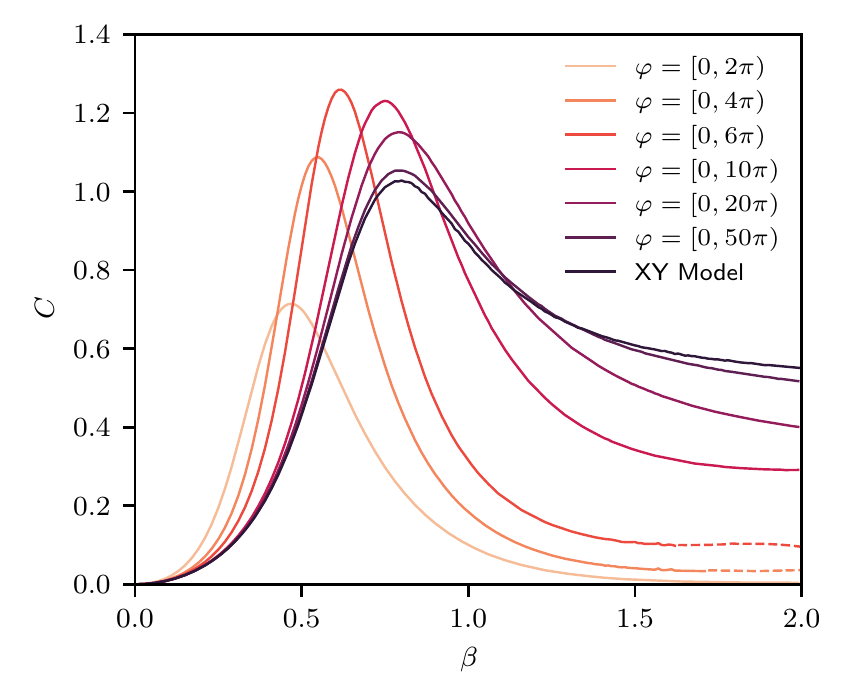}
	\caption{The specific heat for the extended $q$-state clock model with $q=3.141592654$. Results are from Monte Carlo on $4\times 4$ lattices. The different curves correspond to different allowed domains for the spin angles $\varphi$. The specific heat of the $XY$ model is included for reference. As the upper limit of the domain goes to infinity, we get the $XY$ model. Statistical error bars are omitted since they are smaller than the line thickness. Dashed lines indicate regions where we have data but we do not have the uncertainty fully under control. \label{4x4_cv_largek}}
\end{figure}

In Fig.~\ref{4x4_cv_largek} we explore the effect of increasing the cutoff for the model with $q = 3.141592654 \approx \pi$. For example, for the case $\varphi \in [0,2\pi)$, the allowed angles are $2\pi k/\pi = 2k$ with $k = 0, 1, 2, 3$, whereas for the case $\varphi \in [0,4\pi)$, we have $k = 0,1,\ldots, 6$. In all of these cases there remains a $\mathbb{Z}_2$ symmetry, and we expect the model to have an Ising transition at very large $\beta$ for certain cutoffs. As the upper limit of the domain is moved to infinity (i.e. as the cutoff is removed), this Ising transition moves to infinity and the model becomes the $XY$ model. A minor detail is that $3.141592654 = 3141592654/1000000000$ is in fact a rational number, so one would actually get the $3141592654$-state clock model if the cutoff were removed completely. However, it would be indistinguishable from the $XY$ model for practical purposes.

\section{Monte Carlo Results}
\label{sec_appmc}

We used a heatbath algorithm to study the extended $q$-state clock model on a $4\times 4$ lattice. The general structure of the Monte Carlo algorithm is as follows:
\begin{itemize}\itemsep0em
	\item[] \textbf{Equilibration:}
	\item[] \textbf{for} $iequi = 1$ \textbf{to} $nequi$ \textbf{do}
	\item[] $\qquad$heatbath sweep
	\item[] \textbf{end}
	\item[] \textbf{Measurements:}
	\item[] \textbf{for} $irpt = 1$ \textbf{to} $nrpt$ \textbf{do}
	\item[] $\qquad$\textbf{for} $imeas = 1$ \textbf{to} $nmeas$ \textbf{do}
	\item[] $\qquad\qquad$\textbf{for} $idisc = 1$ \textbf{to} $ndisc$ \textbf{do}
	\item[] $\qquad\qquad\qquad$heatbath sweep
	\item[] $\qquad\qquad$\textbf{end}
	\item[] $\qquad\qquad$measure observables
	\item[] $\qquad$\textbf{end}
	\item[] $\qquad$save measurements to file
	\item[] \textbf{end}
\end{itemize}
Parameters used in the primary data production for the extended $q$-state clock model are given in Table~\ref{tab:toymodelparms}. The total number of heatbath sweeps performed (not including equilibration) is $nrpt \times nmeas \times ndisc$. The number of measurements taken is $nrpt \times nmeas$. Ensemble averages and error bars for the energy and magnetization were calculated after binning with $nrpt$ bins each of size $nmeas$. For specific heat and magnetic susceptibility, the measurements were binned and jackknifed.

\begin{figure*}
	\includegraphics[scale=1]{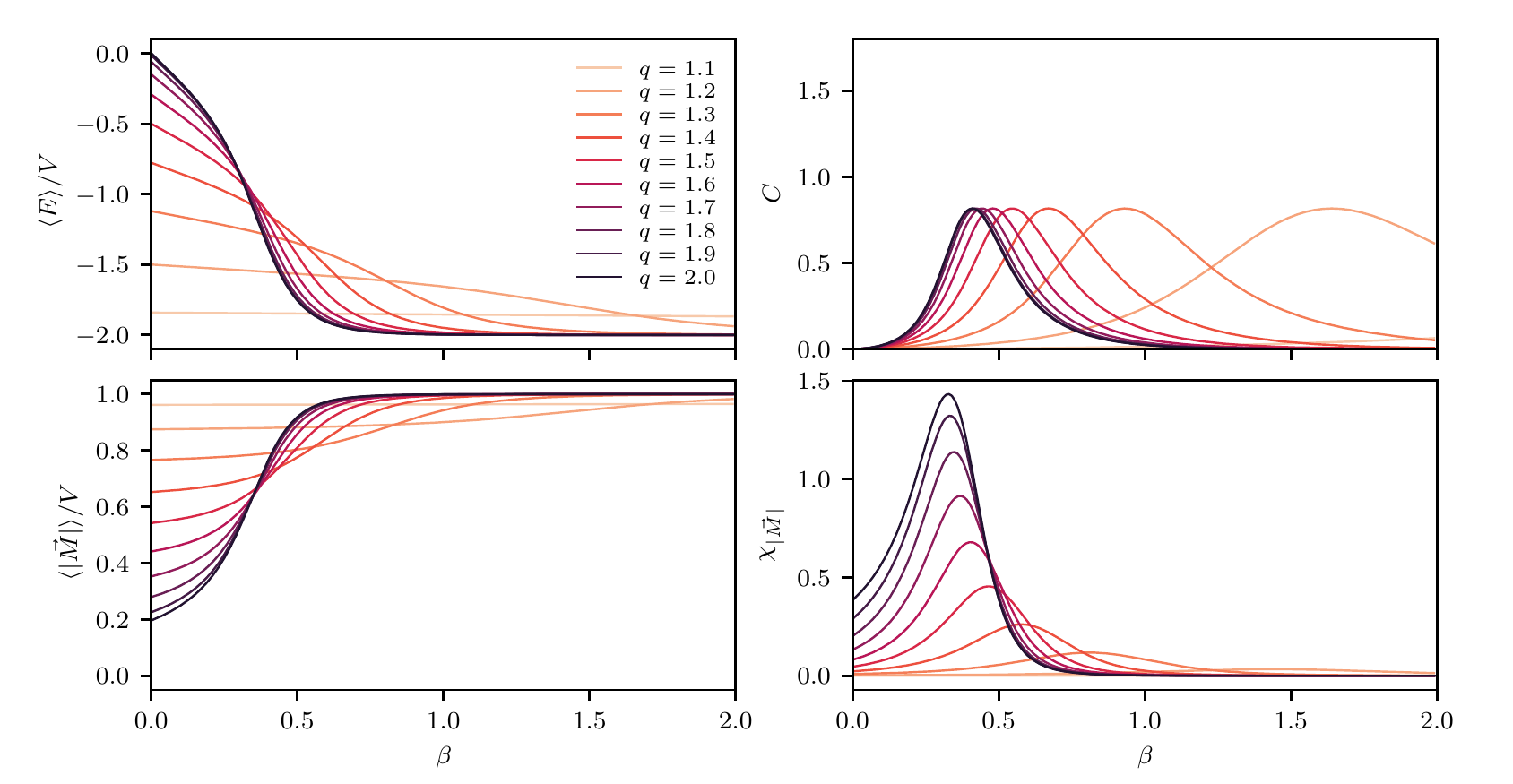}
	\caption{Monte Carlo results for the extended $q$-state clock model on a $4\times 4$ lattice for $1.1 \leq q \leq 2.0$. The top panel shows energy density and specific heat, and the bottom panel shows proxy magnetization and magnetic susceptibility. Statistical error bars are omitted since they are smaller than the line thickness. Dashed lines indicate regions where we have data but we do not have the uncertainty fully under control. \label{zz_ones_all_0004_0p00_23_1_encv}}
\end{figure*}

\begin{figure*}
	\includegraphics[scale=1]{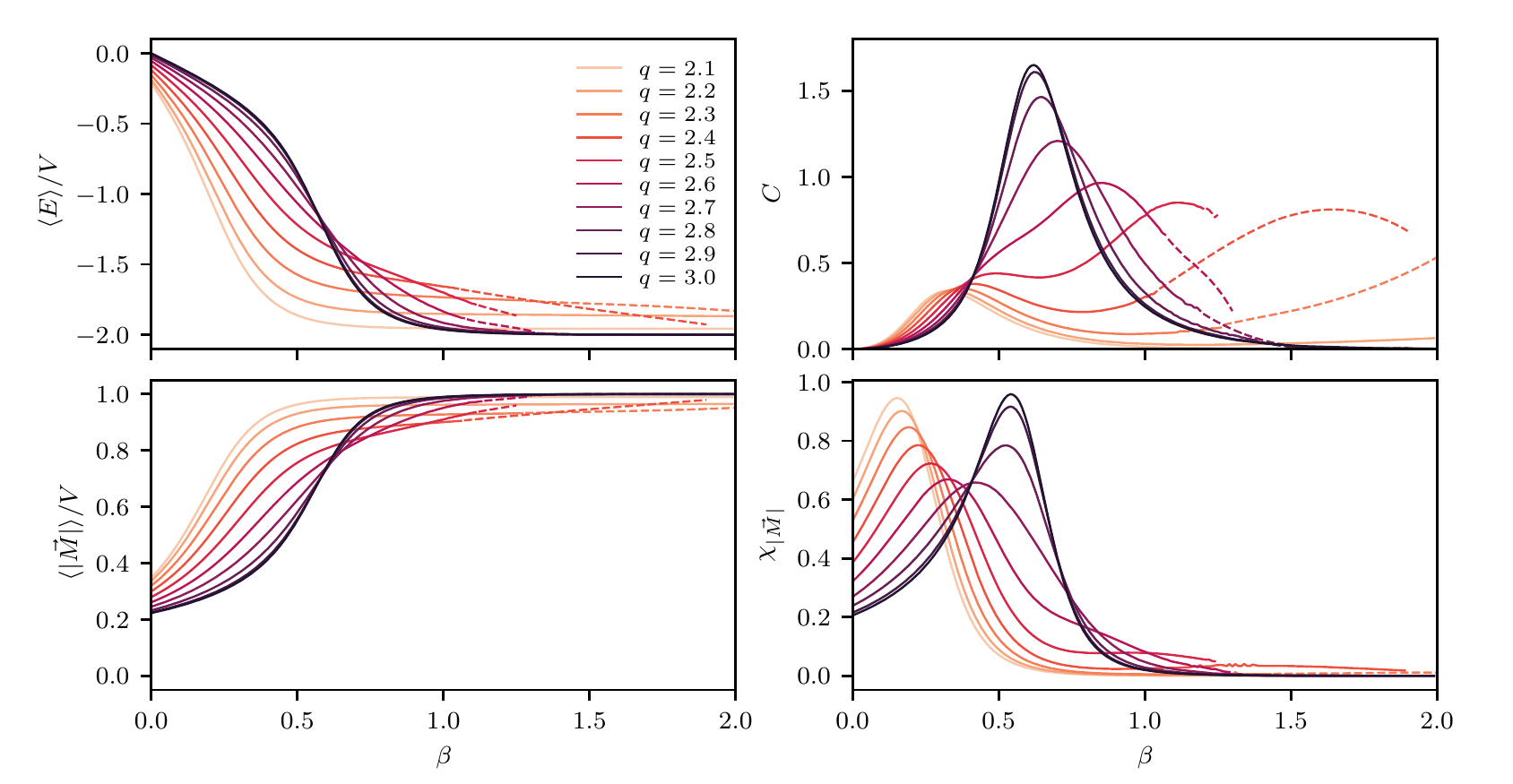}
	\caption{Monte Carlo results for the extended $q$-state clock model on a $4\times 4$ lattice for $2.1 \leq q \leq 3.0$. The top panel shows energy density and specific heat, and the bottom panel shows proxy magnetization and magnetic susceptibility. Statistical error bars are omitted since they are smaller than the line thickness. Dashed lines indicate regions where we have data but we do not have the uncertainty fully under control. \label{zz_twos_all_0004_0p00_23_1_encv}}
\end{figure*}

\begin{figure*}
	\includegraphics[scale=1]{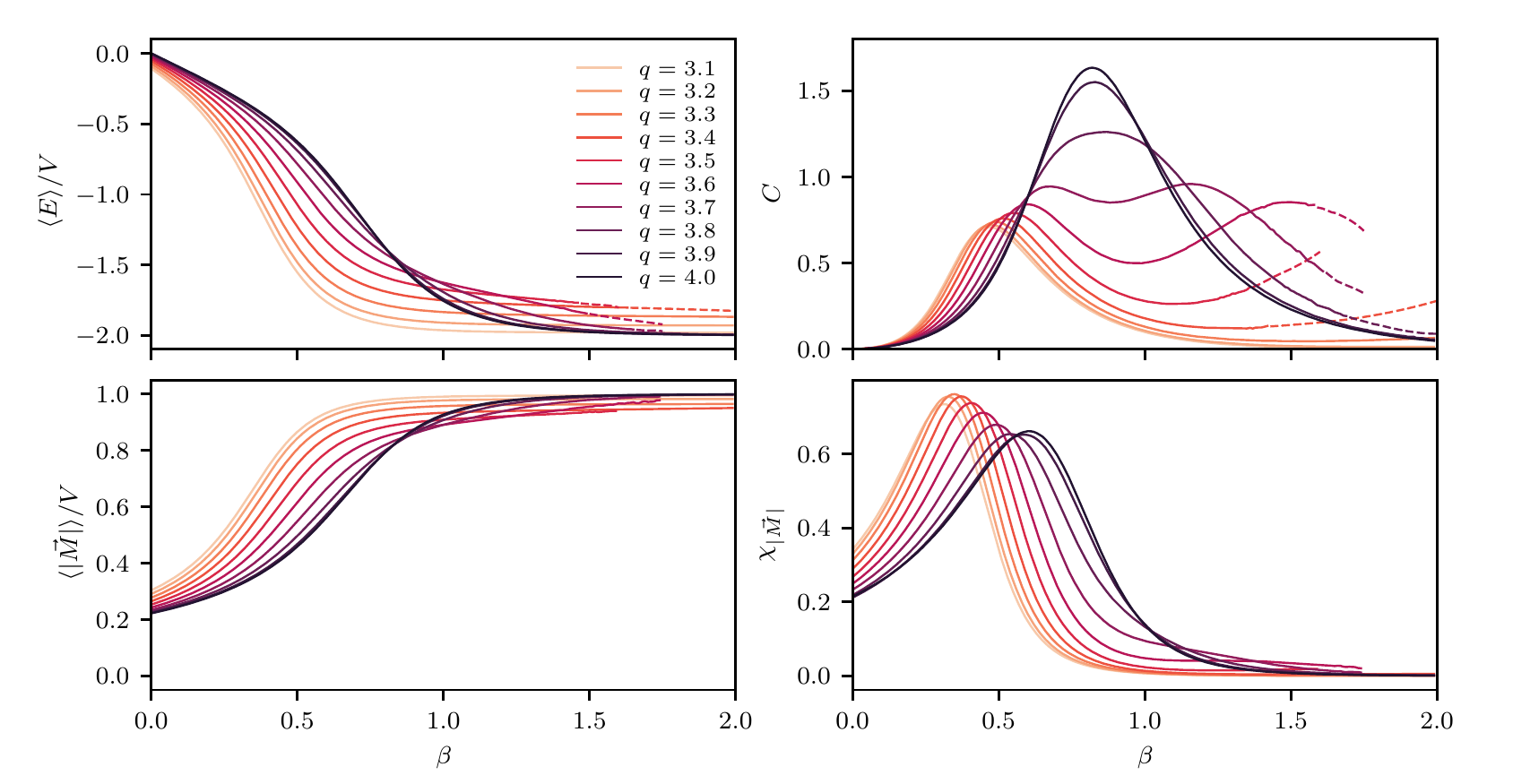}
	\caption{Monte Carlo results for the extended $q$-state clock model on a $4\times 4$ lattice for $3.1 \leq q \leq 4.0$. The top panel shows energy density and specific heat, and the bottom panel shows proxy magnetization and magnetic susceptibility. Statistical error bars are omitted since they are smaller than the line thickness. Dashed lines indicate regions where we have data but we do not have the uncertainty fully under control. \label{zz_threes_all_0004_0p00_23_1_encv}}
\end{figure*}

\begin{figure*}
	\includegraphics[scale=1]{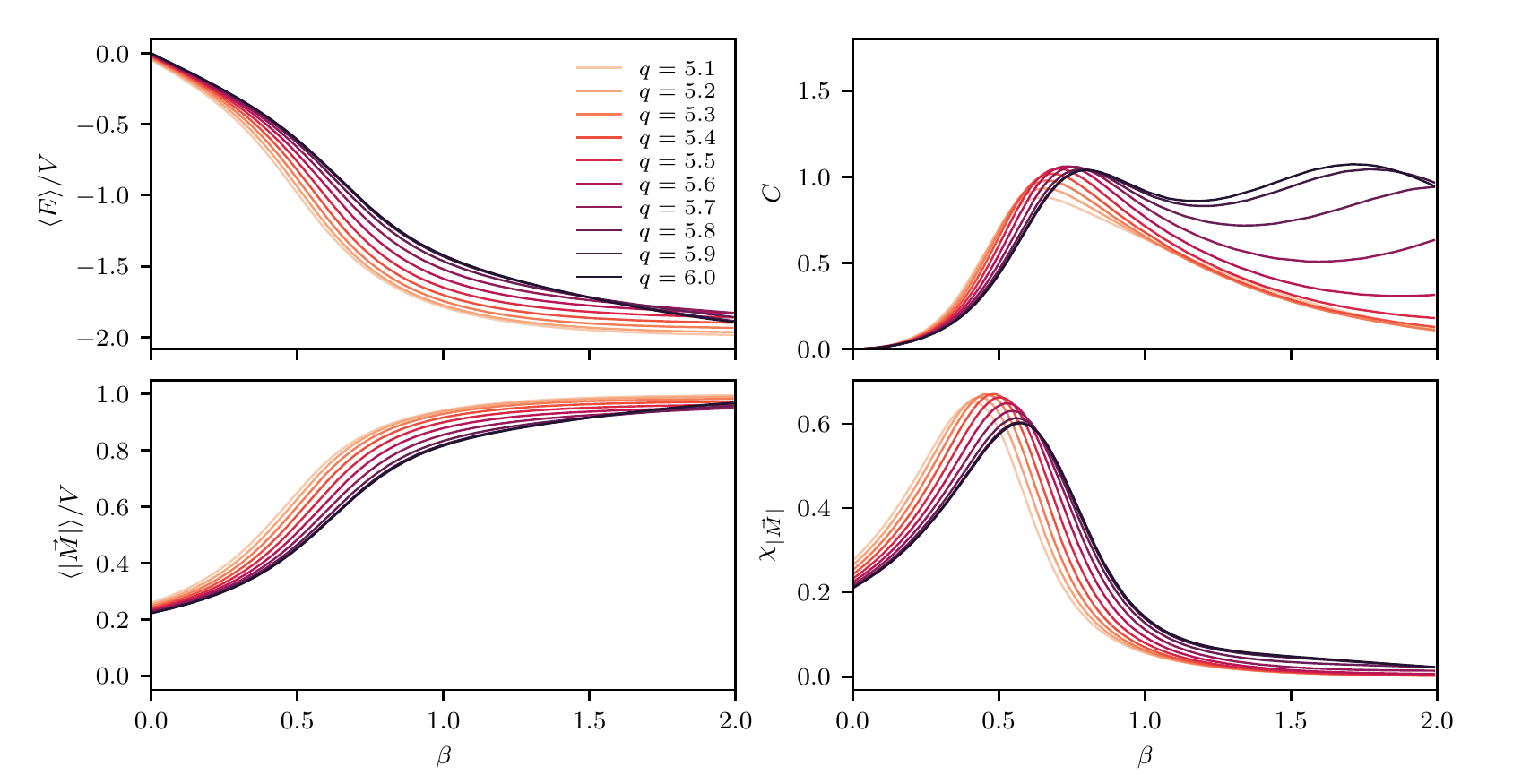}
	\caption{Monte Carlo results for the extended $q$-state clock model on a $4\times 4$ lattice for $5.1 \leq q \leq 6.0$. The top panel shows energy density and specific heat, and the bottom panel shows proxy magnetization and magnetic susceptibility. Statistical error bars are omitted since they are smaller than the line thickness. \label{zz_fives_all_0004_0p00_23_1_encv}}
\end{figure*}

We studied the extended $q$-state clock model on a $4\times 4$ lattice with zero external magnetic field. For each $\beta$, we initialized to a random lattice (hot start) then we performed $2^{15}$ equilibrating sweeps followed by $2^{22}$ measurement sweeps. Each measurement sweep was followed by $2^8$ discarded sweeps. We calculated the energy density and specific heat as defined in Eqs.~(\ref{eq_internalenergy}) and (\ref{eq_specificheat}). We calculated the proxy magnetization and susceptibility as defined in Eqs.~(\ref{eq_proxymag}) and (\ref{eq_proxyc}). Results for $1< q< 6$ are shown in Figs.~\ref{zz_fours_all_0004_0p00_23_1_encv}, \ref{zz_ones_all_0004_0p00_23_1_encv}--\ref{zz_fives_all_0004_0p00_23_1_encv}.

\begin{table}[h!]
	\centering
	\setlength{\tabcolsep}{10pt}
	\begin{tabular}{ll}
		$q$ & $[1.1,6.0]$ with $\Delta q = 0.1$ \\ \hline
		$\beta$ & $[0.0,2.0]$ with $\Delta\beta=0.01$ \\ \hline
		$H$ & 0.0 \\ \hline
		Lattice & $4\times 4$ \\ \hline
		Start type & Random (hot) \\ \hline
		$nequi$ & $2^{15}$ \\ \hline
		$nrpt$ & $2^6$ \\ \hline
		$nmeas$ & $2^{16}$ \\ \hline
		$ndisc$ & $2^8$ \\ \hline
	\end{tabular}
	\caption{The Monte Carlo parameters used in the primary data production for the extended $q$-state clock model. See Figs.~\ref{zz_fours_all_0004_0p00_23_1_encv} and \ref{zz_ones_all_0004_0p00_23_1_encv}--\ref{zz_fives_all_0004_0p00_23_1_encv}). Here, $q$ refers to the $q$-state clock model, $H$ is the external magnetic field, $nequi$ is the number of equilibrating heatbath sweeps used, $nrpt$ is the repetitions of measurement sweeps, $nmeas$ is the number of measurement sweeps, and $ndisc$ is the number of sweeps discarded between each measurement sweep.}
	\label{tab:toymodelparms}
\end{table}

\begin{figure}
	\includegraphics[scale=0.8]{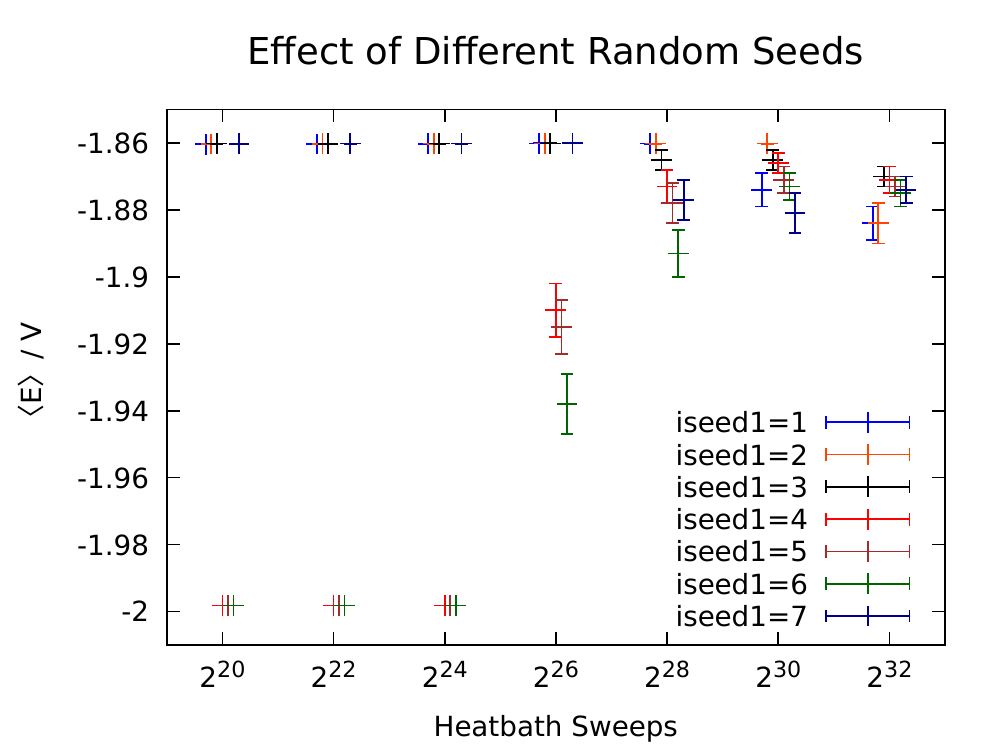}
	\caption{The measured energy density for $q=4.5$ at different number of heatbath sweeps for different random-number-generator (RNG) seeds. The vertical axis is the energy density. The horizontal axis gives the number of heatbath sweeps used (not including equilibration sweeps). The different colors indicate runs with different seeds for the RNG. All runs were performed at $\beta = 2.5$ with hot-started $4\times 4$ lattices. This shows that the configuration space consists of two distinct sectors, and the heatbath algorithm has difficulty appropriately sampling both sectors unless the number of heatbath sweeps is taken very large. \label{ck_mcslowdownck_randrand_beta2p52}}
\end{figure}

In the large-$\beta$ (low temperature) regime, the heatbath algorithm has difficulty appropriately sampling the configuration space. At large-$\beta$, the lattice freezes along a particular magnetization direction. When $q\in\mathbb{Z}$ all magnetization directions are equivalent, however, when $q\notin\mathbb{Z}$, the discrete rotational symmetry is broken and the direction of magnetization matters. The configuration space is split into two sectors with different thermodynamic properties. In one sector, the magnetization is in the direction 0 or $-\tilde{\phi}$, defined in Eq.~(\ref{small_angle}), where relatively large fluctuations may still be possible. In the other sector, the magnetization is in one of the other directions where fluctuations are less likely. To appropriately sample the configuration space one has to use very large statistics or run multiple heatbath streams at the same parameters but with different seeds for the random number generator. In  Fig.~\ref{ck_mcslowdownck_randrand_beta2p52}, we show an example of this phenomenon for $q=4.5$ and $\beta=2.5$ on a $4\times 4$ lattice. In this example, we need $\gtrsim 2^{32}$ heatbath sweeps to adequately sample both sectors. This Monte Carlo slowdown makes it difficult to study larger lattices, and is a strong motivation for using TRG.

\begin{figure}
	\includegraphics[scale=1]{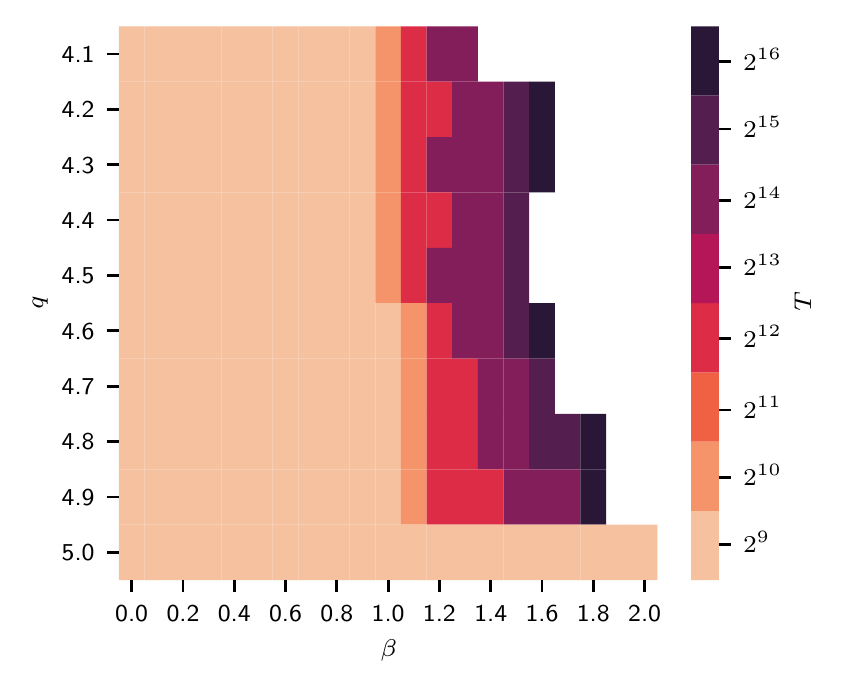}
	\caption{The values of $T$ defined in Eq.~(\ref{eq_ace}) needed to extract the integrated autocorrelation times shown in Fig.~\ref{zz_all_all_0004_0p00_var_1_ac2}. The vertical axis gives the values of $q$ and the horizontal axis gives the values of $\beta$. Blank/white regions in this heatmap indicate cases where $T>2^{16}$ is needed to get a reliable estimate of the integrated autocorrelation time. These were not attempted due to the computational cost. \label{ck_4x4_ace_table}}
\end{figure}

The Monte Carlo slowdown is illustrated by an explosion of the integrated autocorrelation time in the intermediate-$\beta$ regime. For an observable $O$, an estimator of the integrated autocorrelation time is given by
\beq
\label{eq_ace}
\tilde{\tau}_{O,int} = 1+ 2\sum_{t=1}^T \frac{C(t)}{C(0)}
\enq
where $C(t) = \langle O_i O_{i+t}\rangle - \langle O_i\rangle \langle O_{i+t}\rangle$ is the correlation function between the observable $O$ measured at Markov times $i$ and $i+t$. The integrated autocorrelation time $\tau_{O,int}$ is estimated by finding a window in $t$ for which $\tilde{\tau}_{O,int}$ is nearly independent of $t$. The integrated autocorrelation time for $4< q\leq 5$ is shown in the main document in Fig.~\ref{zz_all_all_0004_0p00_var_1_ac2}. The values of $T$ needed to extract these points are given in Fig.~\ref{ck_4x4_ace_table}.

To mitigate the effect of autocorrelation in our results, we discarded $2^8$ heatbath sweeps between each saved measurement. The saved measurements were then binned (i.e. preaveraged) with bin size $2^{16}$ before calculating the means and variances.

\section{Validating TRG with MC}
\label{sec_appvalidation}

Whereas Monte Carlo methods are well understood in the context of classical spin models, TRG is a relatively new approach. We validate TRG results at small $\beta$ and small volume using exact and Monte Carlo results. Exact results can be computed for $q=2$ (Ising model) and $q=4$ (two coupled Ising models). We use Monte Carlo to validate the TRG results for other (including fractional) values of $q$.

\begin{figure}
	\includegraphics[scale=0.99]{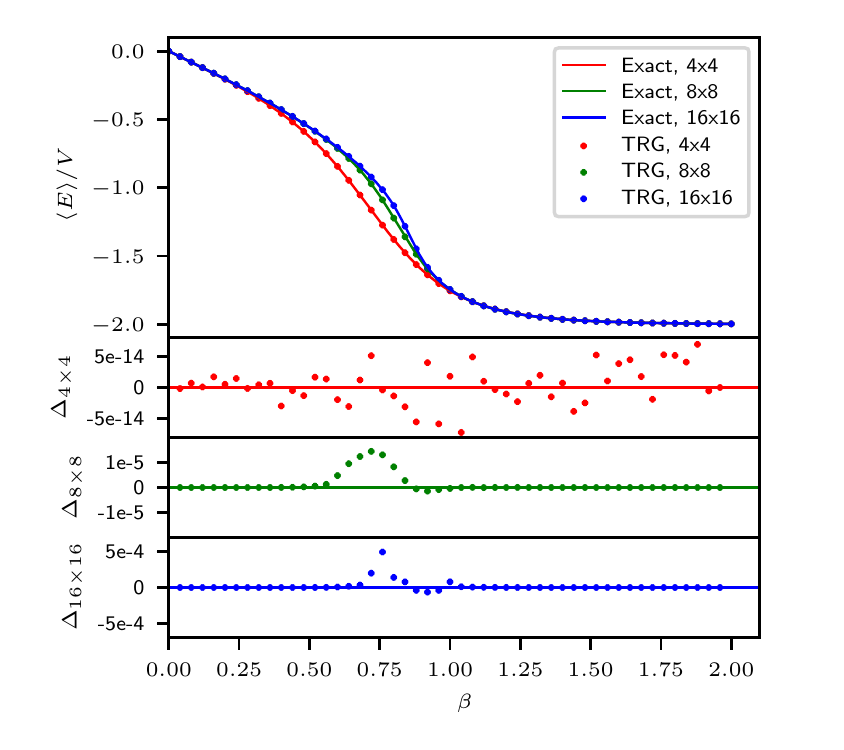}
	\caption{An example comparison of the energy density from exact calculation and that from TRG for $q=4.0$. The top panel shows the energy density for lattices of size $4\times 4$, $8\times 8$, and $16\times 16$. The three panels on the bottom show the difference between exact and TRG results. TRG shows deviations from exact results near the phase transition. Here $D_{\rm{bond}} = 64$. \label{mcvtrg_4p0_exact_en}}
\end{figure}

\begin{figure}
	\includegraphics[scale=0.99]{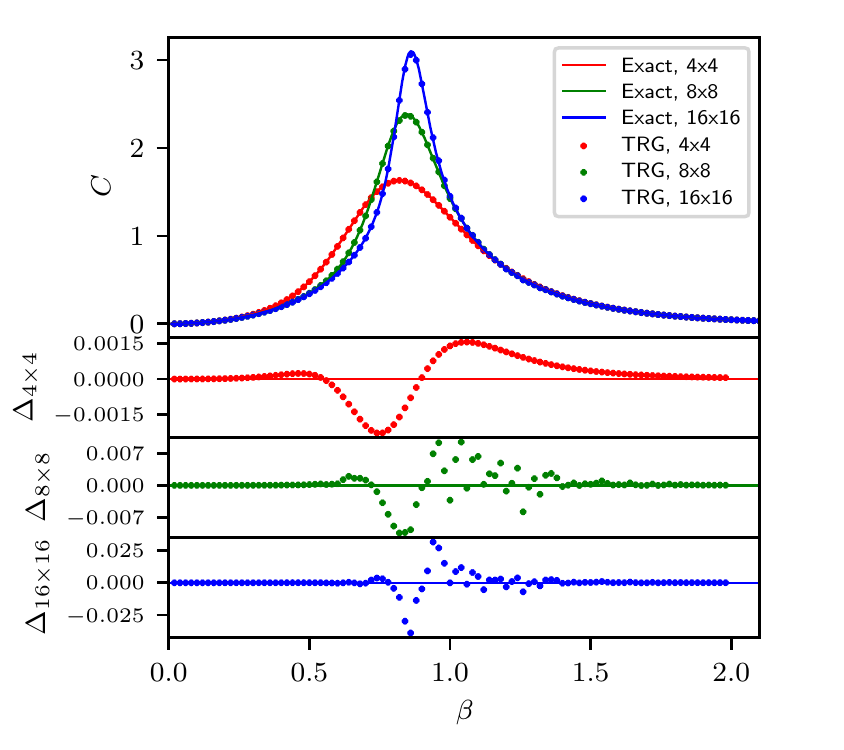}
	\caption{ Same as Fig.~\ref{mcvtrg_4p0_exact_en}, but for the specific heat. TRG shows deviations from exact results near the peak, however, this is mostly due to discretization error from the derivative. \label{mcvtrg_4p0_exact}}
\end{figure}

In Fig.~\ref{mcvtrg_4p0_exact_en}, we show that the energy density from TRG agrees very well with the exact calculation, and that the tiny discrepancy appears only around the critical point. In TRG, the specific heat is calculated by taking a finite difference derivative of the energy. In Fig.~\ref{mcvtrg_4p0_exact}, we compare the specific heat from TRG with the exact values for $q=4.0$. TRG deviates from the exact results near the peak of the specific heat, but this deviation is mostly due to the discretization error from the derivative.

\begin{figure}
	\includegraphics[scale=0.99]{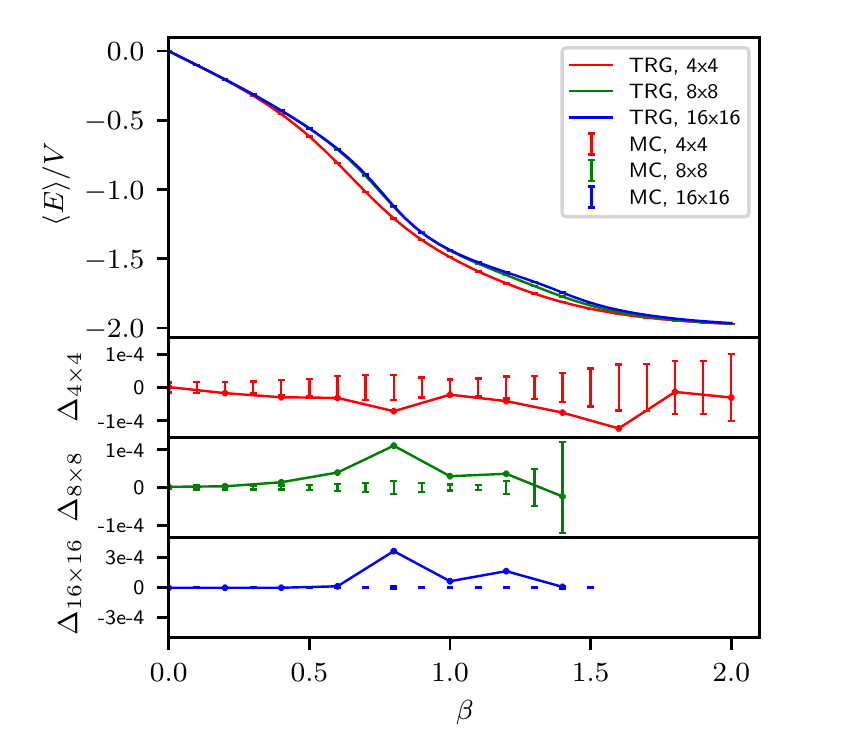}
	\caption{An example comparison of TRG and Monte Carlo measurements of the energy density for $q=4.9$. The top panel shows the energy density for lattices of size $4\times 4$, $8\times 8$, and $16\times 16$. The three panels on the bottom show the difference between Monte Carlo and TRG results. Here, Monte Carlo is taken to be the baseline. Here $D_{\rm{bond}} = 64$. \label{mcvtrg_4p9_en}}
\end{figure}

\begin{figure}
	\includegraphics[scale=0.99]{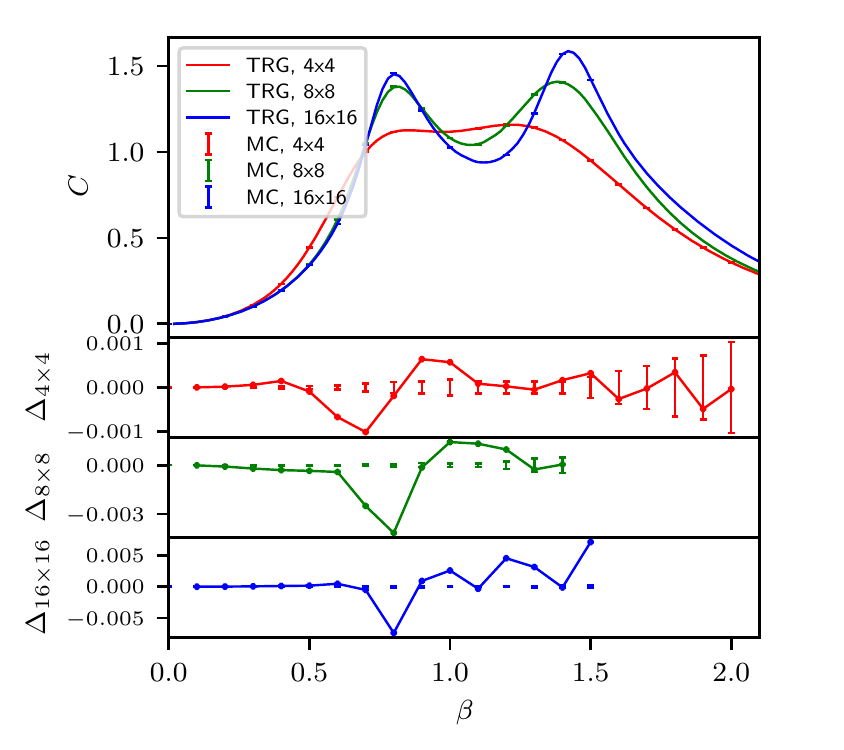}
	\caption{ Same as Fig.~\ref{mcvtrg_4p9_en}, but for the specific heat. TRG shows deviations from Monte Carlo results near the peak, however, this is mostly due to discretization error from the derivative. \label{mcvtrg_4p9}}
\end{figure}

Exact solutions are not known for fractional-$q$, so we validate TRG by comparing with results from Monte Carlo at small $\beta$ and small volume. For example, Fig.~\ref{mcvtrg_4p9_en} shows that the discrepancy between the energy density from TRG and that from Monte Carlo is only of order $10^{-4}$. In Fig.~\ref{mcvtrg_4p9}, we compare the specific heat from TRG with that of Monte Carlo for $q=4.9$. TRG deviates from the Monte Carlo results near the peak of the specific heat. However, this deviation is again almost entirely due to discretization error from the derivative.

\end{document}